\documentclass[12pt, preprint,numberedappendix]{emulateapj}
\usepackage{rotating}
\usepackage{booktabs,threeparttable}
\usepackage{morefloats}
\usepackage{gensymb}
\usepackage{amsmath,amsthm,amssymb} % for math
\usepackage[utf8]{inputenc}
\usepackage[english]{babel}
 \usepackage[usenames, dvipsnames]{color}

\usepackage{epstopdf}
\shorttitle{Slow Winds in TTS: Forbidden Line Emission}
\shortauthors{Simon et al.}

\newcommand{\neii}{[Ne~{\sc ii}]}
\newcommand{\oi}{[O~{\sc i}]}
\newcommand{\oii}{[O~{\sc ii}]}
\newcommand{\sii}{[S~{\sc ii}]}

\begin{document}

%\onecolumn

%% LaTeX will automatically break titles if they run longer than
%% one line. However, you may use \\ to force a line break if
%% you desire.

\title{Tracing Slow Winds from T Tauri Stars via Low Velocity Forbidden Line Emission}

\author{M. N. Simon\altaffilmark{1},
I. Pascucci\altaffilmark{1,2,3},
S. Edwards\altaffilmark{4,2},
W. Feng\altaffilmark{5},
U. Gorti\altaffilmark{6,7},
D. Hollenbach\altaffilmark{6},
E. Rigliaco\altaffilmark{8,2},
J. T. Keane\altaffilmark{1}}

\altaffiltext{1}{Lunar and Planetary Laboratory, University of Arizona, Tucson, AZ 85721, USA}
\altaffiltext{2}{Visiting Astronomer, Keck Observatory}
\altaffiltext{3}{Earth in Other Solar Systems (EOSS) team}
\altaffiltext{4}{Five College Astronomy Department, Smith College, Northampton, MA 01063, USA}
\altaffiltext{5}{School of Earth and Space Exploration, Arizona State University, Tempe, AZ 85287, USA}
\altaffiltext{6}{SETI Institute, Mountain View, CA 94043, USA}
\altaffiltext{7}{NASA Ames Research Center, Moffett Field, CA 94035, USA}
\altaffiltext{8}{Institute for Astronomy, ETH Zurich, Wolfgang-Pauli-Strasse 27, CH-8093 Zurich, Switzerland}

\begin{abstract}
%Broad and blueshifted optical forbidden lines are tracers of outflowing gas from accreting T Tauri stars. 
Using Keck/HIRES spectra ($\Delta v \sim$ 7 km/s) we analyze forbidden
lines of \oi{} 6300 \AA, \oi{} 5577 \AA~and \sii{} 6731 \AA~from 33 T Tauri stars covering a range of disk evolutionary stages. After removing
a high velocity component (HVC) associated with microjets, we study the
properties of the low velocity component (LVC). The LVC can be
attributed to slow disk winds that could be magnetically (MHD) or
thermally (photoevaporative) driven. Both of these winds play an important
role in the evolution and dispersal of protoplanetary material.

LVC emission is seen in all 30 stars with detected \oi{} but only in 2 out of eight
with detected \sii{}, so our analysis is largely based on the properties of
the \oi{} LVC. The LVC itself is resolved into broad (BC) and narrow
(NC) kinematic components. Both components are found over a wide range of accretion rates and their
luminosity is correlated with the accretion luminosity, but the NC is proportionately stronger than the BC in transition disks.

The FWHM of both the BC and NC correlates with disk inclination, consistent with Keplerian broadening from radii of 0.05 to 0.5 AU and 0.5 to 5AU, respectively. The velocity centroids of the BC suggest
formation in an MHD disk wind, with the largest blueshifts found in sources with closer to
face-on orientations. The velocity centroids of the NC however, show no dependence on disk inclination. The origin of this component is less clear and the evidence for photoevaporation is not conclusive. 
\end{abstract}

\section{Introduction}
Low excitation forbidden lines of \oi{} and \sii{} are some of the defining spectroscopic characteristics of the low-mass, pre-main sequence stars known as T Tauri stars \citep[TTS,][]{herbig1962}. Their broad, blueshifted emission profiles were originally interpreted as arising in winds with receding flows occulted by the circumstellar disk \citep{App1984, edwards1987}, and the correlation between their luminosity and the luminosity of infrared emission from the disk demonstrated that the forbidden line emission was powered by accretion \citep{cabrit1990}. Studies of large samples of TTS conducted by \cite{hamann1994}, \citealt{HEG95} (hereafter HEG), and \cite{hirth1997} showed that forbidden lines in these stars are characterized by two distinct components: a high velocity component (HVC) and a low velocity component (LVC).

The HVC forbidden emission, typically blueshifted by 30 to 150 km/s, was demonstrated to be formed in microjets, small-scale analogs to the parsec-long collimated jets emerging from more embedded Class I young stellar objects (YSOs) \citep{ray2007, hartigan94}. A correlation between mass loss rates, derived from the luminosity of the HVC in the strongest \oi{} 6300 {\AA} forbidden line and the accretion luminosity and/or disk accretion rate has been the foundation of the accretion-outflow connection in young accreting stars (HEG, \citealt{cabrit2007}). The origin of the outflows traced by the HVC is not yet fully understood, but is likely tied to mass and angular momentum loss in the accretion disk and/or the accreting star through magnetized magnetohydrodynamic (MHD) winds \citep{ferreira2006}.

The origin of the LVC forbidden emission is even less well understood. HEG found the LVC to be present in all TTS with near infrared (NIR) excess at K-L (Class II sources, e.g. \citealt{lada1984}), typically with small blueshifts $\sim$ 5 km/s. They considered the possibility the LVC might arise in a slow wind on the surface of a disk in Keplerian rotation with the LVC surface brightness decreasing as $\sim r^{-2.2}$. This possibility was investigated more thoroughly by \cite{kwan1995} who used line luminosities and line ratios to evaluate the physical conditions in the wind, and estimated disk wind mass loss rates as $\sim 10^{-8} M_\odot$/yr. The possibility that the LVC emission might trace thermally driven disk winds powered by photoevaporative heating due to extreme ultraviolet (EUV) radiation from the central accreting star was investigated by \cite{font2004}. However, the EUV heated flows produced very little neutral oxygen and thus could not account for the LVC observations.

More recently, the growing realization that photoevaporative flows are likely an important means of disk dispersal during the era of planet formation has led to renewed interest in additional sources of heating for photoevaporation, i.e. X-ray and far ultraviolet (FUV) radiation. These have the potential of producing significant amounts of neutral oxygen in the flow and accounting for the LVC forbidden emission \citep{hg09,ercolano2010, ercolano2016, gorti2011}. 

Another new development is the acquisition of high resolution spectra of \neii{} at 12.8\,\micron{}. To date, 24 high-resolution profiles of \neii{} in TTS have been acquired, showing, like the optical \oi{} profiles, a mixture of high velocity and low velocity components. Low velocity components of \neii{}, blueshifted from  -2 to -12 km/s, are seen in 13/24 of the high resolution spectra, and have been interpreted as direct tracers of photoevaporative flows (see the recent review by \citealt{alexander2014} and references therein). Although the ionization required to produce substantial \neii{} emission in a photoevaporative flow could arise either from EUV or X-ray heating, \cite{pascucci2014} compared the measured \neii{} luminosities with upper limits on the EUV radiation reaching the disk and demonstrated that, at least in three systems, this emission probes the X-ray rather than EUV-ionized surface. Due to the fact that X-rays penetrate deep in the disk and drive flows that are mostly neutral,  blueshifted \neii{} emission signals mass loss rates that are considerably higher than if the heating is from EUV, possibly as high as  $\sim 10^{-8}$\,M$_\sun$/yr. Such mass loss rates are comparable to those estimated by \citet{kwan1995} from the LVC optical lines. If these mass loss rates are characteristic of photoevaporative flows, their similarity to TTS disk accretion rates (e.g. \citealt{alcala2014}), suggest that photoevaporation may play a major role early in the evolution and dispersal of protoplanetary material. Photoevaporation can drastically change the disk surface density by opening gaps in planet-forming regions that widen with time, thus setting the timescale over which [giant] planet formation must occur. 
The implication this would have on the final architecture of planetary systems, the chemistry of the disk, planet interactions (such as the delivery of volatiles to planets located in the inner solar system), and the final mass and location of giant planets specifically, would be critical to enhancing our understanding of planet formation more generally.

A better understanding of the empirical properties of the forbidden LVC in TTS is needed to assess whether it arises in photoevaporative flows. Two recent works, \citealt{rigliaco13}, (hereafter R13) and \citealt{natta2014}, (hereafter N14) have begun this process and provide the motivation for the present study. R13 introduced two changes in interpreting the LVC. Firstly, R13 revisited the forbidden line data from HEG with modern estimates for the accretion luminosity. \cite{gullbring98} demonstrated that the mass accretion rates in HEG were too large by nearly an order of magnitude due to uncertain bolometric corrections to the continuum excess measured in the R band, and from the simplistic assumption that the accretion luminosity was produced in a boundary layer at the stellar surface. R13 re-derived the accretion luminosity for 30 of the HEG stars from H$\alpha$ emission lines, observed simultaneously with the forbidden lines, which appeared in \cite{beristain01}. They found correlations between the LVC luminosity from HEG with the improved accretion luminosities as well as with published values for FUV and stellar luminosities, but not with literature values for the X-ray luminosity. This led to the suggestion that {\it if} the LVC is from a photoevaporative flow, FUV heating may be more important than X-ray radiation in generating \oi{} emission, although if \neii{} emission were also present then X-rays would need to be a contributor as well. A further implication would be that photoevaporative flows might be prevalent throughout the T Tauri phase, decreasing in proportion to the disk accretion rate. 

Secondly, R13 demonstrated, with a small set of new high resolution spectra, that the LVC itself may have two kinematic components. Focusing on only two \oi{} LVC that were well separated from any HVC emission, the profiles were decomposed into  a broad and a narrow contribution, with the suggestion that the broad feature may be formed in bound material in the disk and rotationally broadened, while the narrow feature may be associated with material from further out in the disk, possibly in a photoevaporative flow.  

The study of N14 provided further refinement of the definition of the LVC for 44 TTS in Lupus and $\sigma$ Ori. In HEG the LVC was defined very simply, assigning any emission within 60 km/s of line center to the LVC and outside of that to the HVC, using profiles from echelle spectra with a velocity resolution of 12 km/s. N14 improved on this by applying Gaussian fitting techniques to separate HVC from LVC emission, although their low spectral resolution of 35 km/s meant that a clear separation between these two components was not always possible, nor could they resolve the LVC into a broad and narrow component as done in R13. Nevertheless, they found good correlations between the total LVC luminosity and both the stellar luminosity and the accretion luminosity, but not the X-ray luminosity. They interpret the LVC as arising in a slow wind ($<20$ km/s) that is dense (n$_{\rm H}>10^8$\,cm$^{-3}$), warm (T $\sim 5,000 -10,000$\,K), and mostly neutral. 

In this paper, we continue the investigation of the empirical properties of the LVC forbidden lines in TTS in order to elucidate their origin. While the possibility that they may provide a direct tracer of photoevaporative flows that are responsible for protoplanetary disk clearing is intriguing, other potential contributors are the base of MHD centrifugal winds and heated, bound gas in the disk itself. Our study differs from previous ones in two major ways. First, our sample of TTS spans a larger range of evolutionary stages when compared to the sample of HEG (Sects.~\ref{sect:sample} and \ref{sect:accr}). Secondly, using Keck/HIRES we reach a spectral resolution that is more than five times higher than N14 and about two times higher than HEG (see Section~\ref{sect:obs}). This high spectral resolution enables us to define the kinematic structure of the LVC, where about half the detected sources show two kinematic components and the remaining LVC are separated into either broad or narrow profiles (Sects.~\ref{sect:Profiles} and \ref{sect:res_LVC}).  We discuss the relation of these components to MHD and photoevaporative winds in Section~\ref{sect:Discussion}.

\section{Observations and Data Reduction}\label{sect:Obs_and_Redu}

For this project, we focus on the kinematic properties of several forbidden lines in T Tauri stars: \oi{}  6300.304 {\AA},  \oi{}  5577.339 {\AA}, and \sii{} 6730.810 {\AA}. We also searched for \oii{} 7329.670 {\AA} but did not detect it in any source. In addition, we use H$\alpha$ as a tracer of the accretion luminosity and disk accretion rate \citep{alcala2014}. In the following subsections we describe our sample (\ref{sect:sample}), the observations (\ref{sect:obs}), the methodology to create forbidden line profiles free from telluric and photospheric absorption (\ref{sect:line_profiles}), and the evaluation of  line equivalent widths and upper limits (\ref{sect:EWs}).

\subsection{Sample}\label{sect:sample}
		
All of our science targets, with the exception of TW Hya, belong to the well-characterized star-forming region of Taurus-Auriga (age $\sim$1 Myr and average distance 140 pc; \citealt{kenyon08}). Our sample represents the spread of disk properties and disk accretion rates, but not the statistical distribution of mass, age, or other properties in this star formation region. The required high S/N precluded observing late M dwarfs and brown dwarfs, and disk inclination was not one of the criteria used to select our targets. Our sample is presented in Table \ref{tab:source_properties} and consists of 33 mostly single and bright (B$\ge$16) T Tauri stars (TTS) with disk spectral energy distributions encompassing: 26 Class II, full disks with optically thick inner regions; 5 transition disks, with absent or low NIR to mid-infrared (MIR) excess emission from the inner disk but large far-infrared (FIR) emission; and, 2 Class II/III evolved disks with weak NIR to FIR excess.  Additional information on this sample can be found in \cite{pascucci2015}, who analyzed the Na D and K profiles (at 5889.95 and 7698.96 {\AA}, respectively) from the same spectra presented here. The spectral types (SpTy), extinctions  ($A_{\rm v}$), stellar bolometric luminosities ($L_{\ast}$), and stellar masses ($M_{\ast}$) in Table \ref{tab:source_properties} are all taken from \cite{HH14} who derived them in a homogeneous way from spectrophotometric data, while taking into account excess continuum emission (veiling) and extinction when deriving the stellar properties.\footnote{The only source that did not have a stellar mass reported in \cite{HH14} is HN Tau. The mass reported in Table \ref{tab:source_properties} is taken from \cite{keane2014}.}

Table \ref{tab:source_properties} also gives disk inclinations for 22 stars taken from the literature based on spatially resolved disk images, shown by \cite{App2013} to be the most reliable means of determining system orientations. For most sources uncertainties are reported to be $\sim$ 10\%. However, for one source, DR Tau, nominally reliable techniques for estimating inclination range from close to face-on to almost edge-on and suggest that the inner and outer disk ($>$10\,AU) have different orientations \citep{banzatti2015}. Resolved millimeter continuum images point to a highly inclined outer disk ($\sim$$70^{\circ}$, \citealt{AW07} and $\sim$$35^{\circ}$, \citealt{isella09}). However, MIR interferometric visibilities coupled with spectral energy distribution (SED) fitting, as well as modeling of the spectroastrometic signal in the CO ro-vibrational line, suggest a much smaller inclination for the inner disk ($\sim 20^{\circ}$, \citealt{scheg09} and $\sim9^{\circ}$, \citealt{pontoppidan2011}). Additional evidence for a small inclination to the inner disk comes from exceptionally deep and broad sub-continuum blueshifted absorption at both He I 10830 {\AA} and H$\alpha$ \citep{edwards2003}, requiring absorption along a line of sight through a wind that emerges radially from the stellar poles.  From these considerations, we adopt a disk inclination of $20^{\circ}$ for DR Tau, since, as will be shown in Section \ref{sect:inclination}, the LVC forbidden line emission is likely to arise within just $\sim$5\,AU of the star. For an additional 10 targets with no reliable inclination reported in the literature, we calculate potential disk inclinations based on forbidden line widths, as described in Section \ref{sect:inclination} and noted in Table \ref{tab:source_properties} with a dagger. 

\color{black}
\subsection{Observations}\label{sect:obs}
	
	We observed all targets using the Keck/HIRES spectrograph \citep{vogt94} with the C5 decker and a 1.1\arcsec x 7\arcsec slit, which covers a wavelength range of 4,800-9,000 \AA~ at a nominal resolution of 37,500. \cite{pascucci2015} independently calculated the spectral resolution achieved by HIRES in this setting and found a slightly better resolution of $\sim$45,000 corresponding to 6.6 km/s. The targets were observed in two campaigns with the same instrumental setting, one in 2006 and another in 2012. Two of the targets, UX Tau A and IP Tau, were observed in both campaigns. In addition to the science targets, we observed 5 late type stars that are used as photospheric standards and a set of O/B type stars which are used as telluric standards. Spectra were acquired in the standard mode which places the slit along the parallactic angle to minimize slit losses. Slit position angles with respect to disk position angles are discussed in Appendix~\ref{PA_appendix} and the table therein.
	
The data reduction was carried out using the highly automated Mauna Kea Echelle Extraction (MAKEE) pipeline written by Tom Barlow. In addition to bias-subtraction, flat-fielding, spectral extraction and wavelength calibration, the pipeline automatically subtracts the sky. An example of a spectrum before and after sky subtraction is discussed in Appendix~\ref{PA_appendix} and the figure therein. Further details about the data reduction and source exposure times are given in \cite{pascucci2015}.

\subsection{Corrected Forbidden Line Profiles}\label{sect:line_profiles} 
	
\begin{figure}[h] %  figure placement: here, top, bottom, or page (H,T,B,P RESPECTIVELY)
  \centering
 \includegraphics[width=0.5\textwidth]{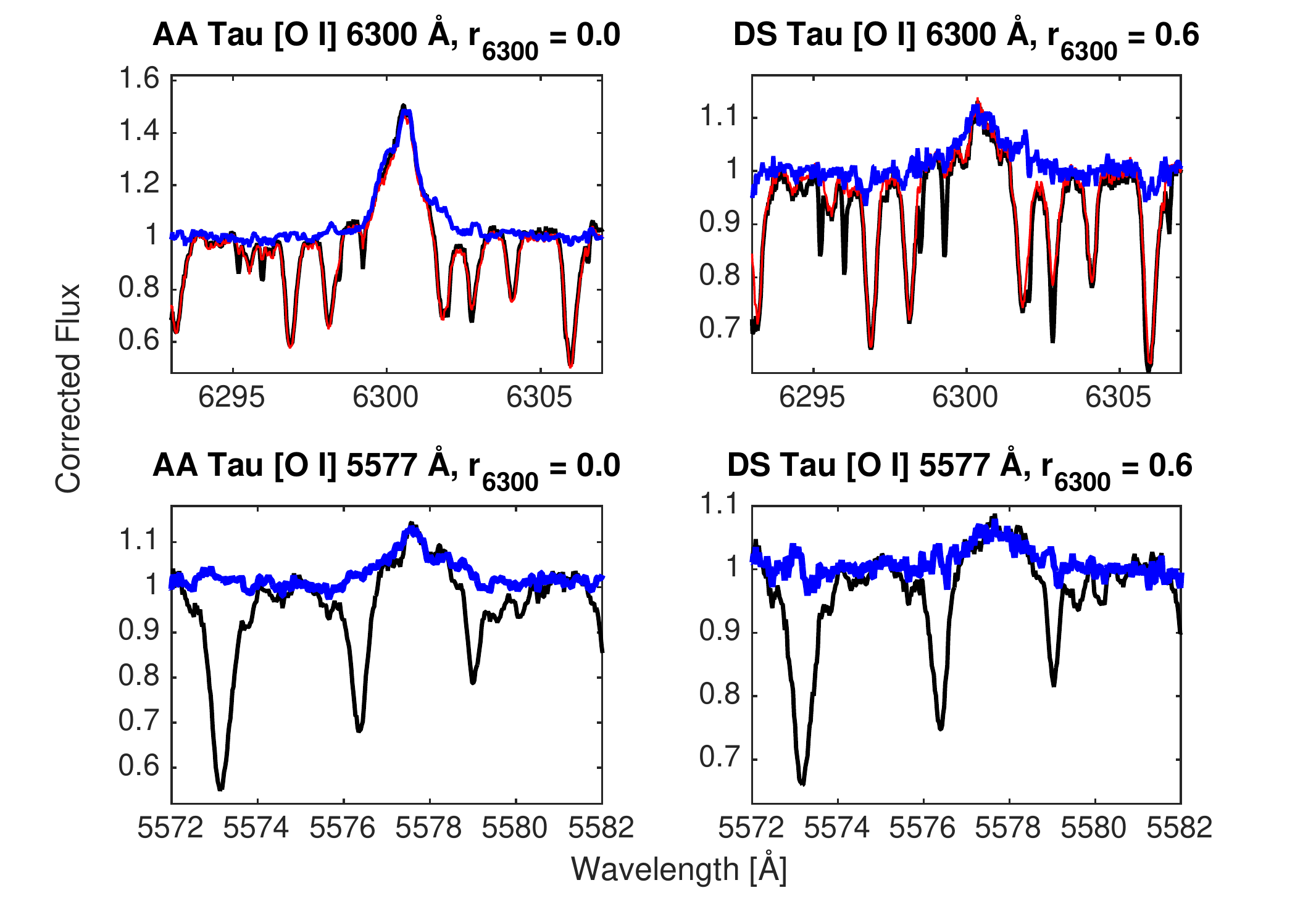}
 \caption{An example of the technique used to correct \oi{} line profiles for two sources: AA Tau and DS Tau. The original spectrum (in black) includes the emission feature plus telluric and photospheric absorption lines that have not yet been subtracted. Once the telluric lines are removed the spectrum in red is produced. The spectrum in blue depicts the final corrected line profile after the telluric and photospheric lines have been removed. For the \oi{} 5577 {\AA} feature, no telluric correction is required.}
   \label{fig:Resid_example}
\end{figure}

	In order to identify weak intrinsic emission in the forbidden lines we first remove any telluric and/or photospheric absorption contaminating the spectral order of interest. Telluric absorption is prevalent in the orders containing \oi{} 6300 {\AA} and \oii{} 7330 {\AA}, but minimal for \oi{} 5577 {\AA} and \sii{} 6731~{\AA}. We remove the atmospheric features by matching the telluric lines in an O/B standard star to those in the target spectrum and then dividing the target spectrum by the telluric standard. In order to remove photospheric lines, we follow a procedure motivated by the approach of \cite{hartigan89}. First, we select a photospheric standard with a spectral type that closely matches the spectral type of the target star. If need be, we broaden the absorption lines of the photospheric standard to match the width of the lines present in the target spectrum. We then apply a cross-correlation technique to shift the photospheric standard to align with the photospheric lines in the target spectrum, and if necessary, we add a flat continuum to the photospheric standard to match veiled photospheric lines in the young accreting stars. The veiling ($r_{\lambda}$) is defined as the ratio of the excess to the photospheric flux, and $r_{6300}$ is included in Table \ref{tab:OI_5577_6300}, with values ranging from 5.6 (DR Tau) to 0 (5 sources). The final corrected profile is created by dividing the target spectrum by the veiled photospheric spectrum of the standard. This method for correcting the line profiles is illustrated in Figure~\ref{fig:Resid_example} for the \oi{} emission features at 6300 and 5577 {\AA} for two sources.  One of them, DS Tau, has a moderate veiling, while the other, AA Tau, has zero veiling. No corrections are made for H$\alpha$, which has no telluric absorption and is a strong emission feature in all sources.
	
	Four of the five photospheric standards are luminosity class V and one is the weak lined T Tauri star (WTTS) V819 Tau, with no infrared excess from a disk. As discussed elsewhere (see \citealt{HH14}), WTTS are ideal candidates for matching photospheric features in TTS, and V819 Tau was used whenever it was a good match to the spectral type of the target.  The five photospheric standards and their spectral types are: HR~8832 (K3), HBC~427 (K6), V819~Tau (K8), GL~15a (M2), V1321~Tau  (M2). The standard applied to each target and line of interest is included in Table \ref{tab:OI_5577_6300}. 
	
Final corrected line profiles are presented in Figures~\ref{fig:HVC_6300_panel1} and 
\ref{fig:HVC_6300_panel2} for \oi{} 6300\,\AA{} for all 30 stars, including the 3 non-detections (see next section). Final corrected profiles for the other two forbidden lines are shown only for the detections, in Figure~\ref{fig:HVC_5577} for \oi{} 5577\,\AA{} and in Figure~\ref{fig:HVC_SII} for \sii{} 6731\,\AA. They are plotted as corrected flux above the continuum versus radial velocity. The velocity is relative to the stellocentric frame, as determined from photospheric line centroids. The stellar radial velocities of our sources are given in Tables 5 and 6 of \cite{pascucci2015} and have a 1$\sigma$ uncertainty of $\sim$1\,km/s.

\begin{figure*}[h] %  figure placement: here, top, bottom, or page (H,T,B,P RESPECTIVELY)
   \centering
   \includegraphics[width=4.5in]{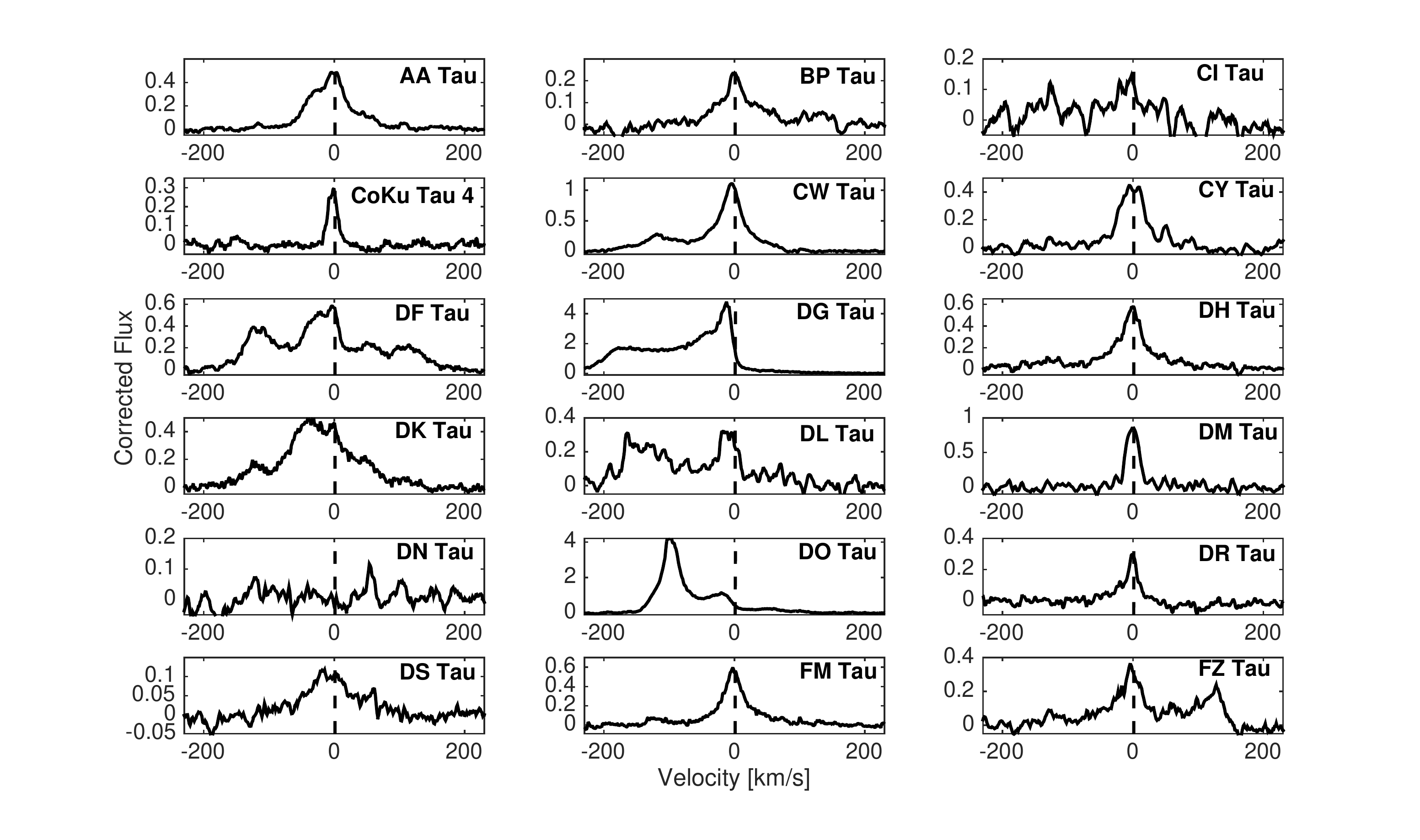}
   \caption{Corrected \oi{} 6300 {\AA} profiles with the stellar continuum subtracted for half of the sample.}
   \label{fig:HVC_6300_panel1}
\end{figure*}
\begin{figure*}[h] %  figure placement: here, top, bottom, or page (H,T,B,P RESPECTIVELY)
   \centering
    \includegraphics[width=4.5in]{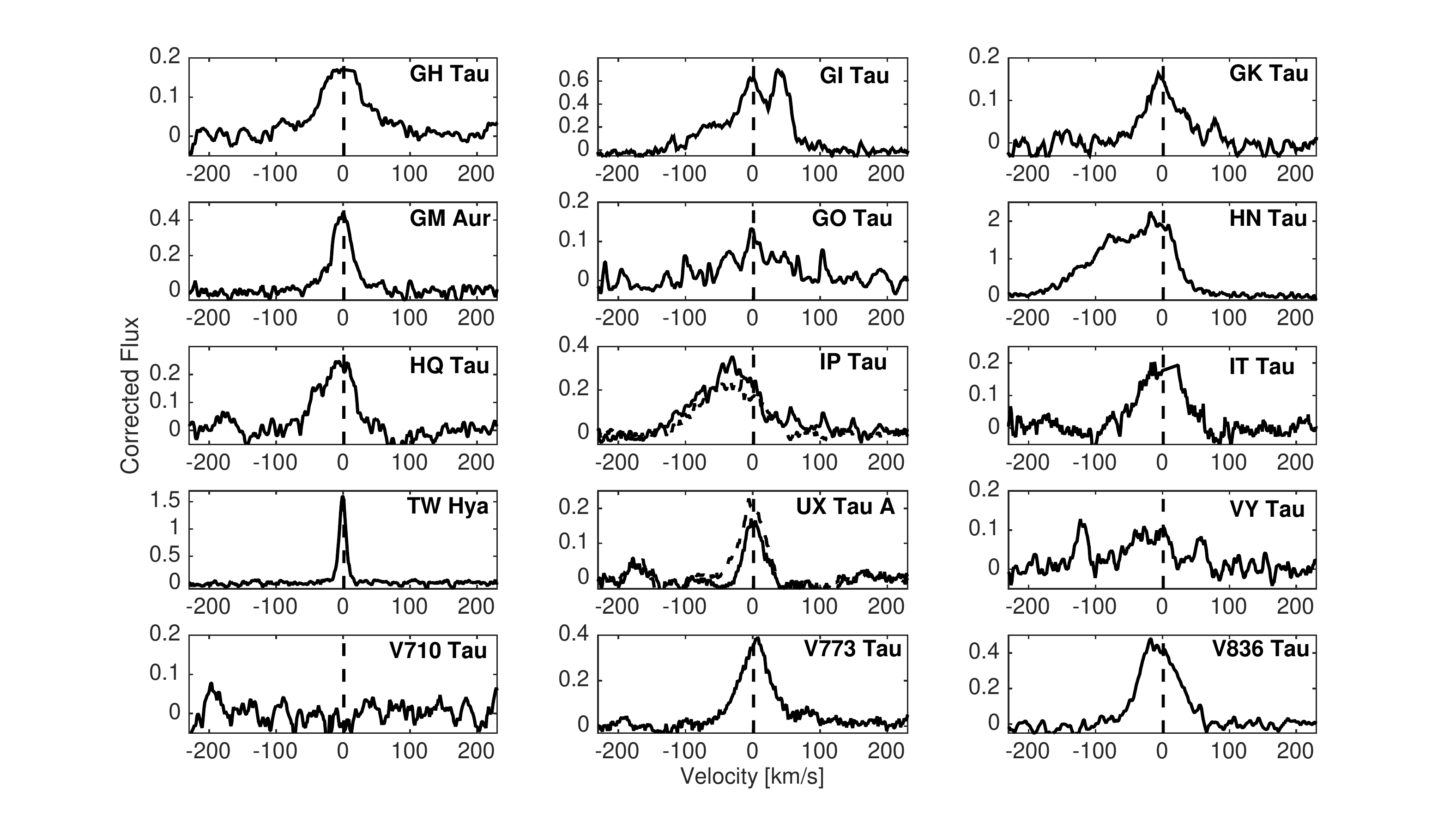} 
   \caption{Continuation of Figure 2: Corrected \oi{} 6300 {\AA} profiles with the stellar continuum subtracted for the second half of the sample. For IP and UX Tau A, the profiles from both observing campaigns are plotted. The solid line corresponds to the spectrum we analyzed in the paper (2006 for IP Tau and 2012 for UX Tau A).}
   \label{fig:HVC_6300_panel2}
\end{figure*}
\begin{figure*}[p] %  figure placement: here, top, bottom, or page (H,T,B,P RESPECTIVELY)
   \centering
   \includegraphics[width=4.5in]{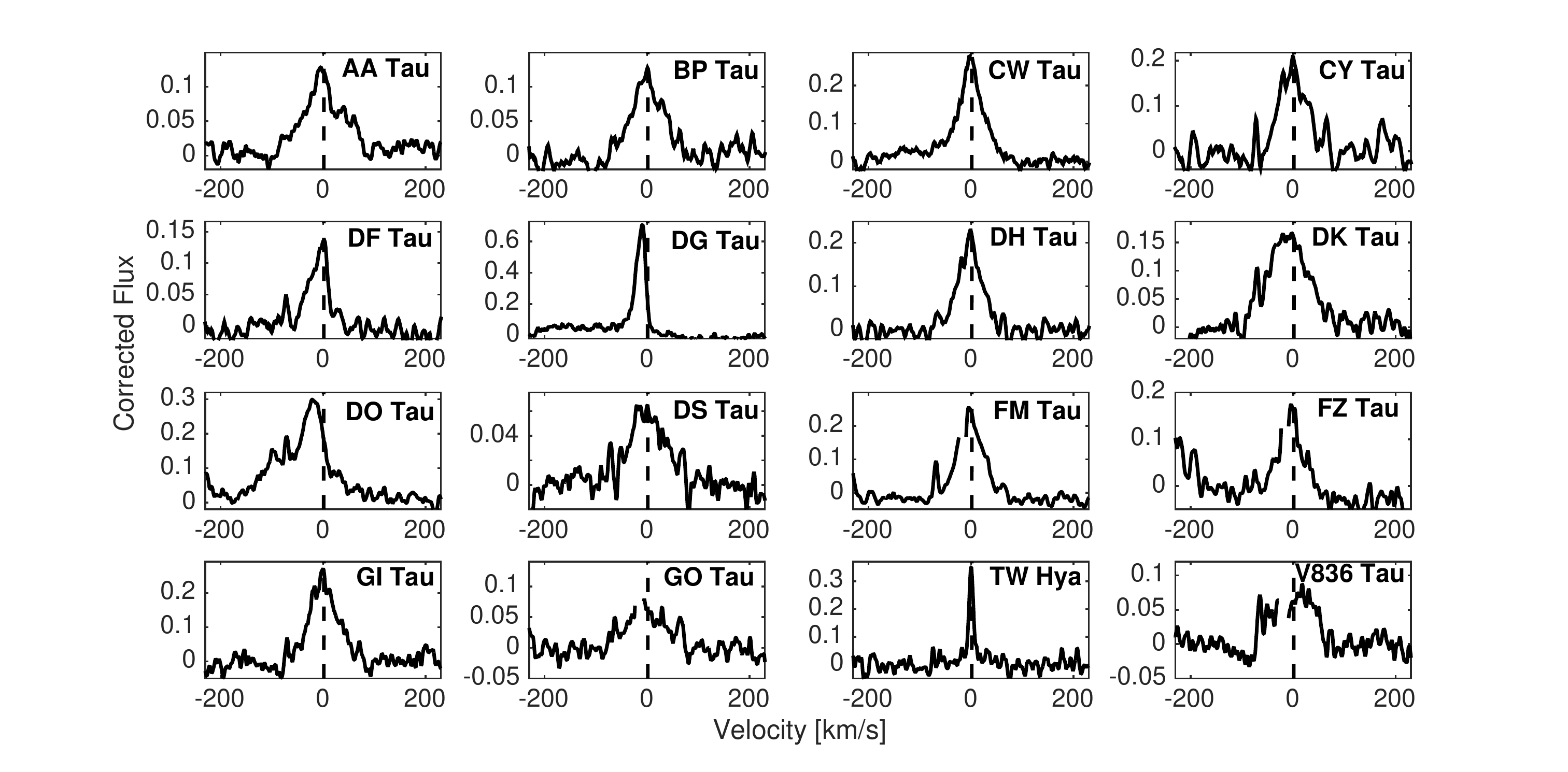} 
   \caption{Corrected \oi{} 5577 {\AA} profiles for the 16 sources with detections.
 }
   \label{fig:HVC_5577}
\end{figure*}
\begin{figure}[h] %  figure placement: here, top, bottom, or page (H,T,B,P RESPECTIVELY)
   \centering
   \includegraphics[width=0.5\textwidth]{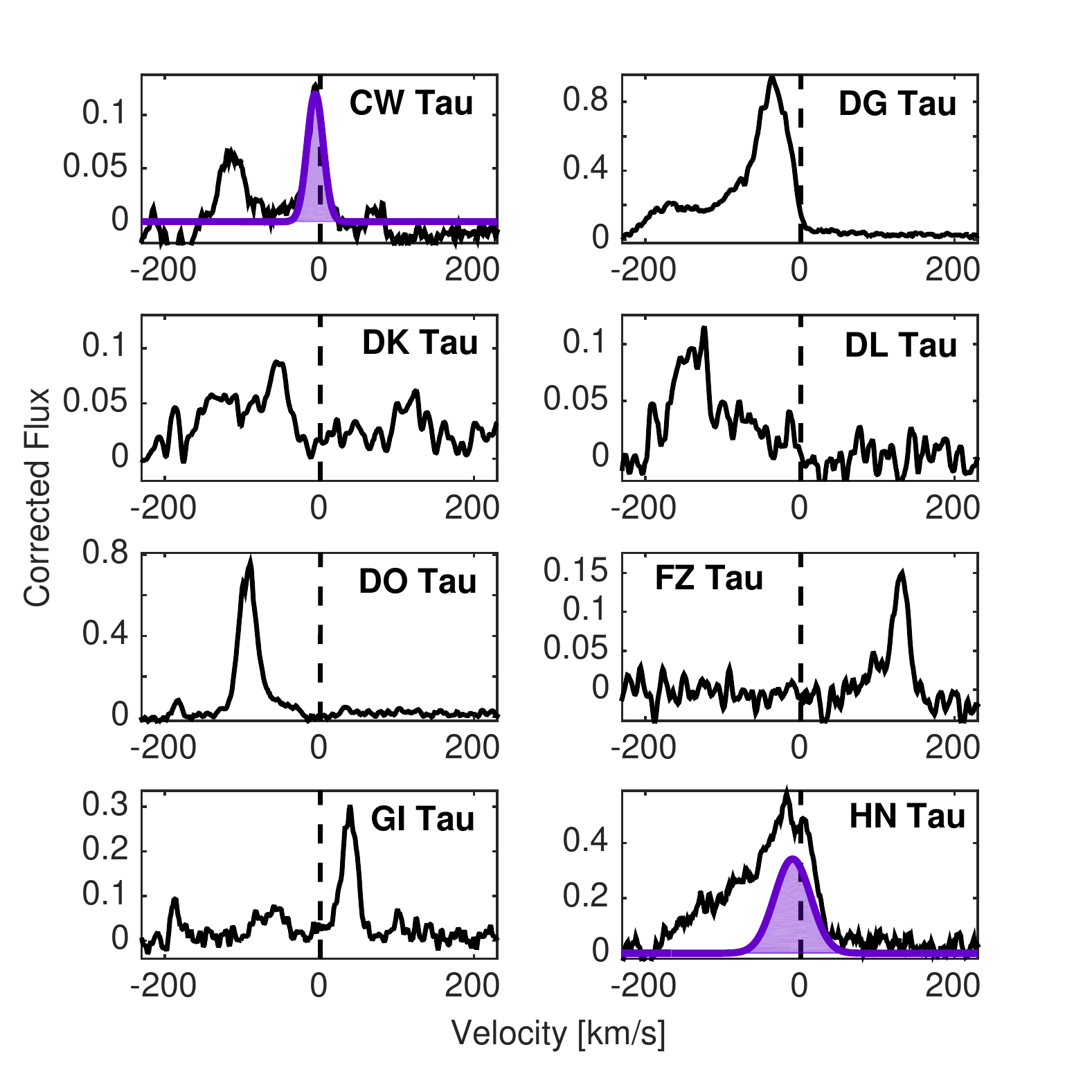} 
   \caption{Corrected \sii{} 6731 {\AA} profiles for the 8 sources with detections. The LVC is only detected in two sources (CW Tau and HN Tau), and is shaded in purple.}
   \label{fig:HVC_SII}
\end{figure}

\subsection{Equivalent Widths}\label{sect:EWs}

	A number of our sources have weak or absent forbidden lines. The procedure used to identify a detection is to calculate the standard deviation of the residuals (RMS) from a linear continuum fit to a region $\sim$ 3 {\AA} outside the line of interest. Any emission near the wavelength of interest, with a peak higher than 3 times the RMS and equal or broader in width than the 6.6 km/s of an unresolved line, is deemed a detection. Based on this method we find detections in the \oi{} 6300 {\AA} transition for 30/33 TTS. Non detections are DN Tau, VY Tau and V710 Tau, all with measured veilings very near or at zero. Both detections and non-detections are included in Figures~\ref{fig:HVC_6300_panel1} and \ref{fig:HVC_6300_panel2} for \oi{} 6300 {\AA}. Additionally, we detect \oi{} 5577 {\AA} for 16 stars and \sii{} 6731 {\AA} for 8 stars. 
Detections are shown for these lines in Figures~\ref{fig:HVC_5577} and \ref{fig:HVC_SII}. The \oii{} transition at 7330 {\AA} is never detected.
		
	For detected lines we calculate the line equivalent width (EW)  by integrating over the wavelength range where there is emission above the continuum. In addition to measuring the line EW, we compute its uncertainty using a Monte Carlo (MC) approach (e.g. \citealt{pascucci08}). We do this by adding a normally distributed noise to each spectrum with the noise being the RMS on the continuum next to the line of interest. We compute the EWs of 1,000 individual spectra generated by the MC approach and assign as the uncertainty the standard deviation of the distribution of EWs. 
	
	For lines we determine to be non-detections, we compute a 3${\sigma}$ upper limit (UL) on the EW assuming a Gaussian unresolved profile, $UL= (3 \times RMS) \times \sigma_{\circ} \times \sqrt{2\pi}$ where the RMS is calculated from the standard deviation of the residuals on a $\sim$ 3 {\AA} continuum, $\sigma_{\circ}$ in {\AA} is defined as $(\lambda \times \Delta v) / (c \times 2.355)$ where $\lambda$ is the wavelength of the line of interest, $\Delta v$ is the FWHM of an unresolved line (6.6 km/s) and $c$ is the speed of light in km/s. 
	
	Table \ref{tab:OI_5577_6300} presents the EW for the detected forbidden lines and the upper limits for the non-detections for our 33 sources.

\section{Accretion Luminosities and Disk Accretion Rates} \label{sect:accr}

In accreting TTS, most of the UV and optical excess continuum emission derives from energy released through accretion, attributed to magnetospheric accretion shocks on the stellar surface \citep{Hartmann98}. Ideally accretion luminosities are determined from flux calibrated spectra that include the Balmer discontinuity, where the spectrum of the continuum UV-excess can be modeled with accretion shocks \citep{calvet98, HH08}. Since our echelle spectra are not flux-calibrated, accretion luminosities can be estimated using well calibrated relationships relating line and accretion luminosities first demonstrated by \cite{muzerolle98} and recently summarized in \cite{alcala2014}. This approach is superior to the older technique of using the emission excess (veiling) at one wavelength in the Paschen continuum, which requires applying an uncertain bolometric correction. Here we will use the calibration on the H$\alpha$ line luminosity from \cite{alcala2014} because it is based on a large sample of low-mass stars and simultaneous UV-excess measurements of the accretion luminosity.

With our echelle spectra, emission line equivalent widths ($EW_{\lambda}$) can be converted into a line luminosity using the following relation:
\begin{equation}
L_{line} = 4 \pi \ d^2 \ f_{\lambda} \ (EW_{\lambda}) (1 + r_{\lambda})
\label{eq:Lline}  
\end{equation} 

\setlength{\parindent}{0cm} where $d$ is the distance to the science target, $f_{\lambda}$ is the photospheric continuum flux density near the line of interest, and the factor (1+ $r_{\lambda}$) converts the observed equivalent width to one that is veiling corrected, i.e. measured in units of the stellar continuum. For this work, we chose H$\alpha$ for an accretion diagnostic since it is detected in all of our targets. We adopt distances from \cite{HH14}, and use their published extinction corrected continuum flux density at 7510 {\AA}, in conjunction with the ratio of the flux densities at 6600 {\AA} and 7510 {\AA} in the Pickles Atlas for stars matched in spectral type to each science target \citep{pickles1998} to determine the stellar flux density near H$\alpha$. The veiling $r_{\lambda}$ at 6300 {\AA} listed in Table \ref{tab:OI_5577_6300} is very close to that at H$\alpha$ and was adopted here. While this approach assumes the stellar continuum has not varied, it does account for any variability in veiling between our observations and that of \cite{HH14} in setting the continuum flux adjacent to the line.

\setlength{\parindent}{0.5cm}The H$\alpha$ line luminosities (L$_{H\alpha}$), are then converted to accretion luminosities, L$_{acc}$, using the relation derived by \cite{alcala2014}:

	\begin{equation}
	\log{(L_{acc}/L_{\sun})} = (1.50 \pm 0.26) + (1.12 \pm 0.07) \times \log{(L_{H\alpha}/L_{\sun})} 
	\end{equation}
\setlength{\parindent}{0cm}The calibration based on L$_{H\alpha}$ can be compromised in sources with large blueshifted and redshifted absorption, masking what otherwise would be a larger emission EW. However, since it is observed in all our targets, we preferred it to other indicators such as He~I. 
\setlength{\parindent}{0.5cm} Finally, the accretion luminosities are converted into mass accretion rates using the magnetospheric model developed by \cite{gullbring98} and the following equation:
\begin{equation}
	\dot M_{acc}=\frac{L_{acc}~R_{\ast}}{GM_{\ast}(1-\frac{R_{\ast}}{R_{in}})}
	\end{equation}
where $R_{\ast}$ and $M_{\ast}$ are the radius and mass of the star, $G$ is Newton's gravitational constant, and $R_{in}$ is the inner truncation radius of the disk.  $R_{in}$ is generally unknown, but it is usually assumed to be $\approx 5R_{\ast}$, the co-rotation radius (e.g. \citealt{gullbring98, shu1994}). For all of the sources, we use the stellar masses and radii from \cite{HH14}.

All accretion parameters, including EW$_{H\alpha}$, $L_{H\alpha}$, $L_{acc}$ and $\dot M_{acc}$ are listed in Table \ref{tab:acc_properties}. The table also includes columns for the line luminosity of \oi{} 6300 {\AA}, first for the whole line, then for the LVC and then for the NC of the LVC, all with $f_{\lambda}$ determined from the Pickles Atlas near 6300 {\AA}. The definition of the latter two will be described in the next section.

We estimate a typical uncertainty for $L_{acc}$ as follows. Sixteen of our targets have multi-epoch R-band photometry measurements \citep{herbst1994} where the standard deviation of the photometric points over the mean R-magnitude is at most 0.1, which indicates that variability in the continuum is probably not a major source of uncertainty for most stars. The extinctions derived by \cite{HH14} also have a small uncertainty, only 0.15 dex. Therefore, the major uncertainty in $L_{acc}$ will be the calibration of $L_{H\alpha}$ and $L_{acc}$. Errors in the slope and y intercept of this relation are included in Equation~2. Adding these uncertainties in quadrature gives a final uncertainty on $L_{acc}$ of $\sim$ 0.3 dex. Variations in TTS H$\alpha$ EW and veiling, attributed to variations in accretion luminosity, are typically less than a factor of two. Two of our targets, UX Tau A and IP Tau, were observed in both 2006 and 2012 with changes in H$\alpha$ EW by factors 1.2 and 1.5, respectively, with negligible changes in veiling. This level of variability corresponds to a variation of $\sim$ 0.2 dex in $L_{acc}$.

\begin{figure}[h] %  figure placement: here, top, bottom, or page (H,T,B,P RESPECTIVELY)
   \centering
 \includegraphics[width=0.5\textwidth]{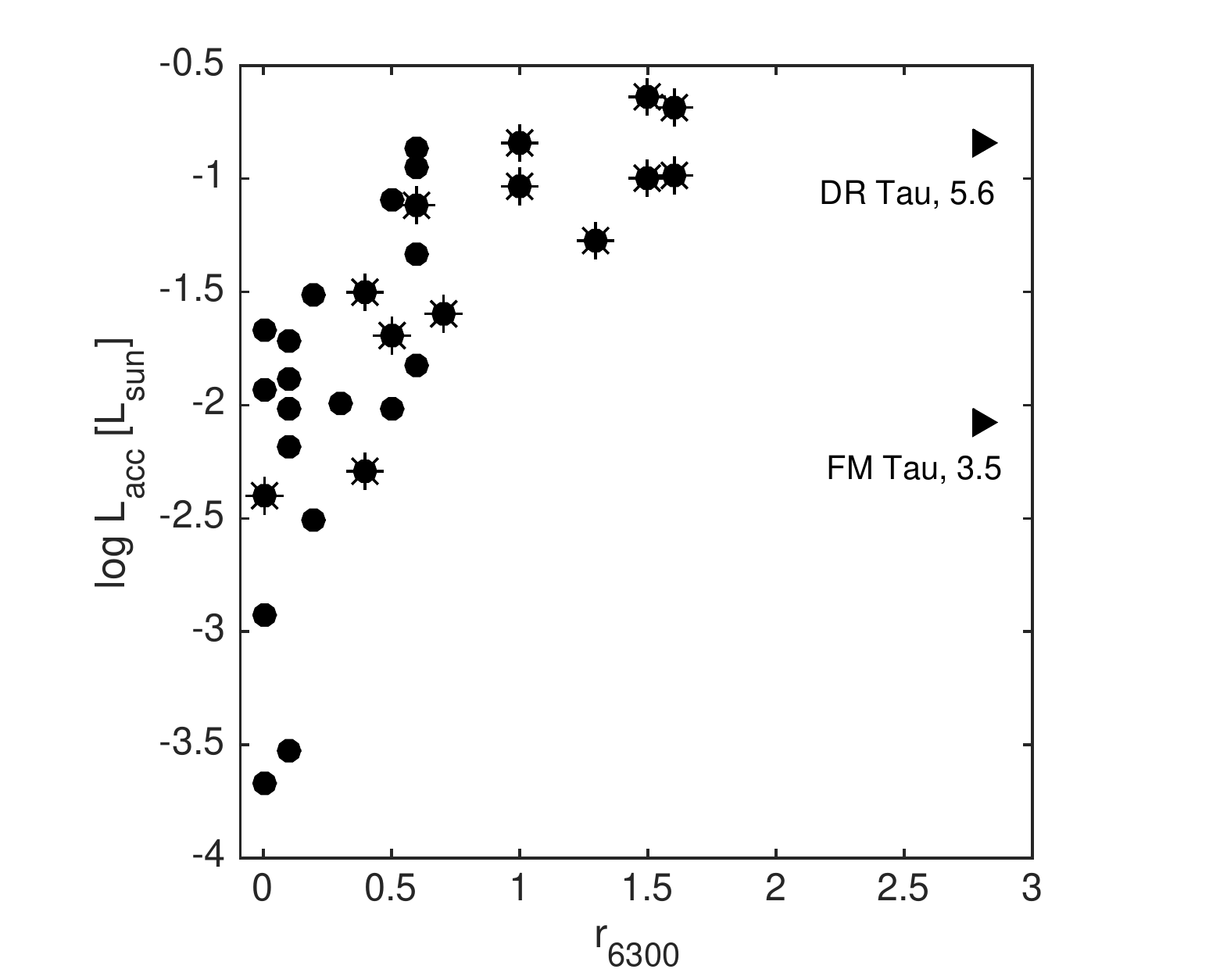}
   \caption{ There is a general trend of increasing $L_{acc}$ with increasing veiling, defined by a clear lower boundary for $L_{acc}$ at a given veiling with considerable scatter above that boundary. The extreme veiling sources DR Tau and FM Tau, with $r_{6300}$ = 5.6 and 3.5, respectively, exceed the plot boundary. Symbols with radial spokes denote sources with a high velocity component at \oi{} 6300\,\AA. }\label{fig:Lacc_veiling}
\end{figure}

The range in H$\alpha$ EW for our sample runs from a low of 1.2 {\AA} to a high of 230 {\AA}, translating into a span of more than three orders of magnitude in $L_{acc}$ and $\dot M_{acc}$, with accretion luminosities from $10^{-3.7}$ to $10^{-0.6}$ $L_\odot$ , and mass accretion rates from $10^{-10.7}$ to $10^{-7.5}$ $M_\odot$/year. \cite{manara2013} demonstrated that for $L_{acc}$ $<10^{-3.0} L_\odot $ there is a possibility for chromospheric emission to dominate the line luminosity. Since the continuum veiling is a diagnostic of accretion, deriving from excess emission in the accretion shock on the stellar surface, the relation between $r_{6300}$ and $L_{acc}$, shown graphically in Figure~\ref{fig:Lacc_veiling}, offers additional insight on whether a star is accreting. In general these quantities are correlated, in the sense that there is a well defined lower boundary in $L_{acc}$ at a given veiling with considerable scatter above that boundary\footnote{Note that the two sources with the highest veiling, DG Tau and FM Tau, with $r_{6300}$ = 5.6 and 3.5 respectively, do not fall into the relation between veiling and L$_{acc}$ defined by the other stars. Exceptionally high veilings for these sources were also found by HEG. Either the relation between veiling and accretion luminosity breaks down at high veilings \citep{gahm2008} and/or the accretion luminosities based on H$\alpha$ are severely underestimated due to strong wind absorption features in sources with high disk accretion rates and high veilings.}. For example, the objects with low veiling ($r_{6300}$ of 0 or 0.1) have inferred accretion luminosities spanning two orders of magnitude.  We conclude from this that it is not possible to assign an unambiguous accreting/non-accreting status based on the level of the H$\alpha$ EW or absence of veiling, sometimes used as thresholds for defining classical (accreting) or weak (non-accreting) TTS. Since all of our targets have disks in various evolutionary states, we will treat them all identically in converting H$\alpha$ into accretion luminosities. We note that three sources have H$\alpha$ based accretion luminosities near or below the threshold where accretion cannot be distinguished from chromospheric activity (CoKu Tau 4, VY Tau, and V710 Tau). Two of these three also have no detection of even the strongest forbidden line, \oi{} 6300 \AA. The third non-detection at \oi{} 6300 \AA~is DN Tau, with H$\alpha$ EW $=$ 13.5 {\AA}, $r_{6300}$ $=$ 0, and an inferred log $L_{acc}$ of -1.93. As discussed in Section \ref{sect:HEG_compare}, this appears to the only example of a star that is accreting but does not show forbidden emission.

%Since the continuum veiling is a diagnostic of accretion, deriving from excess emission in the accretion shock on the stellar surface, the relation between $r_{6300}$ and $L_{acc}$, shown graphically Figure~\ref{fig:Lacc_veiling}, can shed insight on this issue. In general these quantities are correlated, in the sense that there is a well defined lower boundary in $L_{acc}$ at a given veiling with considerable scatter above that boundary. For example, the objects with low veiling ($r_{6300}$ of 0 or 0.1) have inferred accretion luminosities spanning two orders of magnitude.  We conclude from this that it is not possible to assign an unambiguous accreting/non-accreting status based on the level of the H$\alpha$ EW or absence of veiling, sometimes used as thresholds for defining classical (accreting) or weak (non-accreting) TTS. Since all of our targets have disks in various evolutionary states, we will treat them all identically in converting H$\alpha$ into accretion luminosities, although we note that the 3 non-detections at 6300\,\AA{} all have low H$\alpha$ EWs (13.5\,\AA{} for DN~Tau, 4.2\,\AA{} for VYTau, and 1.9\,\AA{} for V710~Tau) and the latter two may not actually be accreting.

\section{Deconstructing Forbidden Line Profiles}\label{sect:Profiles}
	
The aim of this paper is to better understand the LVC of forbidden line emission in TTS.  In this section we describe the fitting technique used to define and separate HVC and LVC emission, and then describe the kinematic properties of the LVC emission.

\subsection{Gaussian Fitting}\label{sect:fitting}	

	The focus of this paper is the behavior of the LVC of the forbidden line profiles. Although in some instances the HVC and LVC are well resolved (e.g. CW Tau, DO Tau) in most cases they are blended (e.g. DK Tau, HN Tau). In the earlier HEG study the separation between HVC and LVC was made simply by assigning emission further than 60 km/s from the stellar velocity to the HVC.  With almost a factor of two higher spectral resolution, we attempt to separate blended HVC and LVC using Gaussian fitting. To this end, we fit the profiles interactively using the Data Analysis and Visualization Environment (DAVE) that runs as a Graphical User Interface (GUI) in IDL. This program was developed by the National Institute of Standards and Technology (NIST) Center for Neutron Research \citep{azuah2009}. In order to find the best fit parameters for each emission feature, we identify the minimum number of Gaussians required to describe each profile, specifying an initial estimate of the centroid velocity ($v_c$) and full width at half maximum (FWHM) for each component using the Peak ANalysis (PAN) feature of DAVE.  PAN then performs many iterations to minimize the reduced chi-squared and outputs the best fit centroid velocities, the FWHM, and the areas under the Gaussian fits. Errors in centroid velocities, measured relative to the stellar photosphere, cannot be less than $\pm$1\,km/s, which is the uncertainty in the stellar radial velocity (see \citealt{pascucci2015} for more details).

The number of Gaussians required to fit each line profile depends on the profile's shape and an RMS estimate of the goodness of fit. The fitted components, individually and summed, are superposed on all 30 detected \oi{} 6300\,\AA{} profiles in Figures~\ref{fig:full_fit1} and \ref{fig:full_fit2}. Eleven profiles are well fit with a single Gaussian component and another ten with two Gaussians. The remaining nine profiles required 3 components (5 stars), 4 components (3 stars) and in once case, FZ Tau, five components.  

\begin{figure*}[h] %  figure placement: here, top, bottom, or page (H,T,B,P RESPECTIVELY)
   \begin{center}
   \includegraphics[width=4.5in]{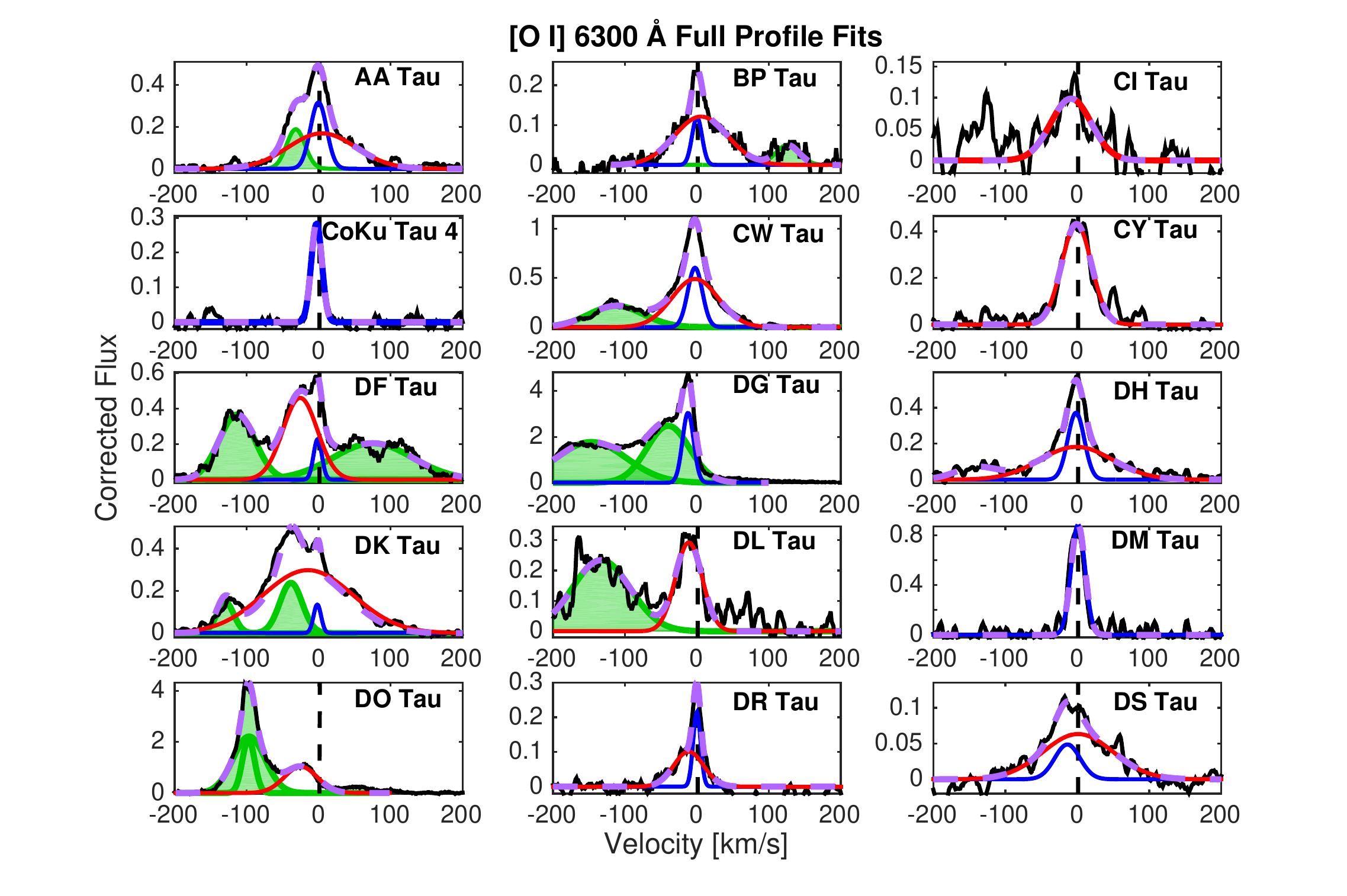}
   \caption{Gaussian component fits for half the stars with a detected \oi{} 6300\,\AA{} line. Areas shaded in green meet the criterion for HVC emission. The LVC fits may be comprised of one or both narrow (blue) or broad (red) components. The sum of all fits is a purple dashed line.}
   \label{fig:full_fit1}
   \end{center}
\end{figure*}
\begin{figure*}[h] %  figure placement: here, top, bottom, or page (H,T,B,P RESPECTIVELY)
   \begin{center}
   \includegraphics[width=4.5in]{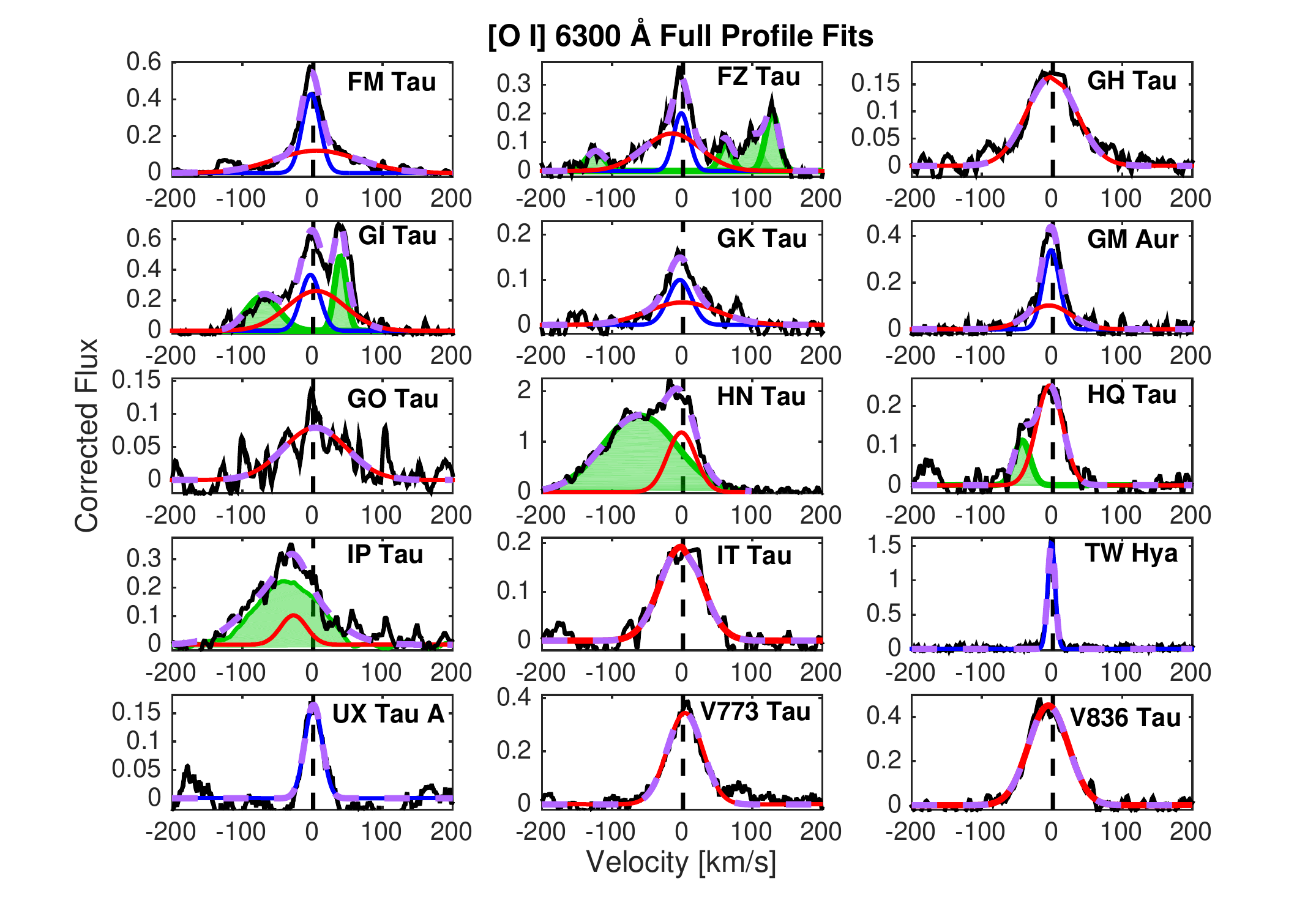}
   \caption{Continuation of Figure~\ref{fig:full_fit1}. Gaussian component fits for the second half of the stars with a detected \oi{} 6300\,\AA{} line. Areas shaded in green meet the criterion for HVC emission. The LVC fits may be comprised of one or both narrow (blue) or broad (red) components. The sum of all fits is a purple dashed line.}
   \label{fig:full_fit2}
   \end{center}
\end{figure*}

\begin{figure}[h] %  figure placement: here, top, bottom, or page (H,T,B,P RESPECTIVELY)
   \begin{center}
   \includegraphics[width=0.5\textwidth]{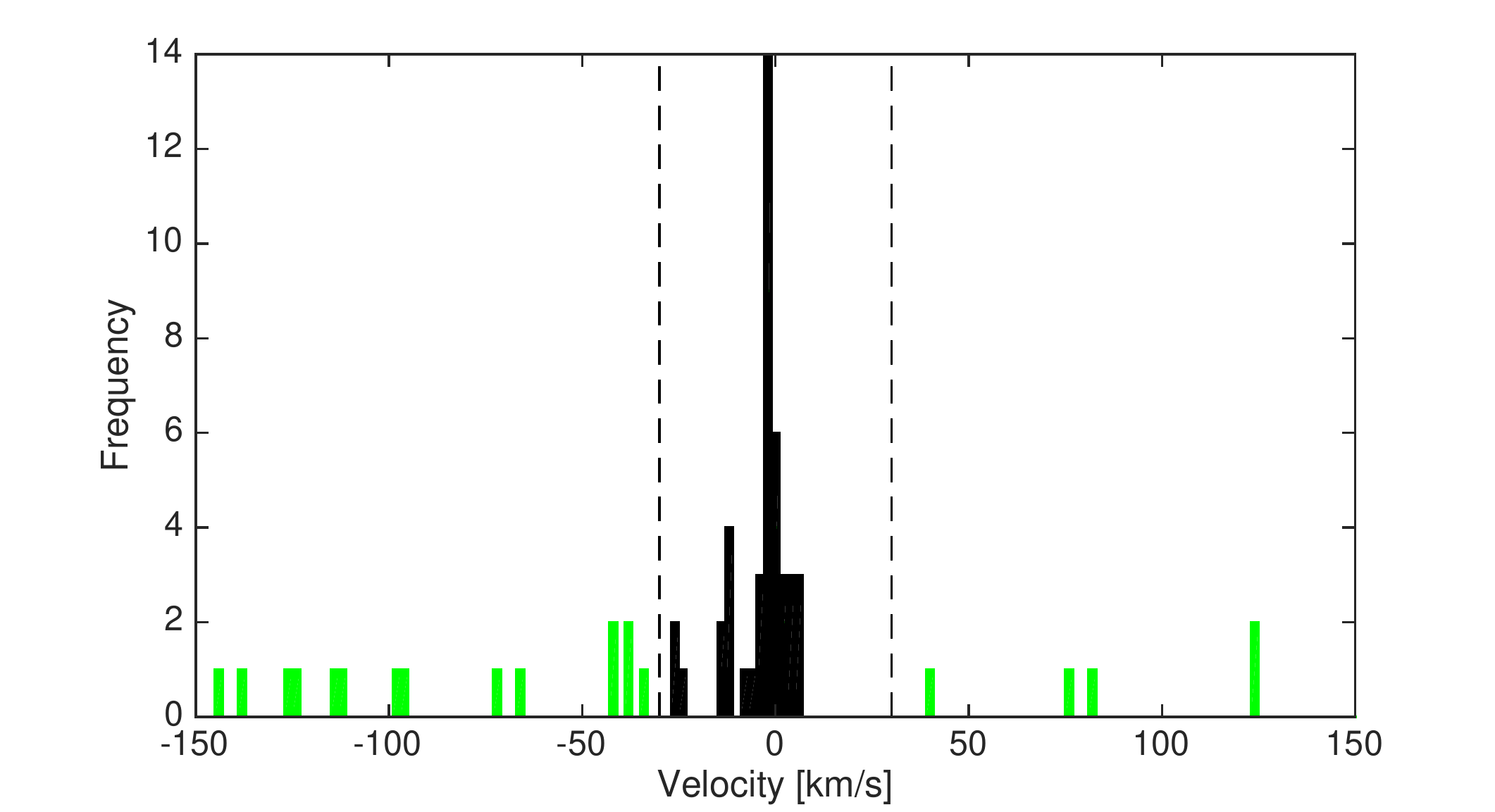}
   \caption{Distribution of velocity centroids for Gaussian components used to describe the \oi{} 6300\,\AA{} profiles. We adopt $\pm$ 30 km/s (dashed lines) as the threshold between HVC (green) and LVC (black) emission.}
   \label{fig:vc_hist}
   \end{center}
\end{figure}

The method we adopt to attribute a component to HVC or LVC emission begins with an examination of the distribution of centroid velocities for all individual components across all stars, shown in Figure~\ref{fig:vc_hist}. The component centroid velocities range from -144 km/s (DG Tau) to +130 (FZ Tau), with blueshifts far more common than redshifts and a high concentration at low velocities.  Based on this distribution, we adopt a centroid velocity of $\pm$ 30 km/s as the threshold between HVC versus LVC emission.  With this method, although all detected \oi{} lines show LVC emission, HVC emission (highlighted in green in Figures~\ref{fig:full_fit1} and \ref{fig:full_fit2}) is seen in only 13 sources at \oi{} 6300 {\AA} and only 3 sources at \oi{} 5577 {\AA}. In contrast all 8 \sii{} detections show HVC emission but only 2 (CW Tau and HN Tau) show weak LVC emission. The HVC fit parameters of centroid velocity and FWHM are listed in Table \ref{tab:HVC_params} for all 3 lines, the LVC fit parameters for the two \oi{}  lines in Table \ref{tab:2_gaus_params} and the LVC fit parameters for \sii{} in Table \ref{tab:SII_properties}. 

\begin{figure}[h] %  figure placement: here, top, bottom, or page (H,T,B,P RESPECTIVELY)
   \centering
 \includegraphics[width=0.5\textwidth]{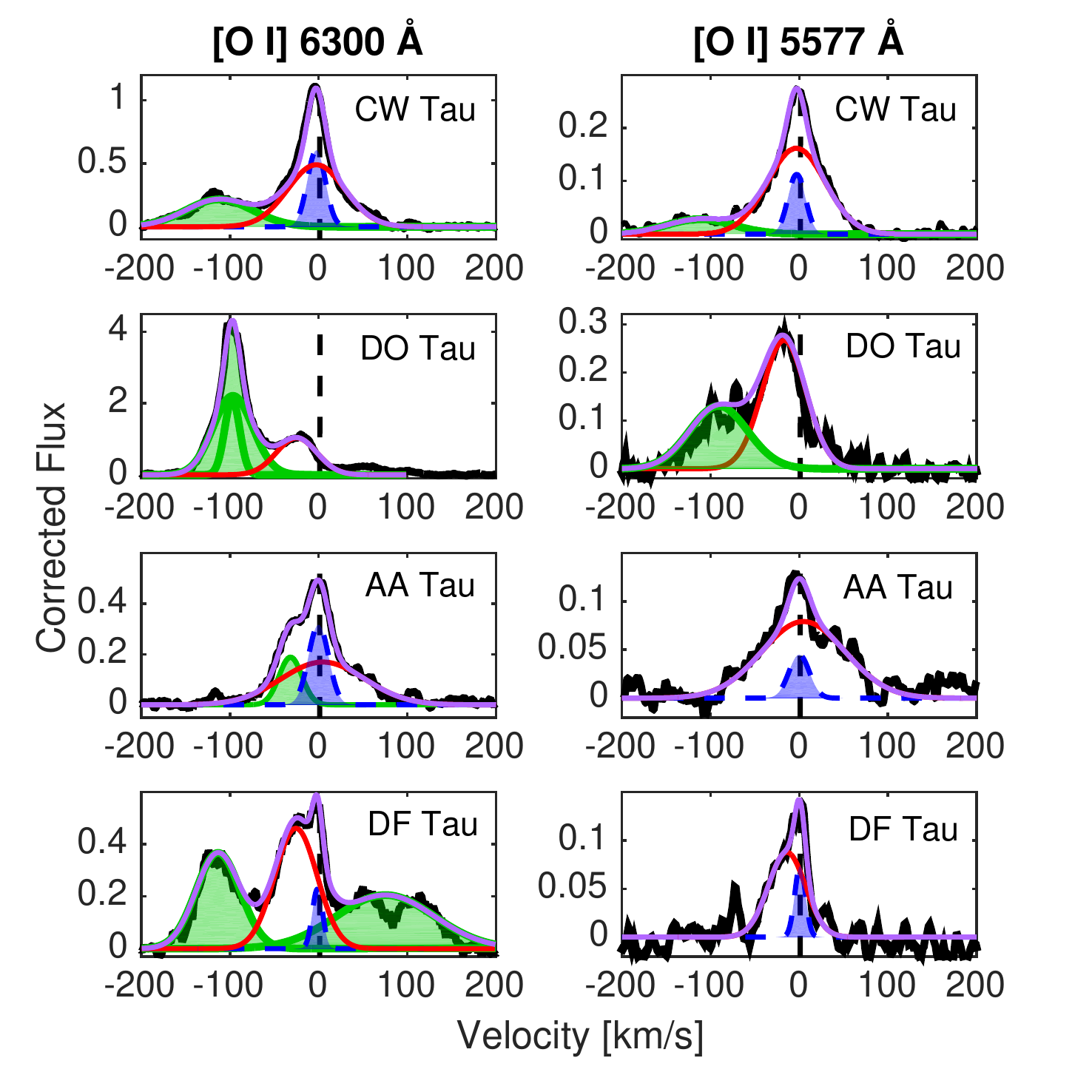}
   \caption{ Examples showing that when both \oi{} lines are present, the LVC components for 6300\,\AA{} (left) and 5577 {\AA} (right) are very similar. The narrow component of the LVC is shaded in blue, and the broad component of the LVC is outlined in red. Areas shaded in green meet the criteria for HVC and the purple line shows the sum of all fits. }
 \label{fig:Pan}
\end{figure}

We find this approach to be reinforced when comparing component fits for the cases when both \oi{} lines are present. We illustrate four examples in Figure~\ref{fig:Pan} including two of the three stars with HVC emission detected at 5577 {\AA} (CW Tau and DO Tau), plus two cases where HVC emission is seen at \oi{} 6300 {\AA} but not 5577 {\AA} (AA Tau and DF Tau). In all cases we find that although the HVC components differ between \oi{} 6300 {\AA} and 5577 {\AA}, their LVC component(s) are similar.  The tendency for  the LVC components to be comparable in the two \oi{} lines, once the HVC is accounted for, gives us confidence that our method to characterize the LVC emission is generally robust. A close comparison of the LVC in the two \oi{} lines will follow in the next section.

\begin{figure}[h] %  figure placement: here, top, bottom, or page (H,T,B,P RESPECTIVELY)
   \centering 
    \includegraphics[width=0.5\textwidth]{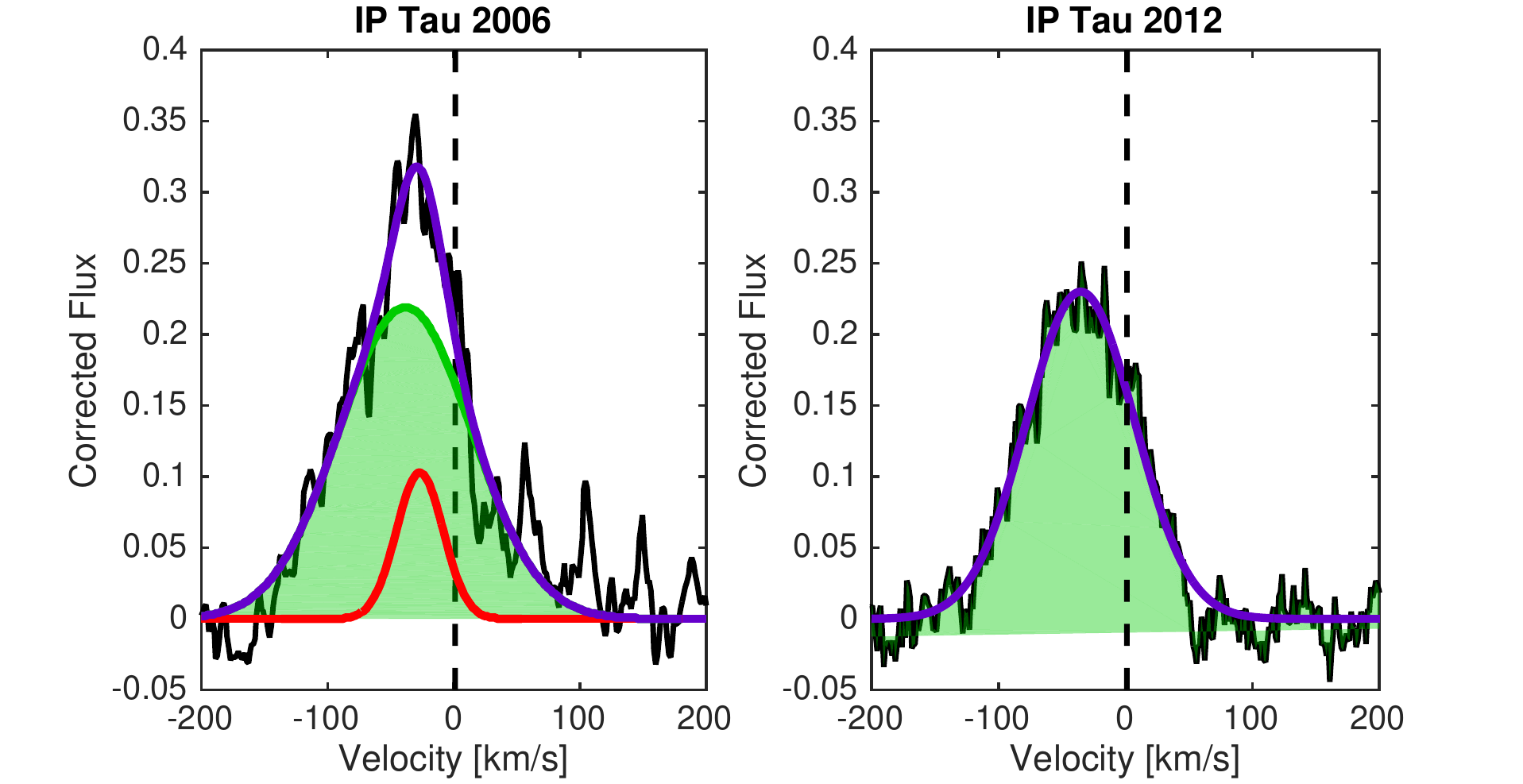}
   \caption{ \oi{} 6300 {\AA} profiles for two observing epochs for IP Tau. In 2006 two gaussian components describe the profile, but in 2012 only one is required. The area shaded in green is the HVC, identical in both epochs. The red line denotes a broad  LVC, seen only in 2006. }
   \label{fig:iptau}
\end{figure}

However, one of the two stars with spectra from both observing epochs, IP Tau, indicates an exception to the finding that LVC emission is always present. We illustrate the change in the morphology of the \oi{} 6300 {\AA} profile between the two epochs in Figure~\ref{fig:iptau}. In 2006 the profile is asymmetric, requiring  two Gaussians to describe it, but in 2012 it is symmetric and fit by one component. Using our definition of HVC and LVC, the single component seen in 2012 is classified as HVC emission, and it is identical in velocity centroid and FWHM to one of the two components describing the 2006 profile. In contrast the second component in 2006 meets the criteria of LVC emission.  Throughout  this work we will use the 2006 spectrum for IP Tau so its LVC profile can be compared with the other stars. We note that the issue of variability of the LVC will be addressed further in Section 5.1, where we find that the LVC is typically constant when compared to profiles found in the literature. Thus cases like IP Tau would be interesting to monitor for further variability.


\pagebreak \subsection{The Low Velocity Component}\label{sect:the_LVC}

\begin{figure*}[h] %  figure placement: here, top, bottom, or page (H,T,B,P RESPECTIVELY)
   \begin{center}
  \includegraphics[width=5in]{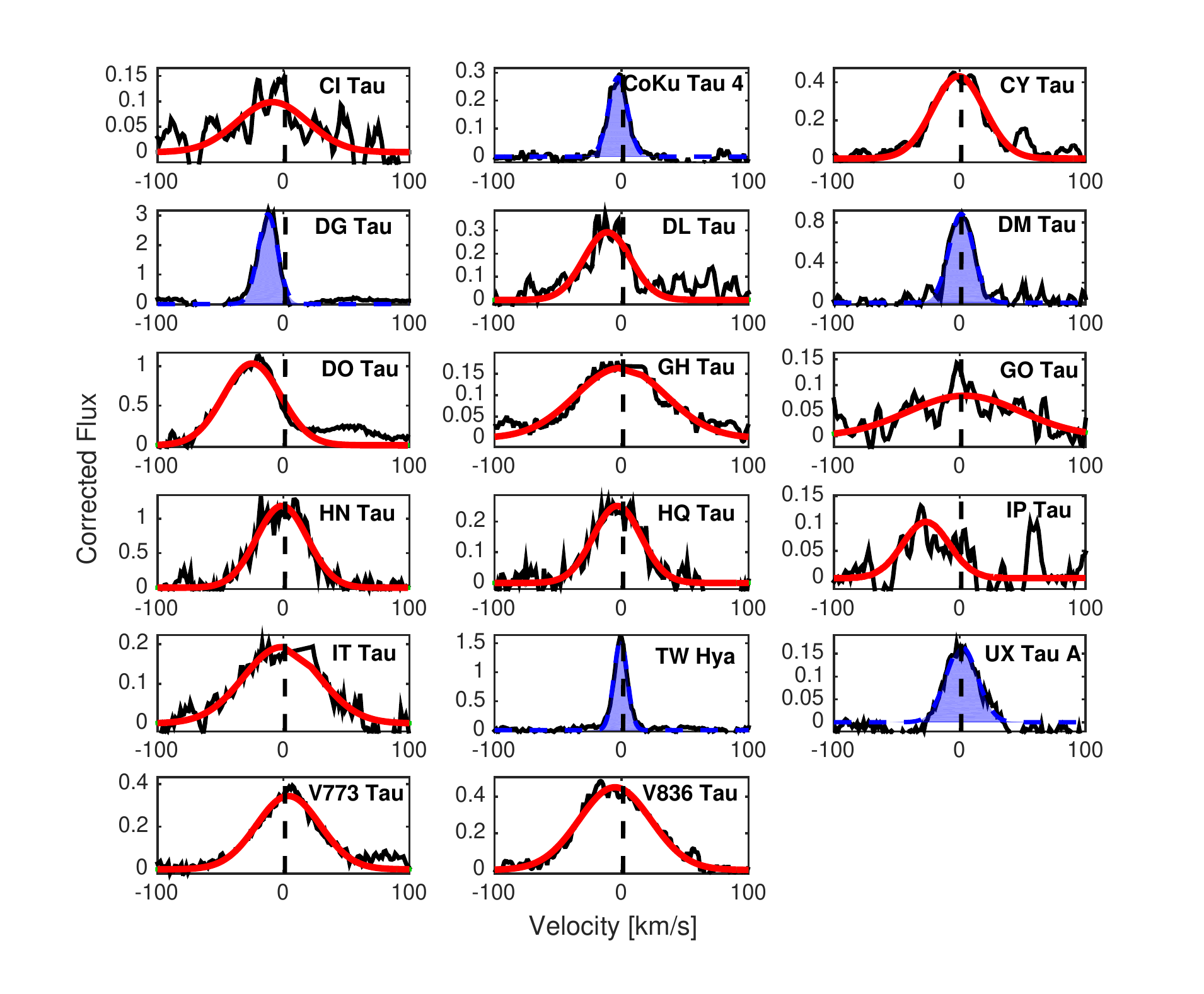} 
   \caption{Residual \oi{} 6300 {\AA} LVC profiles, after the HVC has been removed, for sources that can be fit with one gaussian. Areas shaded in blue represent the narrow component of the LVC, whereas red lines represent the broad component of the LVC as explained in Section~\ref{sect:the_LVC}. }
   \label{fig:LVC_1gaus}
   \end{center}
\end{figure*}

\begin{figure*}[h] %  figure placement: here, top, bottom, or page (H,T,B,P RESPECTIVELY)
   \begin{center}
   \includegraphics[width=5in]{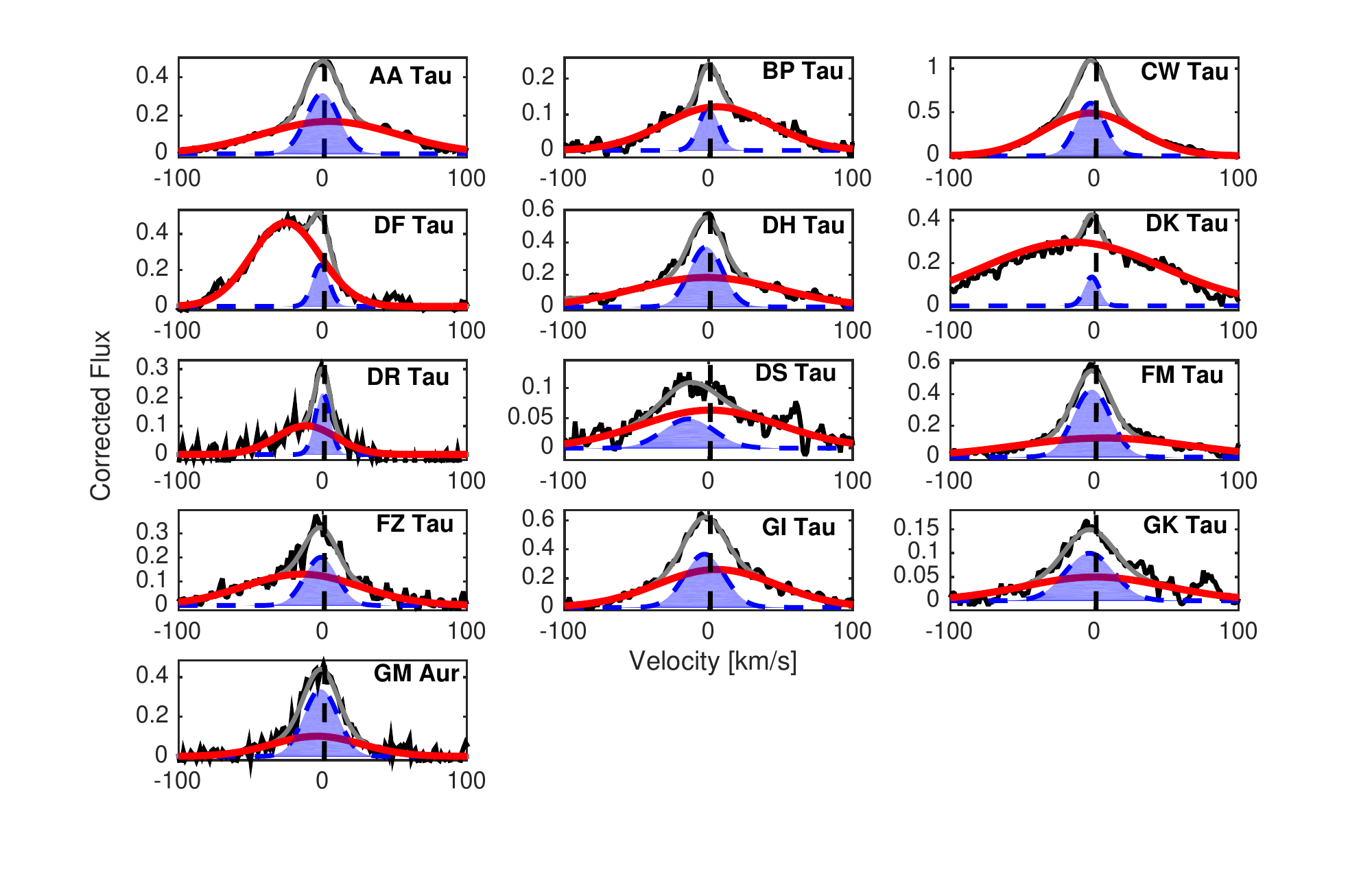} 
   \caption{Residual \oi{} 6300 {\AA} LVC profiles, after the HVC has been removed, for sources that are fit with two gaussians. Areas shaded in blue represent the narrow component of the LVC, whereas red lines represent the broad component of the LVC as explained in Section~\ref{sect:the_LVC}. Gray lines show the total (narrow+broad component) fits.}
   \label{fig:LVC_2gaus}
   \end{center}
\end{figure*}

 In this section we subdivide the LVC into two components, one broad (BC) and one narrow (NC), based on examination of residual LVC profiles generated by subtracting the HVC component fits from the observed profile. The motivation for this further subdivision is that although \oi{} LVC residual profiles from 17 stars are described by one gaussian component, 13 have LVC profiles that require a composite of two gaussians, with one component significantly broader than the other.  The residual LVC \oi{} 6300 {\AA} profiles, with component fits superposed, are presented in two figures, in Figure~\ref{fig:LVC_1gaus} for the 17 single component fits and in Figure~\ref{fig:LVC_2gaus} for the 13 two component fits. 
 

\begin{figure}[h] %  figure placement: here, top, bottom, or page (H,T,B,P RESPECTIVELY)
   \centering
  \includegraphics[width=0.5\textwidth]{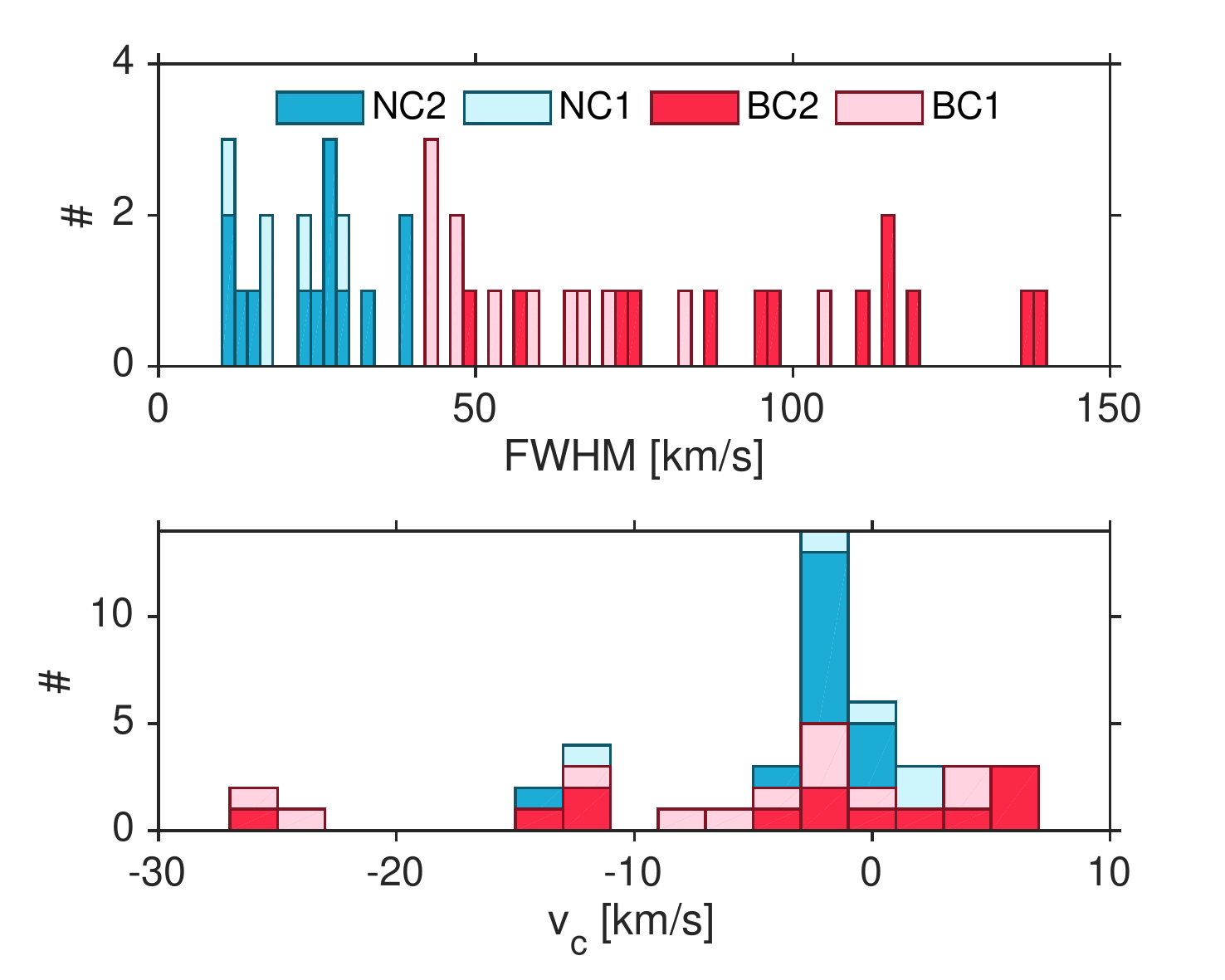}
   \caption{{\it} Upper: Histogram of FWHM for the components found in the \oi{} 6300 {\AA} LVC. The darker shading highlights sources with 2-component fits, with narrower FWHM (blue) clearly separated from broader FWHM (red) components. This leads to the adoption of 40 km/s as the boundary between NC and BC LVC emission. Lighter shading with the same color scheme shows the distribution of the sources with a single kinematic component.  {\it Lower}: Histogram of the  peak centroid values for \oi{} 6300 {\AA} for the components found in the \oi{} 6300 {\AA} LVC. Again, darker shading is for sources with 2-component fits. Bin sizes are 2 km/s. }
   \label{fig:histogram}
\end{figure}

The combination of a broad and narrow component in the 13  composite LVC profiles is shown quantitatively by the distribution of their FWHM, presented in the upper panel of Figure \ref{fig:histogram}. In the figure, the LVC with two-component fits are highlighted with a darker shading than those with one-component fits.  Among the components with the composite profiles, the narrower component, with FWHM from 12 to 39 km/s, a median of 27 \,km/s and a standard deviation 9 km/s, can be contrasted with the broader component, with FWHM from 49 to 140 km/s, a median of 97 \,km/s and a standard deviation of 29\,km/s.  Based on this FWHM separation between the two components, we designate NC LVC  as those with FWHM $\leq$ 40 km/s, and BC LVC as those with FWHM $>$ 40 km/s, and color-code them in the figure with red for BC and blue for NC.  With this subdivision, we can further classify the single-component LVC fits (lighter shading in Figure \ref{fig:histogram}), into 12 BC LVC and 5 NC LVC. In sum, of the 30 TTS with \oi{} 6300 {\AA}  LVC emission, 18 have NC and 25 have BC, with 13 stars showing both components. 

The lower panel of Figure~\ref{fig:histogram} presents the distribution of centroid velocities (all by definition less than 30 km/s) of the NC and BC LVC. Their centroid velocities overlap, together spanning a range from -27 km/s to +6 km/s with both groups peaking at blueshifts of a few km/s.  A K-S test between the NC and the BC centroid velocities gives a $\sim$ 9\% probability that they are drawn from the same parent population, in the sense that the NC and BC velocity distributions are statistically indistinguishable. However, there are a few considerations that suggest otherwise. Although they have similar average centroid velocities, -2.5 km/s for the NC and -3.7 km/s for the BC, the standard deviation for the NC is almost a factor of three smaller than for the BC, 3.6 km/s versus 9.7 km/s. We will explore possible differences in the centroid velocities of the NC and BC in the next section.

\begin{figure*}[h] %  figure placement: here, top, bottom, or page (H,T,B,P RESPECTIVELY)
   \centering
    \includegraphics[width=6in]{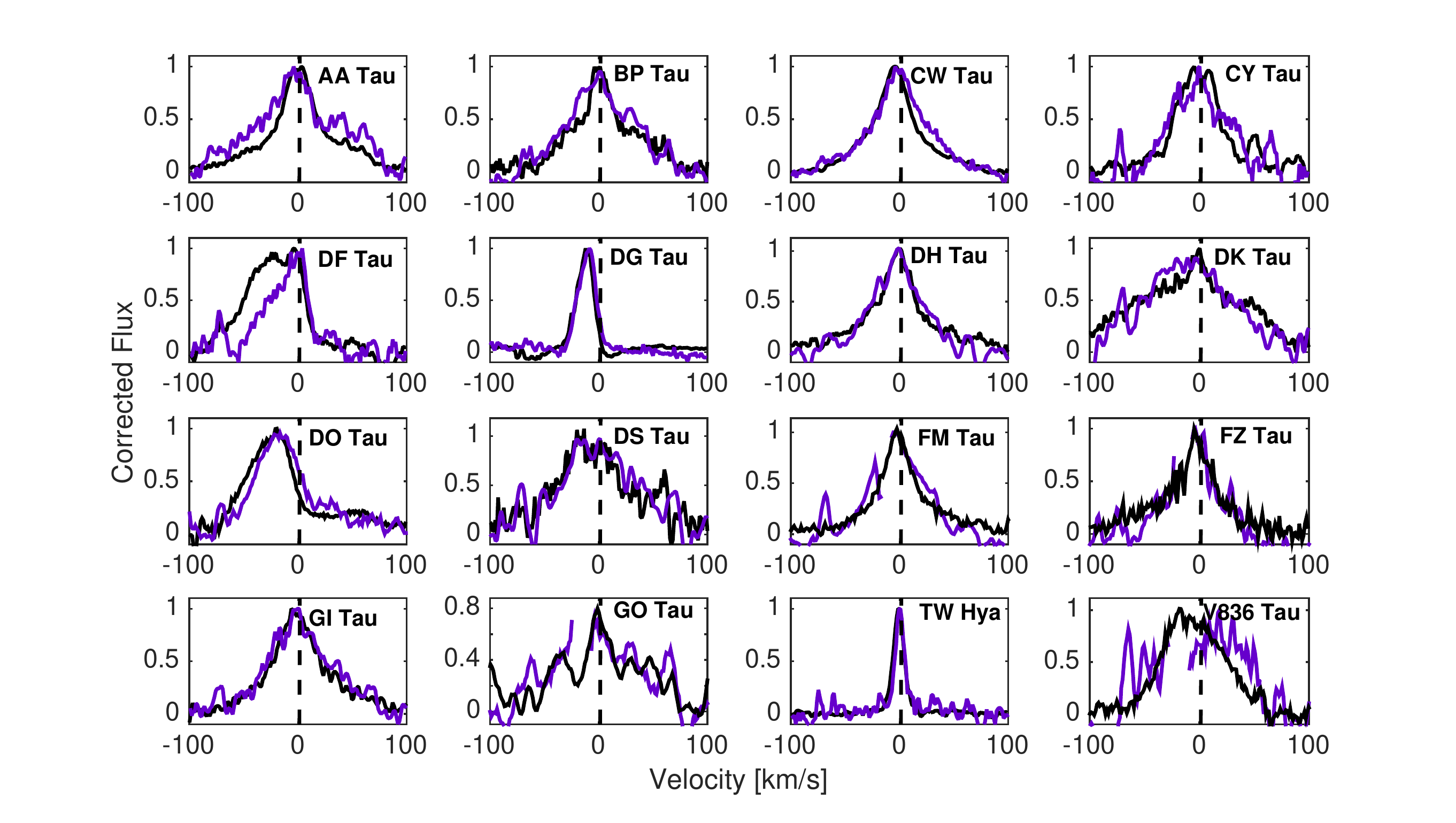}
   \caption{\oi{} residual LVC profiles for 6300 {\AA} (black) and 5577 {\AA} (purple) scaled to a peak of 1 for the 16 sources where both lines are detected.}
   \label{fig:5577_scaled_6300}
\end{figure*}

A comparison of the residual LVC profiles between the two \oi{} lines can made for the 16 stars with \oi{} 5577 {\AA}, where only 3 stars (CW Tau, DG Tau, DO Tau) had HVC that needed to be subtracted. This is illustrated in Figure~\ref{fig:5577_scaled_6300} where both residual LVC are normalized to their respective peaks so the profile structure can be compared. We see that for most stars the LVC structure is very similar between the two lines. This similarity is strengthened when their LVC fit parameters, listed in Table \ref{tab:2_gaus_params}, are compared. For example in AA Tau and BP Tau the centroids and FWHM of both the BC and NC between the two lines is identical and the difference in the superposed profiles is due to a different ratio of BC to NC. This similarity again suggests that the process of subdividing the LVC into BC and NC is robust. 

There are 3 cases where there are differences in the LVC components between the two \oi{} lines. The most extreme case is DF Tau, where the \oi{} 6300 {\AA} BC LVC extends further to the blue than 5577 {\AA}, with centroids of -26 km/s and -14 km/s, respectively. For this star it is possible that there is uncorrected HVC emission at 6300 {\AA} that is blended with the LVC (see Figure~\ref{fig:Pan}) . Reassuringly, this is the only case with such an extreme difference. For the other two cases, CW Tau and DO Tau, the LVC centroid velocities are again more blueshifted at 6300 {\AA} compared to 5577 {\AA} but by only a few km/s. We will discuss these differences further in the next section. However we reiterate that for the majority of the stars the centroids and FWHM of the LVC components are essentially identical, within the errors, for both \oi{} lines, suggesting that this approach is robust.

\begin{figure}[h] %  figure placement: here, top, bottom, or page (H,T,B,P RESPECTIVELY)
   \begin{center}
   \includegraphics[width=0.5\textwidth]{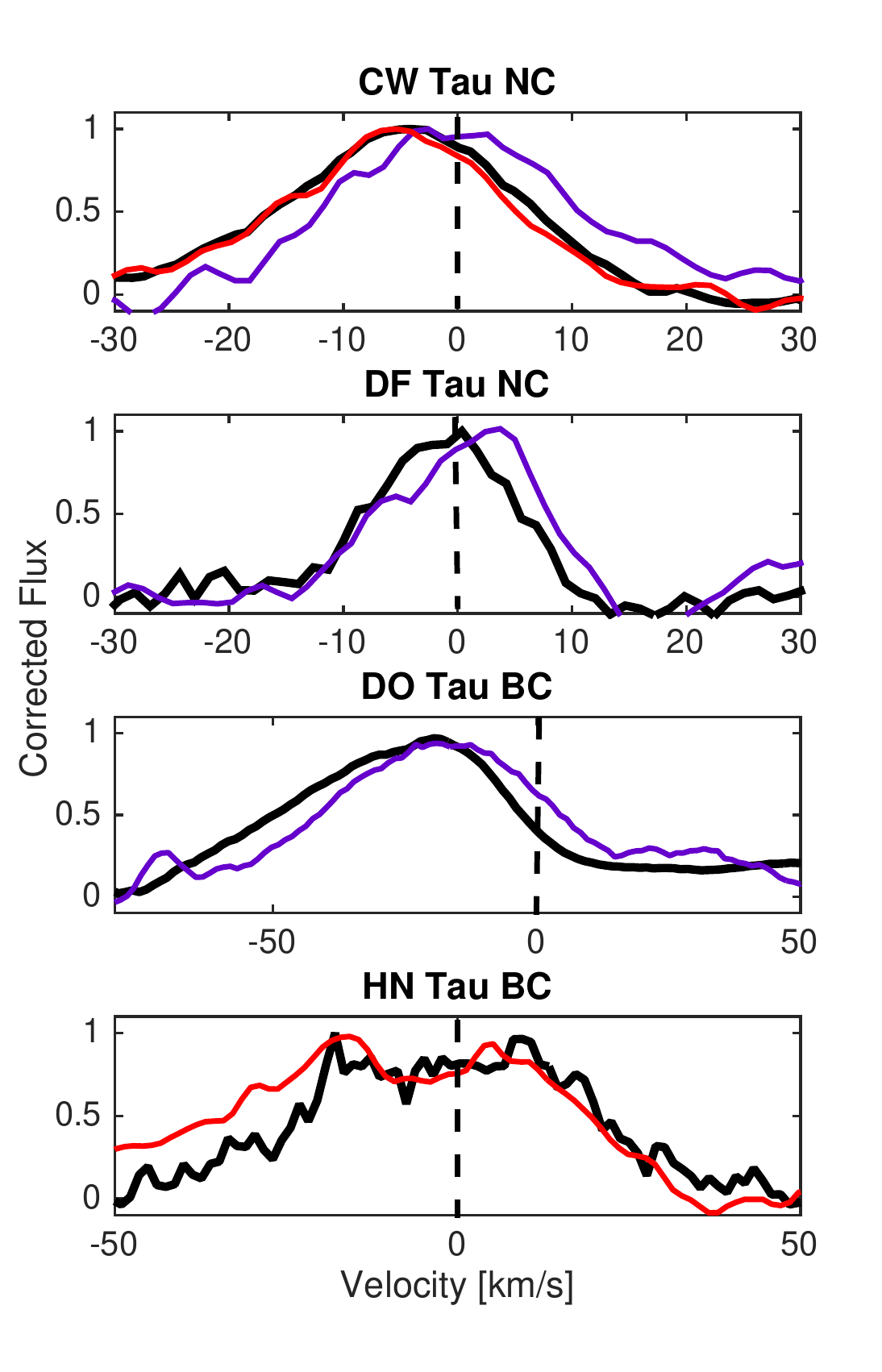} 
   \caption{ Superposed residual LVC profiles, after removal of the HVC, for \oi{} 6300 (black), \oi{} 5577 (purple) and \sii{} 6731 (red) for the 4 sources that show different velocity centroids in the LVC components (see Section~\ref{sect:LVCshifts}). In the cases of CW and DF Tau (top two) the \oi{} LVC had both a BC and a NC, but in this figure the BC is removed. When detected the \sii{} emission is more blueshifted than the \oi{} 6300\,\AA{} which, in turn, is more blueshifted than the \oi{} 5577\,\AA{}.}
   \label{fig:s2lvc}
   \end{center}
\end{figure}
 
\subsection{Velocity Shifts Among LVC}\label{sect:LVCshifts}

Although most stars with both \oi{} lines show similar LVC kinematic properties, 4 sources, all with significant HVC emission, show LVC velocity centroids with velocity shifts among different lines, in the sense that the \oi{} 5577 {\AA} is less blueshifted than the \oi{} 6300 {\AA}, which in turn is less blueshifted than the \sii{} 6731 {\AA}. These small velocity offsets are illustrated in Figure~\ref{fig:s2lvc}, where the forbidden line LVC are superposed and plotted on an expanded velocity scale.

In the case of CW Tau and DF Tau, the velocity differences are seen in the NC of the LVC and in DO and HN Tau the velocity shifts are seen in the BC of the LVC. To best illustrate the NC shifts for CW Tau and DF Tau, where both \oi{} lines require a two-component fit to their LVC, in Figure~\ref{fig:s2lvc} the BC of the LVC has been subtracted, so only the NC are shown. The velocity centroids for the 3 NC LVC for CW Tau are -5.5 km/s for the \sii , -2.63 km/s for the \oi{} 6300\,\AA , and 0.92 km/s for the \oi{} 5577\,\AA. For DF Tau they are -1.3\,km/s and 0.9\,km/s for the 6300 {\AA} and the 5577 {\AA} lines, respectively.

For the two stars with shifts in the BC, there are no NC contributions to the LVC. For DO Tau, the centroid velocity for \oi{} 6300 {\AA} is -25 km/s compared to -17 km/s at 5577 {\AA}. Again, for HN Tau, the centroid for \sii{} is -9.4 km/s and -1.2 km for \oi{} 6300 {\AA}, although in this source the major differences in the two lines is a more extended blue wing at \sii{}. A similar effect of more blueshifted LVC in lines of lower critical density was also found by HEG, suggesting that the lower critical density lines reflect acceleration in a slow wind in a few stars (see also Section~\ref{sect:LineRatios}).

\section{Results for the Low Velocity Component}\label{sect:res_LVC}

As shown in the previous section, LVC emission, seen in all 30 TTS with detected forbidden lines, can be further subdivided into NC and BC kinematic features. We do not find any commonality in the line profiles of the 10 multiple systems in our sample and see no trend with the companion separation (see Tables ~3 and 4 in Pascucci et al. 2015 for stellar separations). As such we will not discuss the possible effect of binaries on the LVC. Instead, we will examine how the emission relates to accretion luminosities, assess the role of disk inclination in determining their FWHM, explore disk surface brightness distributions that can account for the observed profiles, and look at line ratios among the different forbidden lines. We begin by comparing relevant aspects of our results to the previous results.

\subsection{Comparison to HEG}\label{sect:HEG_compare}

\begin{figure*}[h] %  figure placement: here, top, bottom, or page (H,T,B,P RESPECTIVELY)
   \begin{center}
   \includegraphics[width=6in]{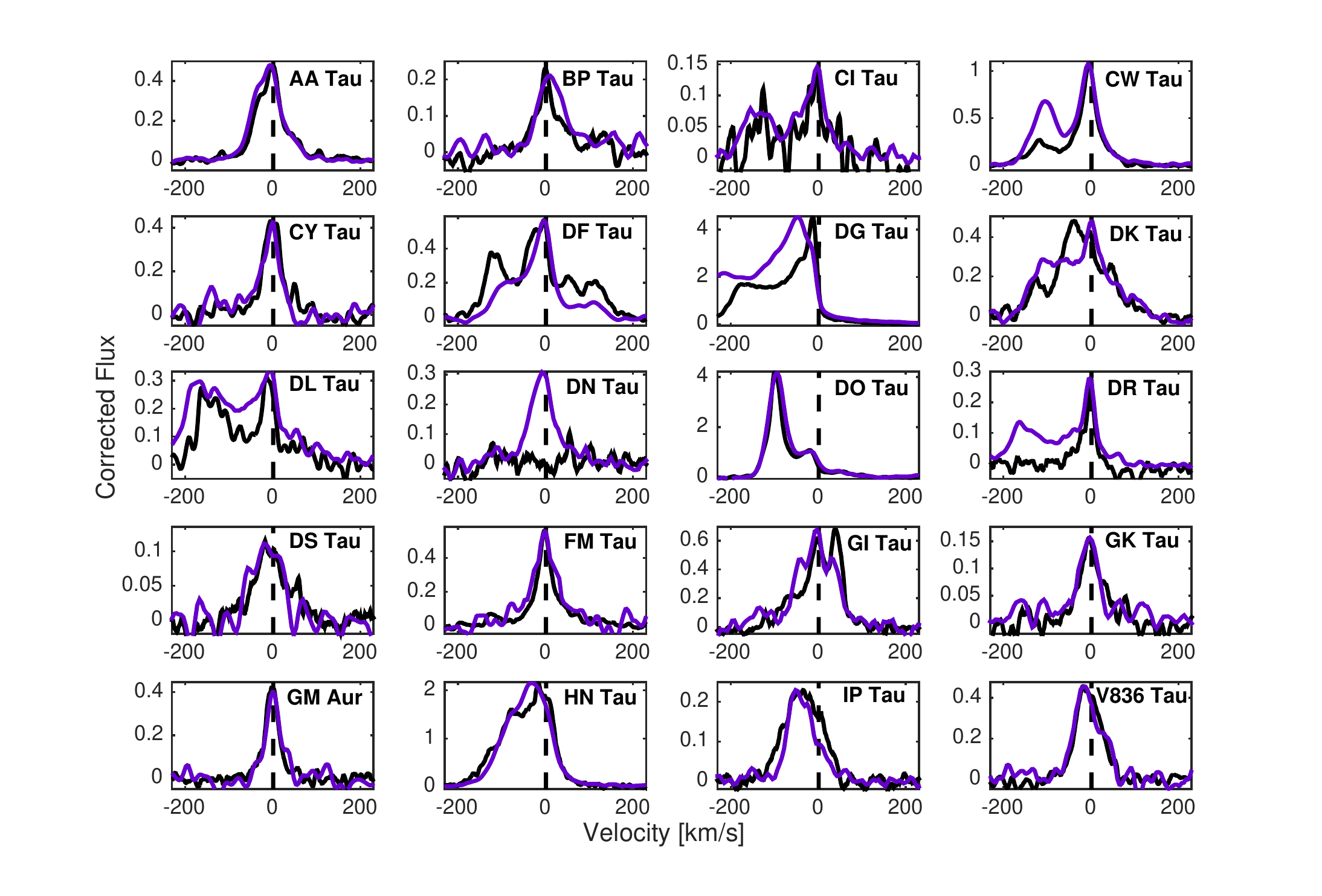}
   \caption{Comparison of  20 \oi{} 6300 {\AA} profiles in common with this work (black) and HEG (purple), scaled to the peak intensities of our spectra.}
   \label{fig:HEG_profile_scaled}
   \end{center}
\end{figure*}

Since both this study and that of HEG focus on TTS in Taurus, there are 20 objects in common whose forbidden line properties can be examined for variability.  In Figure~\ref{fig:HEG_profile_scaled}, we overlay the \oi{} 6300 {\AA} profiles for these 20 sources, with the older profiles rescaled in peak intensity to match the current spectra, providing a comparison of the kinematic structure of the profiles over several decades. Recalling that there is almost a factor of 2 higher spectral resolution in the HIRES spectra (6.6 km/s versus 12 km/s), two things are apparent.  {\it (1) The structure of the LVC is identical in most instances}. One dramatic exception is DN Tau, where the LVC has disappeared between 1995 and 2006. The DG Tau LVC profile is also different, where in 2006 there is a NC peak at -12 km/sec that is not seen in HEG. In the earlier spectrum the lowest velocity peak is at -50 km/s and does not qualify as "Low Velocity" emission. However the red side of the profiles in both epochs are very similar, suggesting the more recent LVC peak may have been present, but much weaker, a few decades ago. 
(2) {\it The HVC is quite different in 4 sources} (CW Tau, DF Tau, DG Tau, and DR Tau), and in the case of DR Tau it has vanished between 1995 and 2006. Since the HVC arises in spatially extended microjets that form time variable knots of shocked gas, such changes are not surprising, especially as differences in slit length and orientation coupled with differences in the width of the stellar point spread functions (PSF) between the two studies can also yield different profiles for extended emission. However, we conclude that the velocity structure of the LVC is generally stable over a timescale of decades in most stars.

\begin{figure}[h] %  figure placement: here, top, bottom, or page (H,T,B,P RESPECTIVELY)
   \begin{center}
   \includegraphics[width=0.5\textwidth]{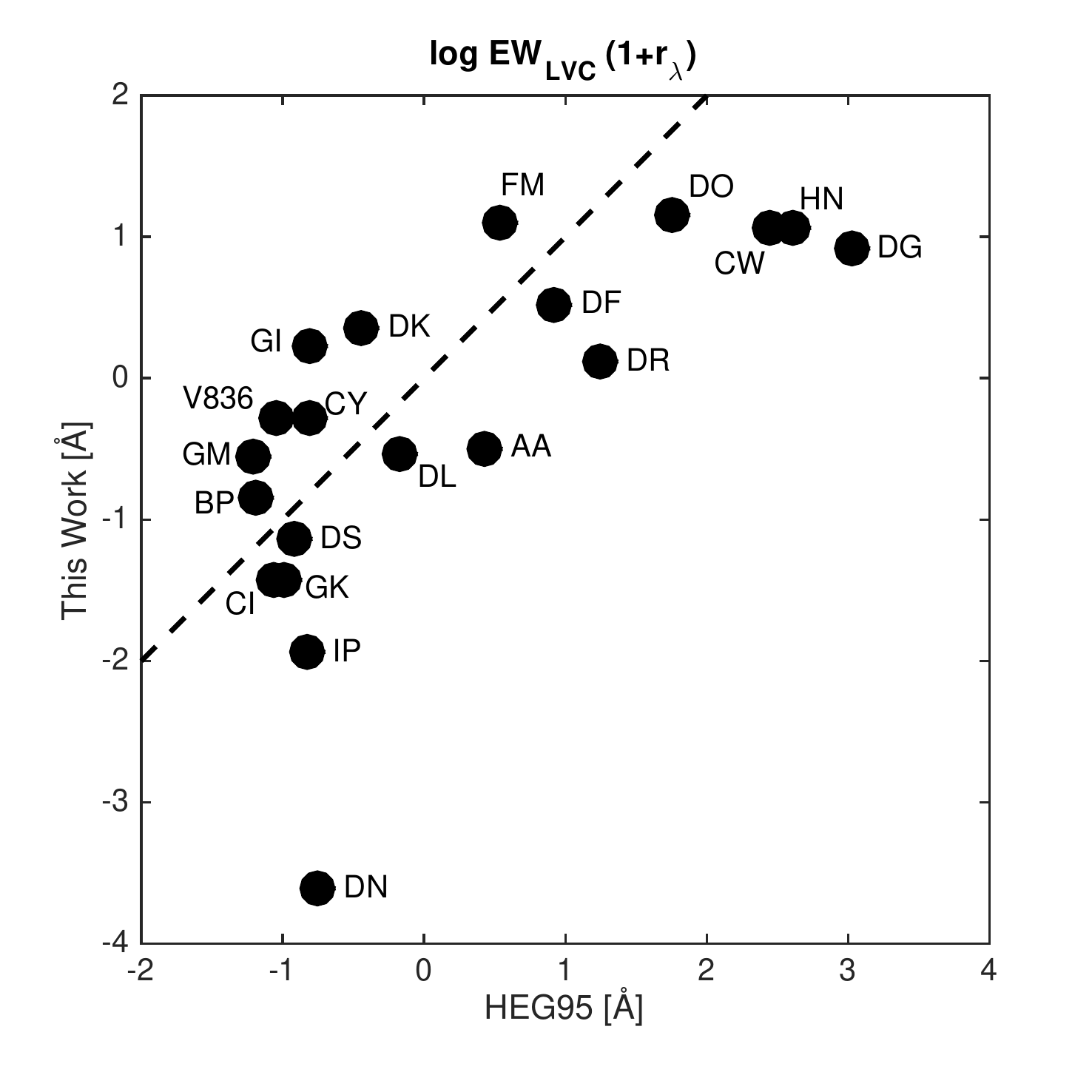}
   \caption{Comparison of veiling corrected EW of the \oi{} 6300 {\AA} LVC in common with this work and HEG. The definition of the LVC is different in the two studies, where HEG includes any HVC emission within 60 km/s of the stellar velocity. The dashed line is for a one-to-one correspondence.}
   \label{fig:EW_veiling_HEG95}
   \end{center}
\end{figure}

We can also examine whether the strength of the LVC emission is comparable between the 2 studies. We do this in Figure~\ref{fig:EW_veiling_HEG95} by comparing the LVC equivalent widths normalized to the photospheric continuum, (EW $\times$ ($1+r_{\lambda}$)), with the caveats that the HEG definition of the LVC was any emission within 60 km/s of the stellar velocity, in contrast to our approach of isolating kinematic components by Gaussian fitting, and the difference in wavelength for which the veiling is reported, $r_{5700}$ in HEG versus $r_{6300}$ here.  This comparison of veiling corrected EW shows comparable emission strength of the LVC for most sources, with the exception of DN Tau, which, as seen in Figure~\ref{fig:HEG_profile_scaled}, has disappeared since 1995. In the other sources, the differences in the veiling corrected equivalent width between the two epochs could be attributed to differences in the definition of the LVC, variation in the stellar continuum and/or variability in the line luminosity itself, or possibly to extended emission in some LVC observed with different stellar PSF or slit orientations, as hinted at in \cite{hirth1997}.

The complete disappearance of  \oi{} 6300 {\AA} emission in DN Tau is surprising, making it the only source known to date that is accreting but has no \oi{} 6300 {\AA} emission. The \oi{} 6300 {\AA} equivalent width found by HEG is 0.43 {\AA} while our upper limit  $\sim$ 0.007 {\AA}, about two orders of magnitude lower than in 1995. The corresponding decrease in the H$\alpha$ equivalent width is only $\sim$ 30\% (see \citealt{beristain01}), and differences in veiling ($r_{5700}$ = 0.1 versus $r_{6300}$ = 0) are small, consistent with the estimated uncertainty. We checked for variability in the continuum with the \cite{herbst1994} catalogue. It was rather stable at R mag $\sim$ 11.5 between 1980 and 1986, and brightened by $\sim$ 0.5 mag through 1995, the last year of recorded data. The profile for DN Tau seen in HEG qualifies as solely LVC, which combined with our current two epochs for IP Tau (see Figure~\ref{fig:iptau}), gives two examples where the LVC has vanished, although in most stars it appears stable over timescales of decades. Additional observations of both DN Tau and IP Tau would be of interest to see if their LVC returns. 

Another interesting comparison is the implication for the velocity of the LVC, noted to be typically blueshifted by $\sim$ 5 km/s by HEG. Although the kinematic structure in the LVC is unchanged in the two studies (Figure~\ref{fig:HEG_profile_scaled}), in the present study of LVC emission we find that only 12/25 of the BC and 9/18 of the NC are blueshifted relative to the stellar photosphere (Table \ref{tab:2_gaus_params}). The process of defining and separating full profiles into HVC, NC, and BC and then subtracting a HVC that is blended with a LVC (e.g. AA Tau) and/or separating a blueshifted BC LVC from a NC LVC (e.g. FZ Tau), means the resultant BC or NC may no longer be identified as blueshifted, although in the full profile the LVC looks to be predominantly blueward of the stellar velocity. However, we find roughly half the LVC components are blueshifted, and as seen in HEG and discussed here in Section~\ref{sect:LVCshifts}, in four stars these blueshifts show velocity gradients among different forbidden lines, with higher blueshifts for lines of lower critical density. We note that these four stars, CW Tau, DF~Tau, DO Tau and HN Tau, have some of the highest veiling corrected \oi{} 6300 {\AA} LVC equivalent widths, as shown in Figure~\ref{fig:EW_veiling_HEG95}. 

\subsection{Comparison to Recent Studies of Luminosity Relations}	
	
The two recent studies of forbidden LVC emission, R13 and N14,  found correlations between the accretion luminosity with both the luminosity of the \oi{} 6300 {\AA} LVC and the stellar luminosity, but not with the X-ray luminosity, for samples primarily from Taurus and Lupus. Additionally, a relation between $L_{acc}$ and $L_{\ast}$ was found by \cite{mendigutia2015} for a large sample of objects in various star forming regions.

Although the correlation in R13 was based mostly on the HEG sample, with which we strongly overlap, we look here at these relations for our data, since R13 used the original 1995 assessment of the LVC luminosity in contrast to our component fitting and here we have a more consistent and reliable conversion of EW of both H$\alpha$ and \oi{} to line luminosities. We use the Astronomy SURVival package (ASURV) developed by \cite{lavalley92} because it includes upper limits in the linear regression and correlation tests. 

\begin{figure}[h] %  figure placement: here, top, bottom, or page (H,T,B,P RESPECTIVELY)
   \begin{center}
   \includegraphics[width=0.5\textwidth]{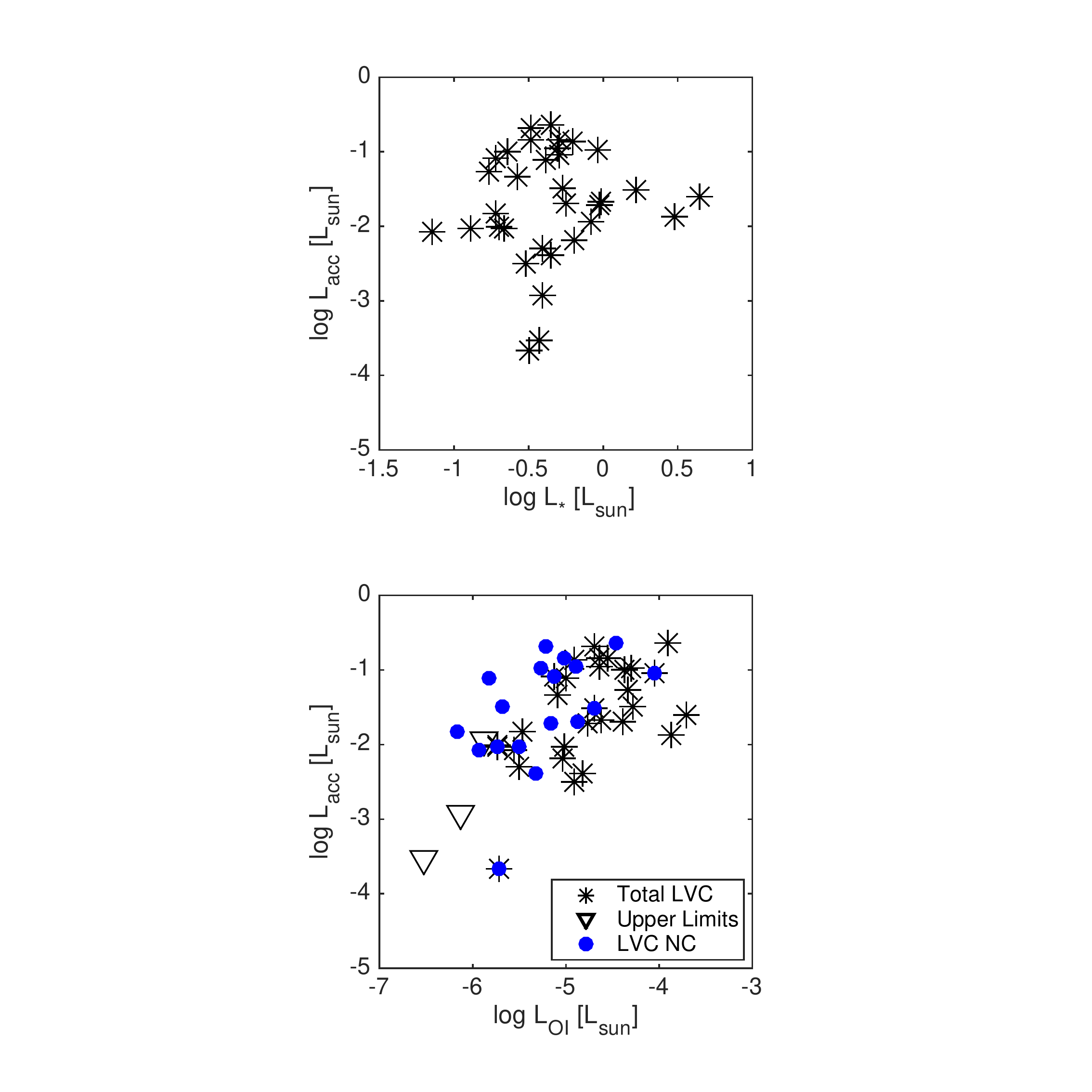}
   \caption{{\it Upper:} There is no correlation between the accretion luminosity and the stellar luminosity. {\it Lower:} The accretion luminosity is correlated with the luminosity of \oi{} 6300 {\AA}, both for the full LVC (asterisks) and for the NC of the LVC (blue circles), when detected.}
   \label{fig:Lacc}
   \end{center}
\end{figure}

In the lower panel of Figure~\ref {fig:Lacc}, we compare $L_{OI}$ for both the complete LVC and only for the NC of the LVC to $L_{acc}$. For the former, the Kendall $\tau$ test lends to a strong correlation, with a probability of 0.07\% that the variables are uncorrelated.\footnote{This test includes three non-detections in $L_{OI}$ (DN Tau, V710 Tau, and VY Tau).} The best fit linear regression is:
	
\begin{equation} 
log L_{[OI]~LVC}= 0.65(\pm 0.13) \times log L_{acc} - 3.84 (\pm 0.23) 
\end{equation} 

\setlength{\parindent}{0cm} when both luminosities are measured in $L_\sun$. This fit is the same as that in R13 and N14. The reason we have a larger uncertainty on the slope and intercept of our linear regression is because our targets cover a narrower range of $L_{acc}$ ($log L_{acc}$ between -3.7 and -0.6) than the samples of R13 ($log L_{acc}$ between -3.2 and 1.8) and N14 ($log L_{acc}$ from -4.8 to -0.3).

\setlength{\parindent}{0.5cm} The lower panel of Figure~\ref {fig:Lacc} also shows the relation between $L_{OI}$ for the NC of the LVC and $L_{acc}$. Again, for the 18/33 sources for which NC LVC is detected there is only a $\sim$ 2\% probability that the NC LVC and accretion luminosity are uncorrelated. Of note in this figure is the fact that the NC of the LVC is found over the full range of $L_{acc}$, and when present, increases proportionately with the accretion luminosity. In contrast to the correlation with $L_{OI}$, $L_{acc}$ shows no correlation with $L_{\ast}$. Although such a correlation has been found in other samples, its absence here is probably because our sample covers a factor of $\sim$100 in $L_{\ast}$ while that of \cite{mendigutia2015} which includes brown dwarfs, spans 8 orders of magnitude.

For the comparison with the X-ray luminosity, 22 of our sources have $L_{X}$ values as reported in Table \ref{tab:source_properties}. As in the earlier studies, we find no correlation with $L_{OI}$ of the LVC, with a Kendall $\tau$ probability not low enough (10.8\%) to indicate that the $L_{X}$ and $L_{OI}$ LVC are correlated.\footnote{This test includes two non-detections in $L_{OI}$ (DN Tau and V710 Tau).} 

\subsection{Fractional Contributions of the BC and NC to the LVC}

\begin{figure}[h] %  figure placement: here, top, bottom, or page (H,T,B,P RESPECTIVELY)
   \centering
 \includegraphics[width=0.5\textwidth]{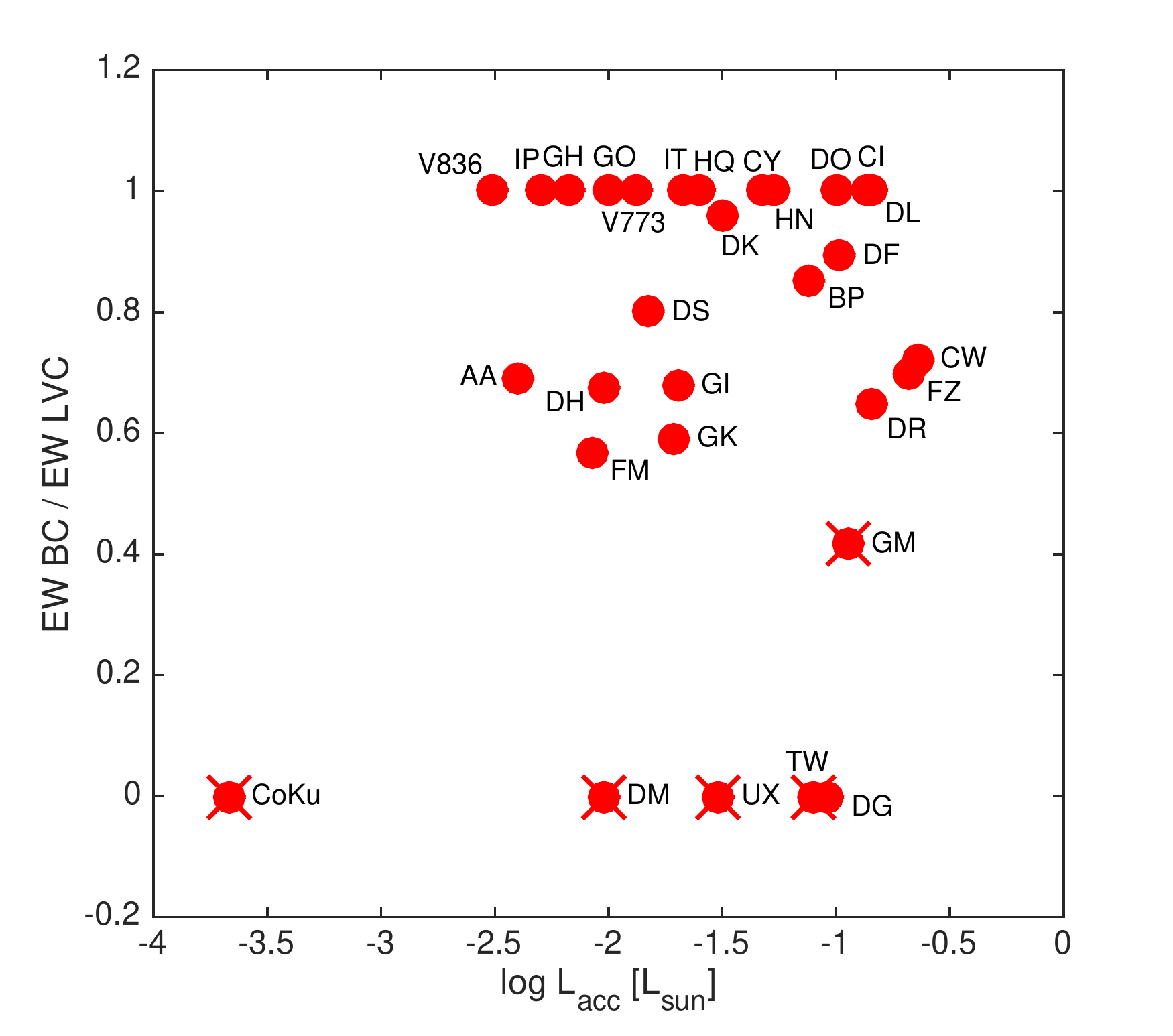} 
   \caption{Proportion of LVC with BC emission plotted as a function of the accretion luminosity. Transition disk sources are marked with X's.} 
   \label{fig:bcfraction}
\end{figure}

Thirteen out of 30 sources with \oi{} 6300 {\AA} emission have LVC comprised of a combination of BC and NC emission. The remaining sources show LVC that are either entirely BC (12/30) or entirely NC (5/30). The proportion of the LVC that is BC (or NC) emission is not strongly dependent on the accretion luminosity,  as shown in Figure~\ref{fig:bcfraction}. For example, the 12/30 sources that show only BC emission cover a wide range of disk accretion rates. The same is true for the 5/30 sources that show only NC emission. However, this figure also shows distinctive behavior for transition disk sources, all 5 of which have LVC emission dominated by the NC, and 4 of those 5 have no BC emission. 

Combining the relations in  Figure~\ref {fig:Lacc} and Figure~\ref{fig:bcfraction} we find that while NC and BC emission are seen over a wide range of accretion luminosities, the luminosity of each component correlates with the accretion, but not the stellar, luminosity.

\subsection {Velocity Centroids of the BC and NC of the LVC}

\begin{figure}[h] %  figure placement: here, top, bottom, or page (H,T,B,P RESPECTIVELY)
   \centering
 \includegraphics[width=0.5\textwidth]{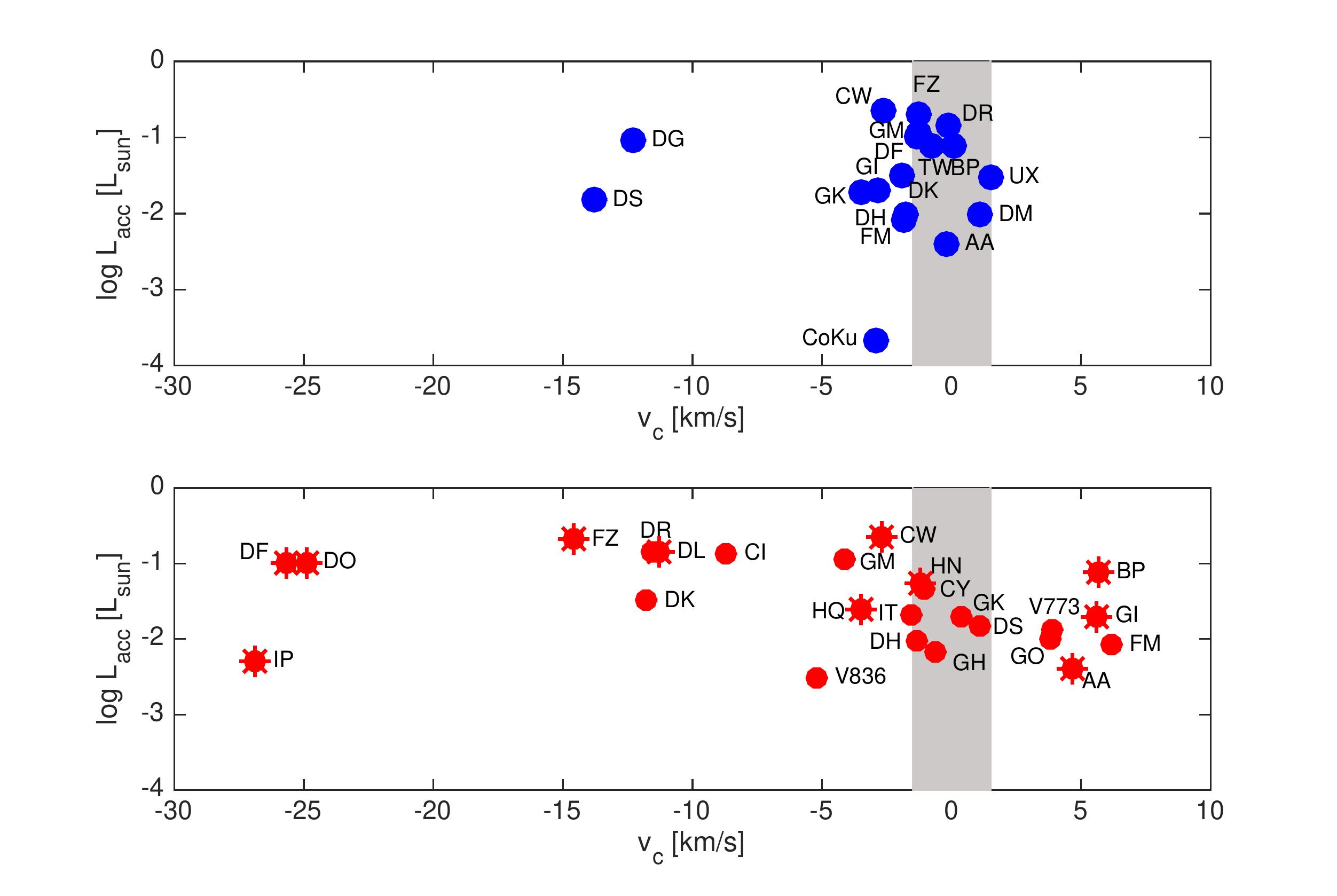} 
   \caption{Relationship between $L_{acc}$ and $v_c$ for the 6300 {\AA} NC of the LVC (upper) and BC of the LVC (lower). Sources with HVC removed are marked with spokes. The gray area around the stellar velocity, $\pm$ 1.5 km/sec, marks the uncertainty in centering the profiles.}
   \label{fig:Lacc_Vc}
\end{figure}

The histogram of the velocity centroids for the two components of the \oi{} 6300 LVC  shows considerable overlap in velocity distributions (see Figure~\ref{fig:histogram}). Here we add another dimension in the comparison of the centroid velocities, with Figure~\ref{fig:Lacc_Vc} showing the relationship between $L_{acc}$ and $v_c$ for the BC and NC. We conservatively use a $\pm$1.5\,km/s from the stellar velocity as the area (depicted in gray in Figure~\ref{fig:Lacc_Vc}) within which velocities cannot be distinguished from being at rest with respect to the star.

For the BC centroids there is a tendency for the stars with higher accretion rates to have blueshifted $v_c$. Specifically, 7/12 stars with log $L_{acc}/L_\odot$ $\geq$ -1.5 have $v_c$ blueshifted from -9 to -26 km/s, while with one exception (IP Tau) all the stars with lower accretion rates have $v_c$ within 6 km/s of the stellar velocity. A Kendall $\tau$ test gives a probability of only 2\% that the BC centroids and  $L_{acc}$ are uncorrelated.
Interestingly, 6 sources have BC that are redshifted between +3.8 to +6.2 km/s (AA Tau, BP Tau, FM Tau, GI Tau, GO Tau, V773 Tau). While blueshifted centroids are readily associated with winds (12 sources), and unshifted centroids with bound disk gas (7 sources), the small redshifts seen in 6 sources can also be consistent with a wind with certain disk geometries. For moderate disk inclinations or strong flaring of the surface, the extended disk height at large radii may obscure (from the observer's perspective) the approaching part of a wind with a wide opening angle from the inner disk while the receding gas from the far side of the disk remains unobstructed (Gorti et al. in preparation).
  
Among the 18 NC detections, there is no trend between $v_c$ and accretion luminosity and the centroid velocities either have small blueshifts or are consistent with the stellar velocity. A Kendall $\tau$ test lends to a 65\% chance that the NC centroids and $L_{acc}$ are uncorrelated. Blueshifted NC centroids ($v_c$ from -1.75 to -13.8  km/s) indicative of slow winds are seen in 9 sources (CoKu Tau 4, CW Tau, DG Tau, DH Tau, DK Tau, DS Tau, FM Tau, GI Tau, GK Tau)  while the remaining 9 are consistent with bound gas (AA Tau, BP Tau, DF Tau, DM Tau, DR Tau, FZ Tau, GM Aur, UX Tau A and TW Hya). There are no redshifted NC.

From these comparisons, although the K-S test comparing the distribution of NC and BC centroids indicated a 9\% chance they were drawn from the same parent population (Section~\ref{sect:the_LVC}), we conclude that there are significant differences in the behavior of the $v_c$ in these two components.

\subsection{FWHM of BC and NC of the LVC: Dependence on Disk Inclination}\label{sect:inclination}

	The two-component nature of the LVC forbidden line emission in TTS is reminiscent of kinematic behavior found in a series of papers looking at high resolution VLT-CRIRES spectra of the 4.6 $\mu$m CO ro-vibrational fundamental lines in Class I and II sources (\citealt{bast2011, pontoppidan2011, brown2013, banzatti2015}). The CO $\nu=$1-0 line sometimes shows a NC/BC structure, while the $\nu=$2-1 transition predominantly traces the BC, which allows it to be isolated when both transitions are observed.  The finding that the FWHM of the broad and narrow CO features may derive from the system inclination leads us to explore this possibility for the FWHM of the two components of the \oi{} 6300 {\AA} LVC. 
	
\begin{figure}[h] %  figure placement: here, top, bottom, or page (H,T,B,P RESPECTIVELY)
   \centering
   \includegraphics[width=0.5\textwidth]{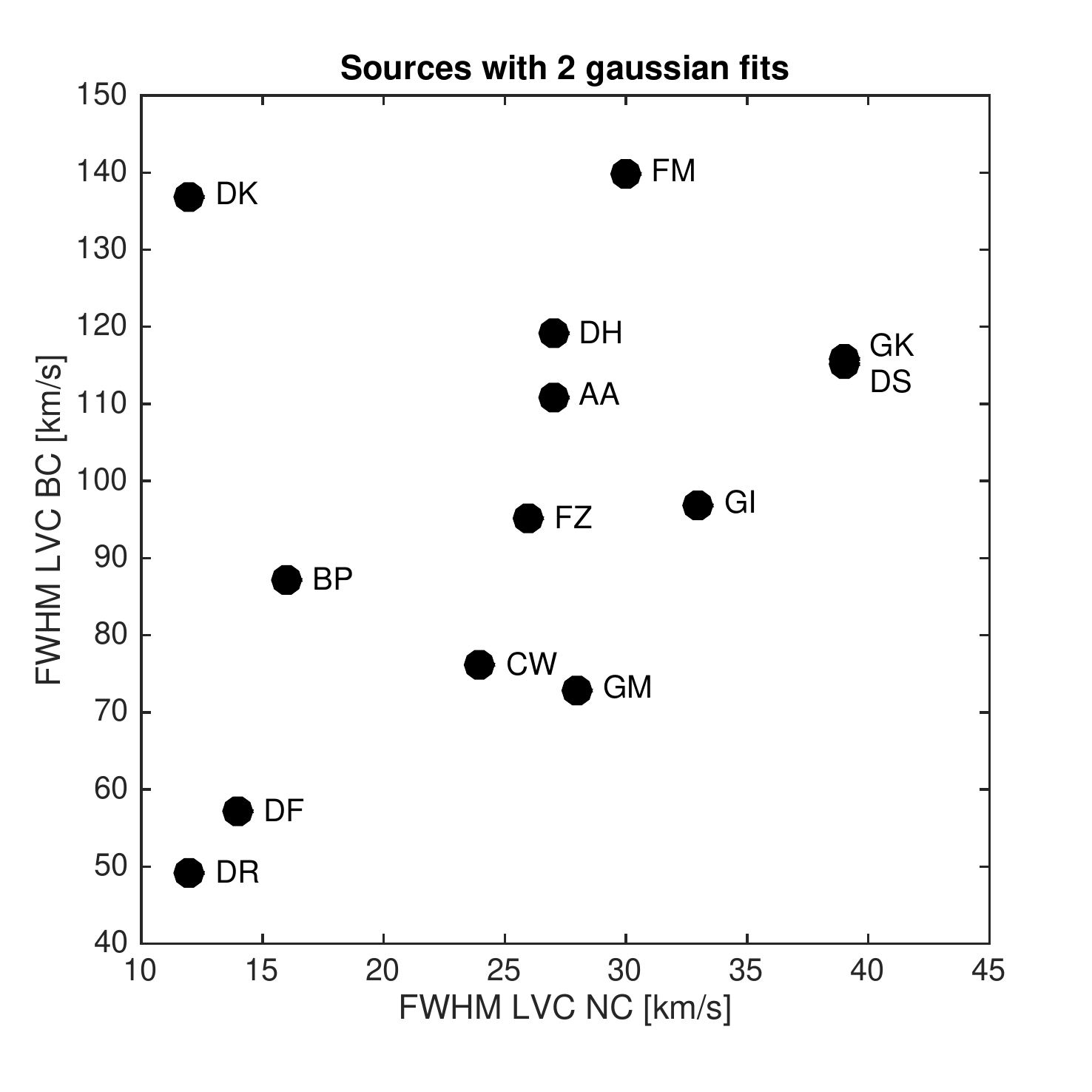} 
   \caption{Relation between the FWHM for the NC and BC of the \oi{} 6300 LVC in the 13 sources where both are present.}
   \label{fig:fwhm}
\end{figure}

	First, we look at the relation between the FWHM of the NC and BC in those 13 sources where both components are seen at \oi{} 6300 {\AA} LVC. Each of these components shows a range in line widths (Figure \ref{fig:histogram}), and if both were rotationally broadened we would expect to see a correlation between them in sources where both are present. We show this relation in  Figure~\ref{fig:fwhm}, where the FWHM of the NC, ranging from 12 to 39 km/sec, is reasonably well correlated with the FWHM of the BC, ranging from 44 to 140 km/sec, in all but one source (DK Tau). The Kendall $\tau$ probability that the two quantities are uncorrelated is indeed only 0.9\% if we exclude DK~Tau, but it increases to 7.5\% when DK Tau is included.
	
\begin{figure*}[h] %  figure placement: here, top, bottom, or page (H,T,B,P RESPECTIVELY)
   \begin{center}
    \includegraphics[width=7in]{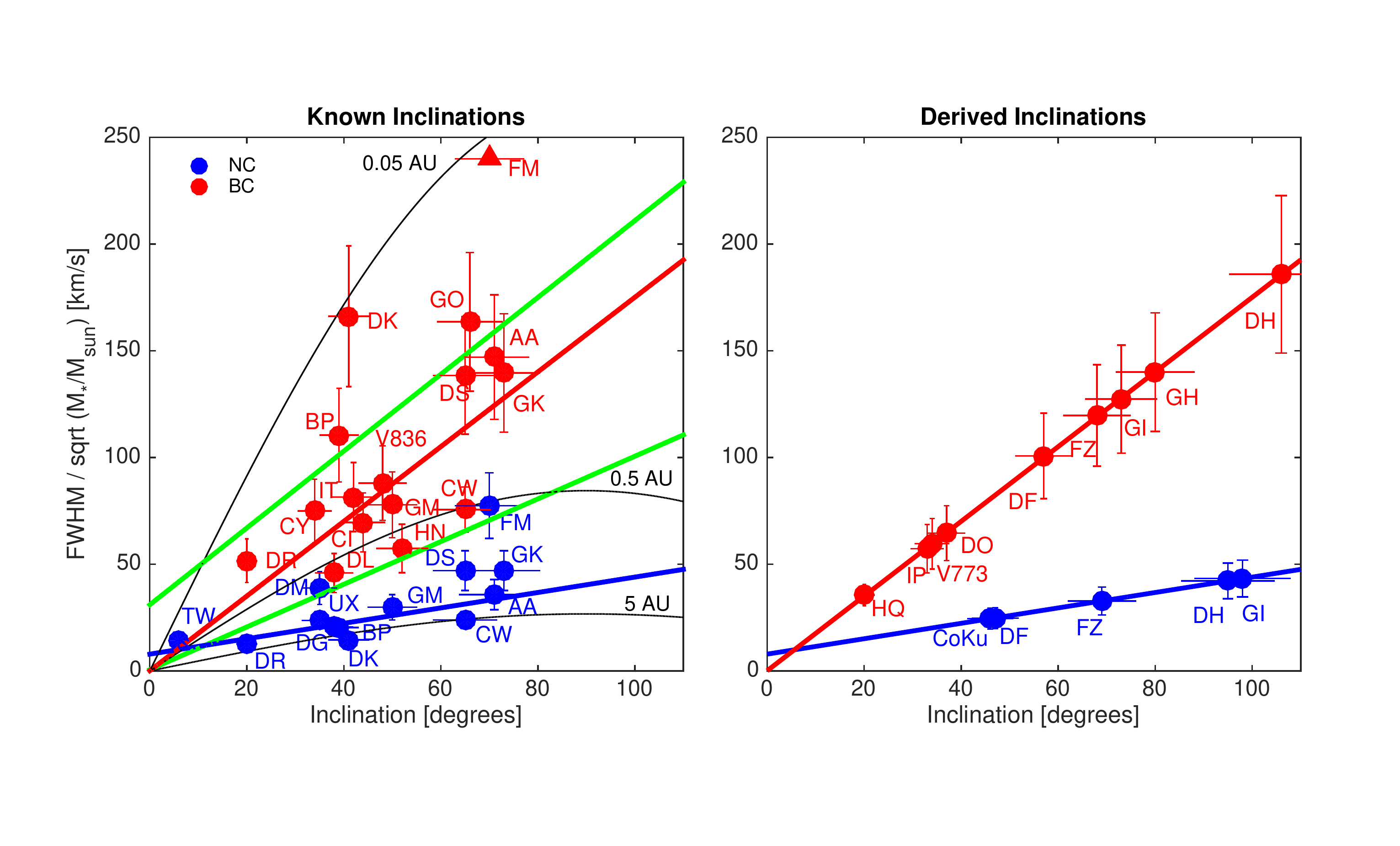}
   \caption{{\it Left}: Relation between the FWHM normalized by the stellar mass and known inclinations, for both NC and BC LVC. Derived linear fits are shown in red (BC) and blue (NC), those found by \cite{banzatti2015} from CO are shown in green. Black lines show line widths as a function of inclination at disk radii of 0.05, 0.5, and 5\,AU. {\it Right}: Inclinations are derived from the observed FWHM for NC and BC observations for sources with unknown inclinations, using the fits from the left panel.}
   \label{fig:inclination_plot}
   \end{center}
\end{figure*}

	We next explore the relationship between the width of the \oi{} 6300 {\AA} LVC NC and BC  with the published values for disk inclination in 22 sources. To directly compare our results with those already published for the CO ro-vibrational  line, we follow \cite{banzatti2015} and plot the FWHM divided by the square root of the stellar mass versus the inclination from spatially resolved disks, rather than the FWHM vs the sine of disk inclination, which would be a better description of the relation expected for Keplerian broadening (left panel of Figure~\ref{fig:inclination_plot}). The figure shows that the two quantities are positively correlated, both for the BC and the NC components. There are two outliers in the BC relation, DK Tau and FM Tau. DK Tau is the source that did not show the expected behavior between NC and BC FWHM in Figure \ref{fig:fwhm} and here we see that the NC is in line with other sources at similar inclinations. The very low mass source FM Tau (0.1 {M$_\odot$}) lies off the scale of the plot, but as we will show in Section~\ref{sect:RadialBrightness}, this can be explained if the emitting region is closer in than for stars of higher mass.
		
 Again, following  \cite{banzatti2015} we will use the positive linear correlation to infer disk inclinations for the remaining sources. In order to include  uncertainties in this process we assume an uncertainty of $\sim$ 10\% in the disk inclination, as reported in the references in Table~\ref{tab:source_properties}.  For the FWHM, we adopt a Monte Carlo approach similar to that discussed in Section \ref{sect:EWs} in relation to the uncertainties on the measured EWs. Using the 17 single component LVC sources, we find an average uncertainty that is $\sim$ 13\% of the measured FWHMs. As for the uncertainties on stellar masses, we assume 10$\%$ for stars with masses $\geq$ 1 M$_\odot$ and 30$\%$ for stars with masses $<$ 1 M$_\odot$ (see \citealt{stassun2014}). Finally, in order to calculate the total uncertainty on the y-axis, we propagate the error on the stellar mass and FWHM, assuming they are independent. 
	
	Adopting these uncertainties and using a linear relationship between FWHM / $\sqrt{M_\ast}$ and disk inclination, we find the following best fits from the 22 sources with measured inclinations:  
		
\begin{equation} 
FWHM{_N}{_C} / \sqrt{M_\ast} = 0.36(\pm 0.07) \times i + 7.87 (\pm 3.72) 
\end{equation} 

\begin{equation} 
FWHM{_B}{_C} / \sqrt{M_\ast} = 1.75(\pm 0.41) \times i + 0.09 (\pm 21.8) 
\end{equation} 

\setlength{\parindent}{0cm} where FWHM is measured in km/s, {M$_\ast$} is measured in {M$_\odot$} and inclination is measured in degrees. These fits demonstrate that although the NC and BC both correlate with disk inclination, they have different slopes, suggesting that they trace different regions. Assuming that the broadening is due solely to Keplerian rotation, we compute the \oi{} disk radii from the velocity at the HWHM. The black solid lines in Figure~\ref{fig:inclination_plot} show that the NC probes disk radii between 0.5 and 5 AU while the BC traces gas much closer in, between 0.05 and 0.5 AU.\footnote{Note that the dashed lines in Figure~3 of \cite{banzatti2015} provide the CO inner disk radius computed from the FWHM. However, the CO disk radii used through their paper are calculated from the velocity at the HWHM.}

\setlength{\parindent}{0.5cm} We can compare these results to those of \cite{banzatti2015} for CO, where our linear fits are shown as red lines in Figure~\ref{fig:inclination_plot} and those for CO are shown in green. We find similar slopes for the BC of both CO and \oi{} but slightly more extended disk radii for the \oi{} than for the CO (the BC CO disk radii range from 0.04 to 0.3\,AU).  
The slopes for the NC of CO and \oi{} are not in agreement, and the inferred formation region for \oi{}, from 0.5 to 5\,AU, is again slightly further from the star than that inferred for the NC of CO (0.2-3\,AU).

\setlength{\parindent}{0.5cm}We can then use the linear fits presented above to derive disk inclinations for the sources with no disk-based inclination values in the literature (right panel of Figure~\ref{fig:inclination_plot}).  For the sources where the LVC has only one component, we calculate the inclination using equations 5 or 6 depending on whether it is classified as NC or BC. For the sources where the LVC has both a broad and a narrow component, we calculate the inclination of each component individually (both are shown in Figure~\ref{fig:inclination_plot}) and take a weighted mean to derive the source inclination. Inclinations derived from these fits to the BC and NC relations are reported in Table \ref{tab:source_properties} and are marked with a dagger.

\subsection{Radial Surface Brightness of the Narrow LVC}\label{sect:RadialBrightness}

In the previous section, we found that the NC of the LVC has a line width consistent with Keplerian broadening at a distance between 0.5 and 5\,AU from the star. Here we will explore the range of radii in the disk required to account for the observed NC profile assuming a simple power law surface brightness fall off.  Of the 18 sources with NC in their LVC, 9 have velocity centroids consistent with bound gas meaning that their profiles can be modeled with a simple Keplerian disk. Of these 9 sources 4 are transition disks (TDs), so we can also test whether the distribution of the gas in disks with dust cavities extending from a few to tens of AU \citep{espaillat2014}, is different from the NC of Class~II sources. 

% We choose NC LVC profiles from three disks with very similar inclinations ($i \sim 35-40^o$) whose NC profiles have velocity centroids consistent with bound gas. Two of them, DM~Tau and UX~Tau A, are TDs with very different stellar masses (0.35 and 1.5\,M$_\odot$ respectively), while the third, BP~Tau, is a Class~II source with a mass in between the other two (0.62\,M$_\odot$, see Table 1). The mass accretion rate of BP~Tau is $\sim10^{-8}$\,M$_\sun$/yr, rather typical of young TTS, while that of DM~Tau and UX~Tau is about an order of magnitude lower (see Table \ref{tab:acc_properties}). In the case of BP~Tau we remove the BC LVC component (see Figure~\ref{fig:HVC_6300_panel1}) before fitting the NC.

Our modeling uses a power law distribution for the line surface brightness of the form $I{_O}{_I}(r)  \propto r^{-\alpha}$, where $r$ is the radial distance from the star and $\alpha$ is varied between 0 and 2.5 (see e.g. \citealt{fedele2011}). The radial profile is converted into a velocity profile assuming Keplerian rotation, with the stellar mass and disk inclination in Table \ref{tab:source_properties} as additional input parameters. The model line is convolved with a velocity width $v = \sqrt{v_{\rm in}^2+v_{\rm th}^2}$ where $v_{\rm in}$ is the instrumental broadening (6.6\,km/s) and $v_{\rm th}$ is the thermal broadening. We assume a temperature of 5,000\,K to compute $v_{\rm th}$ because the \oi{} emission might trace hot collisionally excited gas (\citealt{ercolano2010, ercolano2016}). We then use the {\it mpfitfun} IDL routine\footnote{http://cow.physics.wisc.edu/$\sim$craigm/idl/idl.html} to find the best fit to the observed line profiles where an uncertainty equal to the RMS on the continuum is adopted for each flux measurement. The parameters that are varied in the fitting procedure are the inner and outer radii of the emitting gas and $\alpha$.
 
\begin{figure}[h] %  figure placement: here, top, bottom, or page (H,T,B,P RESPECTIVELY)
   \begin{center}
   \includegraphics[width=0.5\textwidth]{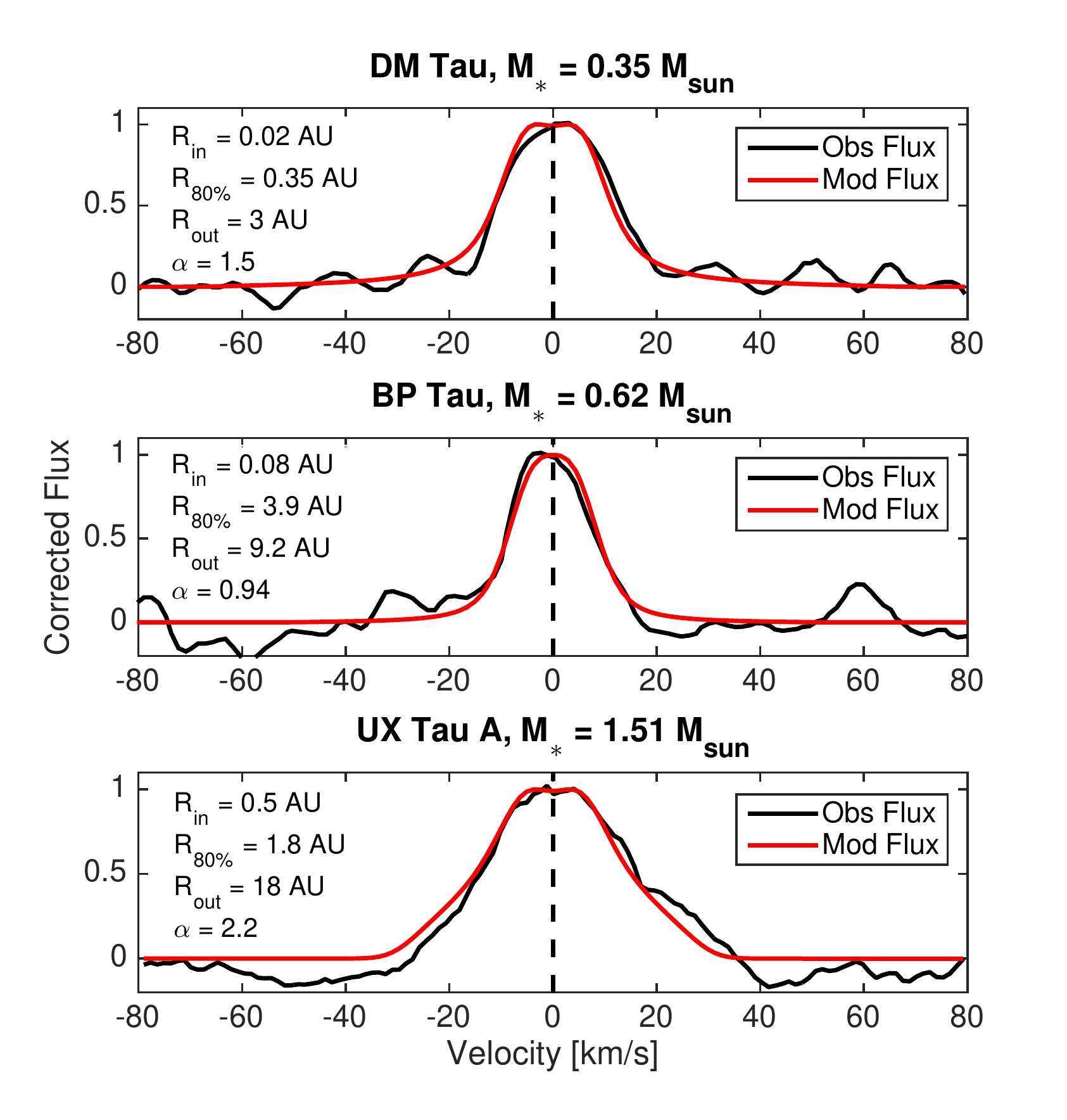} 
   \caption{Comparison of computed profiles in bound disk gas to unshifted NC \oi{} 6300 {\AA} profiles in 3 sources with the same inclination. Two (DM Tau and UX Tau A) are transition disks and one (BP Tau) is a Class II source. Values for the inner and outer disk radii, the radius within which 80\% of the emission arises, and the surface brightness power law are indicated.}
   \label{fig:TD_models}
   \end{center}
\end{figure}

Our best fits to the NC \oi{} 6300 {\AA} profiles of three representative sources with very similar inclinations ($i \sim 35-40^o$) but different disk types and mass accretion rates are shown in Figure~\ref{fig:TD_models}.
Our simple model does a good job in reproducing the observed NC profiles and shows that the radial extent of the gas emitting the NC LVC is from within 1\,AU, to explain the relatively large FWHMs, out to $\sim$10\,AU, to explain the lack of double peaks in the profiles (see Table~\ref{tab:model_results} for a summary of inferred parameters). We see a trend between the inner and outer radii and the stellar mass,  with the smallest \oi{} emitting region located around  the lowest mass star DM Tau and the largest around the most massive $\sim1\,M_\sun$ stars in our sample DR~Tau and UX~Tau A. The power law index of the surface brightness ranges from flat $\alpha$= 0 (for DF~Tau) to $\alpha$= 2.2 (for UX~Tau A and TW~Hya). Using the best fit surface brightness we also compute the radius within which 80\%{} of the NC LVC emission arises (R$_{80\%}$ in Figure~\ref{fig:TD_models} and Table~\ref{tab:model_results}) and  find it to be within $\sim 5$\,AU, in agreement with the \oi{} disk radii estimated from the NC HWHM (see Figure~\ref{fig:inclination_plot}). 
Our steepest power law indexes of $\sim$2.2 are very similar to those inferred for CW~Tau and DQ~Tau by HEG after re-centering their blueshifted \oi{} profiles in the stellocentric frame and assuming that the entire line broadening is due to Keplerian rotation. The extent of the \oi{} emitting region for these two sources is inferred to be between 0.1 out 20\,AU, similar to the ranges we find. This hints that Keplerian broadening may also dominate the profile of wind sources. We plan to model wind profiles in a future paper by adding an unbound component with a prescribed velocity field to the bound Keplerian disk.

In summary, our modeling of LVC NC profiles consistent with bound gas shows  no difference in the radial distribution of the gas for TDs and Class~II sources. Given that the sub-mm dust cavity of UX Tau A and DM Tau are $\sim$ 25\,AU (\citealt{andrews2011, kraus2011, ingleby2013, alcala2014}) and $\sim$ 19\,AU (\citealt{andrews2011, kraus2011, ingleby2013}) respectively, our results also imply that the \oi{} 6300 {\AA} NC traces gas inside the dust cavity.  We find similar results for the other 2 transition disks with NC centroids consistent with bound gas\footnote{We did not model the fifth TD, CoKu Tau 4, since its NC is blueshifted by 3 km/s}, see Table~\ref{tab:model_results}. Thus, we conclude that the \oi{} 6300 {\AA} NC profiles of TDs can be explained by radially extended \oi{} emission, most of which is confined within their dust cavities (see \citealt{espaillat2014} for the dust cavities of TDs). 
%{\bf We stress that the lack of a blueshift in the NC profiles is still compatible with flowing gas because sub-micron dust grains are the main source of opacity (see the case of TW~Hya discussed in \citealt{pascucci2011} and modeling by \citealt{ercolano2010}).} lets find another place for this
%The inferred radial extent is also consistent with the lack of variations in the \oi{} 6300 {\AA} line noted in \cite{pascucci2011} for TW Hya and for GM Aur in Section \ref{sect:fitting}. 

\subsection{Line Ratios}\label{sect:LineRatios}
Ratios of forbidden line equivalent widths can provide insight into the temperature and density of the LVC emitting line region. For example, the ratio of \oi{} 5577/6300  for one of our sources, the transition disk TW Hya, with a value of  $\sim$1/7 \citep{pascucci2011}, has been interpreted by \citet{gorti2011} as tracing the  warm ($\sim$1,000\,K) and bound molecular layer where OH molecules are photodissociated by FUV photons. The LVC \oi{} 5577/6300  ratio for the HEG stars was examined by R13 who found values between $\sim$1 and 1/8 for all sources except two with microjets that display smaller ratios ($\sim$0.07). R13 attributed the larger ratios of 1 to 1/8 also to FUV photodissociation of OH molecules while N14 prefer the alternative possibility of thermal emission from a very hot ($\sim$5,000-10,000\,K) and dense (n$_{\rm H}> 10^8$\,cm$^{-3}$) gas to explain similar ratios in a different sample of stars with disks.

The mean EW \oi{} 5577/6300 ratio for the full LVC for our sample stars is 0.25, similar to those found previously. However, with the decomposition of our high resolution spectra into BC and NC contributions to the LVC we can look to see if these two LVC components have the same ratio. There are only 4 stars where we have sufficient signal to noise in both of these lines to evaluate the \oi{} 5577/6300 for each component: AA~Tau, BP~Tau, CW~Tau, GI~Tau. In each case the BC ratio is a factor of a few higher than the NC ratio of \oi{} 5577/6300. 

\begin{figure*}[h] %  figure placement: here, top, bottom, or page (H,T,B,P RESPECTIVELY)
   \begin{center}
   \includegraphics[width=5in]{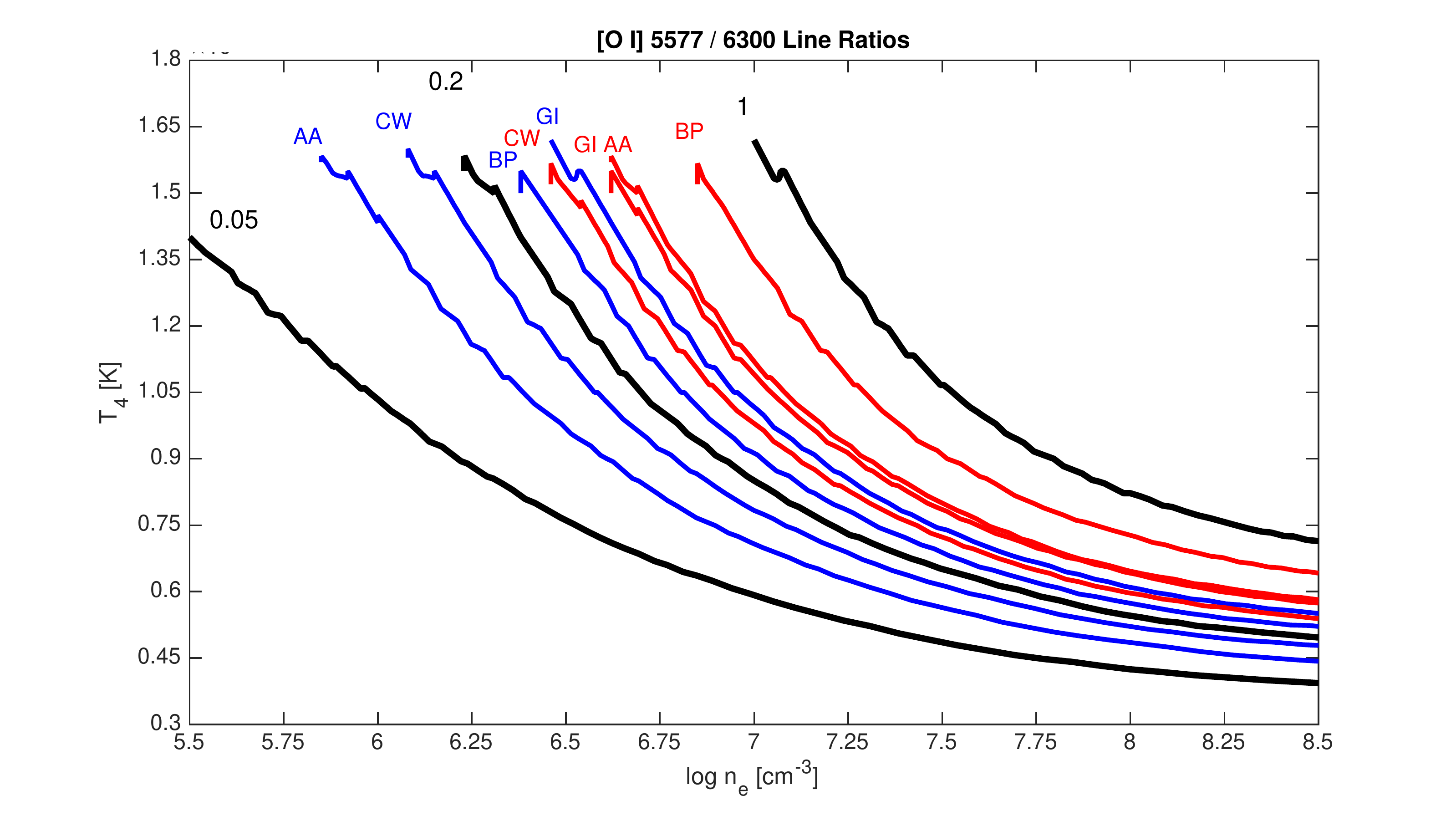} 
   \caption{Contours of the thermal ratio of \oi{} 5577/6300 \,\AA{} as a function of gas temperature (divided by 10,000\,K) and electron density. The line ratios for the four TTS with two kinematic components and high signal-to-noise profiles are shown in red (BC) and blue (NC). The BC has higher ratios than the NC implying a higher electron density for gas at the same temperature. Regardless of the kinematic component, if the lines are thermally excited, gas needs to be hot ($\ge$5,000\,K) and dense to explain the observations.}
   \label{fig:OIline_ratios}
   \end{center}
\end{figure*}

Figure~\ref{fig:OIline_ratios} compares the observed  \oi{} 5577/6300 values for these 4 stars (BC in red and NC in blue) with those predicted by a homogeneous and isothermal gas where the excitation is due solely to electron collisions, see Appendix~\ref{appendix} for details. 
%We locate the observed  BC (in red) and NC (in blue) ratios for the four TTS that have both components and the highest signal-to-noise line profiles and the locus of predicted ratios (in black) for three line ratios (1, 0.2, and 0.05). 
This figure shows that the BC emitting region is about a factor of 3 denser than the NC region if these two components arise from gas at the same temperature.  Alternatively, if they trace similarly dense ($n_e \ge 10^7$\,cm$^{-3}$) gas, the BC gas is $\sim$25\% hotter than the NC gas. In either scenario, gas needs to be hot ($\ge$5,000\,K) to explain the observed ratios if the lines are thermally excited. For gas at 8,000\,K the observed ratios can be explained by $n_e$ ranging from $\sim 5 \times 10^6$ to $ 5 \times 10^7$\,cm$^{-3}$ implying gas densities of a few $10^7-10^8$\,cm$^{-3}$ for an ionization fraction of 0.33, close to the value expected in the \oi{} emitting region in some photoevaporative wind models (see Figure 2 in \citealt{owen2011}). By 5,000\,K electron densities become very high ($\ge 10^8$\,cm$^{-3}$) and, as discussed in the context of TW~Hya \citep{gorti2011}, are unlikely to be present in the hot surface of protoplanetary disks. An alternative interpretation for those sources with no definite blueshift is that the \oi{} emission is not thermal but results from the photodissociation of OH molecules in a cooler ($\sim$1,000\,K), bound, and mostly neutral layer of the disk \citep{gorti2011}.

\begin{figure*}[h] %  figure placement: here, top, bottom, or page (H,T,B,P RESPECTIVELY)
   \begin{center}
 \includegraphics[width=5in]{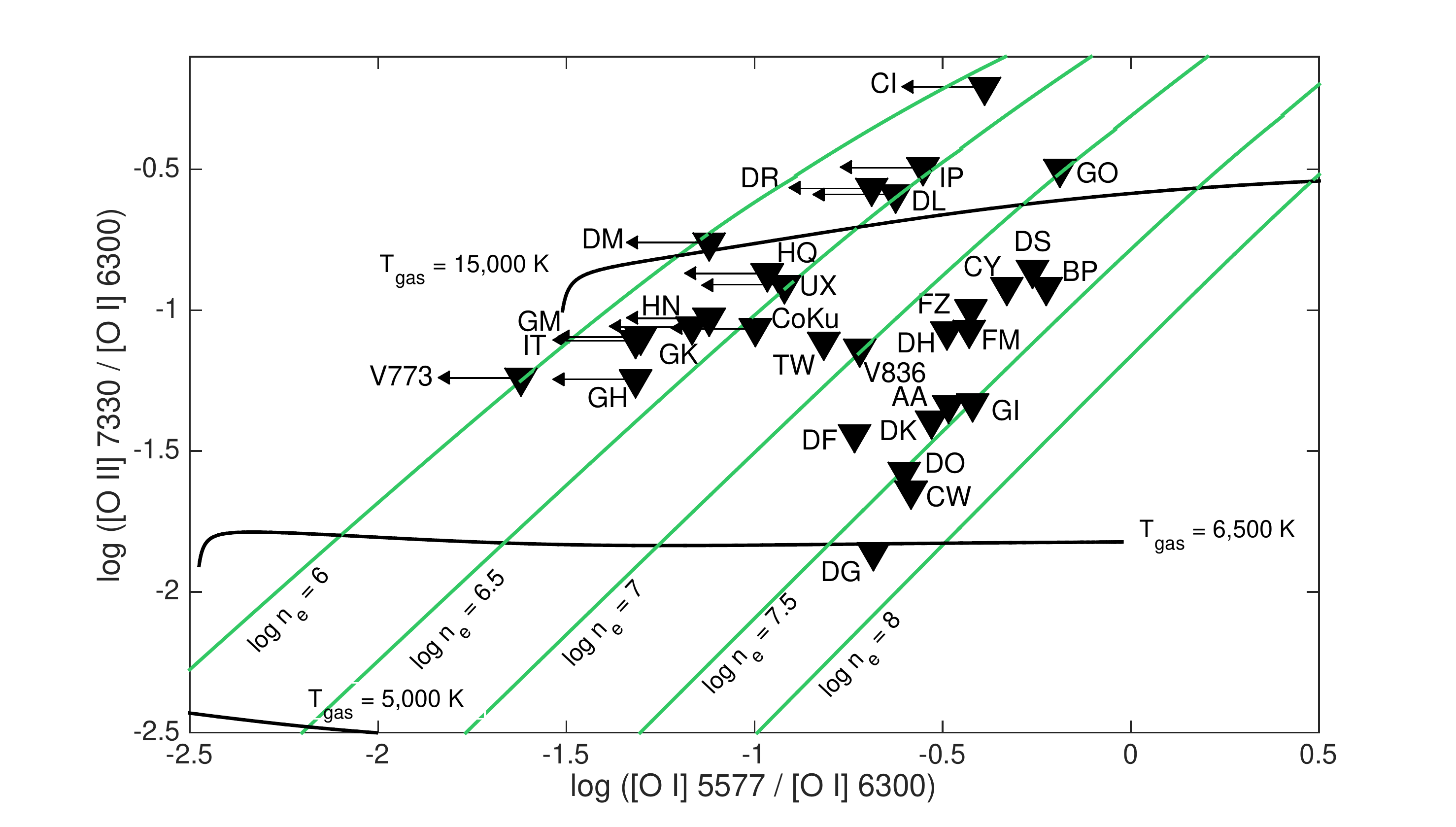}
   \caption{Oxygen line ratios are compared to predicted values assuming collisional excitation by electrons and an ionization fraction of 0.33. The \oii{} 7330\,\AA{} transition is never detected so downward pointing triangles denote upper limits. Sources with no detection at \oi{} 6300 show left pointing arrows for upper limits. Solid black lines denote ratios for gas at constant temperatures while diagonal green lines for gas at the same electron density. Temperatures lower than 6,500\,K are consistent with all \oii/\oi{} upper limits, the most stringent one coming from DG~Tau. Higher temperatures are possible only if the ionization fraction is lower than 0.33. }
   \label{fig:line_ratios}
   \end{center}
\end{figure*}

Although the higher ionization \oii{} 7330 line is not detected in any of our sources, we can use the upper limits to further  constrain the gas temperature and density discussed above.  Figure~\ref{fig:line_ratios} shows the predicted \oii{} 7330/ \oi{} 6300 ratios versus the \oi{} 5577/6300 ratios for the full LVC. As expected, since the critical electron densities for both the \oii{} and \oi{} lines are similar ($\sim 2 \times 10^6$ cm$^{-3}$), the \oii /\oi{} ratio is most sensitive to the gas temperature, given an ionization fraction which again we take to be 0.33. Models with the same electron density (green lines), but different temperatures, run diagonally in Figure~\ref{fig:line_ratios}.  This figure shows that the gas temperature must be lower than 6,500\,K if a single temperature is to explain all of our \oii{} 7330/ \oi{} 6300 upper limits, which would imply high electron densities ($\ge 10^{7.5}$\,cm$^{-3}$) based on the \oi{} 5577/6300 ratios. Higher temperatures are possible only in combination with an ionization fraction lower than 0.33.


As discussed earlier, it is also possible that the \oi{} emission is not thermal but results from the dissociation of OH molecules \citep{gorti2011}. A way to discriminate between thermal vs non-thermal emission would be to obtain high spectral resolution observations of the \sii{} line at 4068.6 \,\AA {} because this line has a critical density of $2.6\times 10^6$\,cm$^{-3}$, very similar to that of the \oi{} line at 6300\,\AA{} (see e.g. \citealt{natta2014}). Similar line profiles for the \sii{} and \oi{} lines would clearly point to thermal emission in a hot dense gas.

\section{Discussion}\label{sect:Discussion}

The major contribution of this work is demonstrating that the low velocity forbidden emission in TTS has kinematic properties that can be described as coming from a combination of a broad and a narrow line formation region. Most of our analysis is based on characterizing the behavior of the BC and NC as though they are two physically distinct emission regions. However, based on the current data we cannot assess whether they both arise from the same phenomenon from different radial ranges in the disk or whether they arise in two different formation scenarios. The most likely formation scenarios for the LVC  are mass loss in the outer part of a centrifugally driven disk wind, mass loss in a photoevaporative flow, and bound gas in the disk. We first review the highlights of our findings here and then look at the merits of attributing the LVC to these scenarios.

The BC of the LVC is very common, seen in 25/30 stars spanning the full range of disk accretion rates. In contrast, the NC of the LVC  is somewhat less common, detected in only 18/30 stars, but again is found over the full range of disk accretion rates. Of the 5 stars which show only NC emission, 4 are transition disks and the 5th transition disk (GM Aur), has the highest percentage of NC emission in the LVC among the sources with BC emission. If the BC comes primarily from within 0.05 to 0.5 AU, as suggested by the relation between its FHWM and disk inclination, then its absence in transition disks is likely due to a paucity of significant disk gas in this region. This conclusion is in line with what has been proposed for TW~Hya from detailed modeling of emission lines covering a large range of radial distances \citep{gorti2011}\footnote{We note that DG Tau, a high accretion rate object which appears to have a full disk and strong HVC emission from a micro jet, is the fifth object with no BC emission. It is possible that our method cannot isolate the LVC BC given the intensity and large velocity range covered by the HVC and/or we are not directly seeing this region because DG~Tau is embedded in significant nebulosity as inferred from dust scattering.}. Four of our sources (BP~Tau, DF~Tau, FZ~Tau, and GI~Tau) have a redshifted HVC that might also hint to depletion in the inner disk, in this case of the dust component, which is the main source of opacity. However, their infrared indices, as reported in Furlan et al.~(2011), place them in the full disk portion. Also their LVC are not reduced with respect to sources with no redshifted HVC (compare BP~Tau to AA~Tau), thus viewing through a disk hole cannot explain the redshifted HVC emission. 

While the range of observed FWHM for the BC and the NC can be explained as a result of Keplerian broadening from different radii in the disk (between 0.05 to 0.5\,AU for the BC and 0.5 to 5\,AU for the NC), the different behaviors of their centroid velocities are the most useful in trying to understand their connection to disk winds.

\subsection{Role of Winds in the Low Velocity Component}\label{Sect:Comparison}

\begin{figure*}[h] %  figure placement: here, top, bottom, or page (H,T,B,P RESPECTIVELY)
   \begin{center}
   \includegraphics[width=6in]{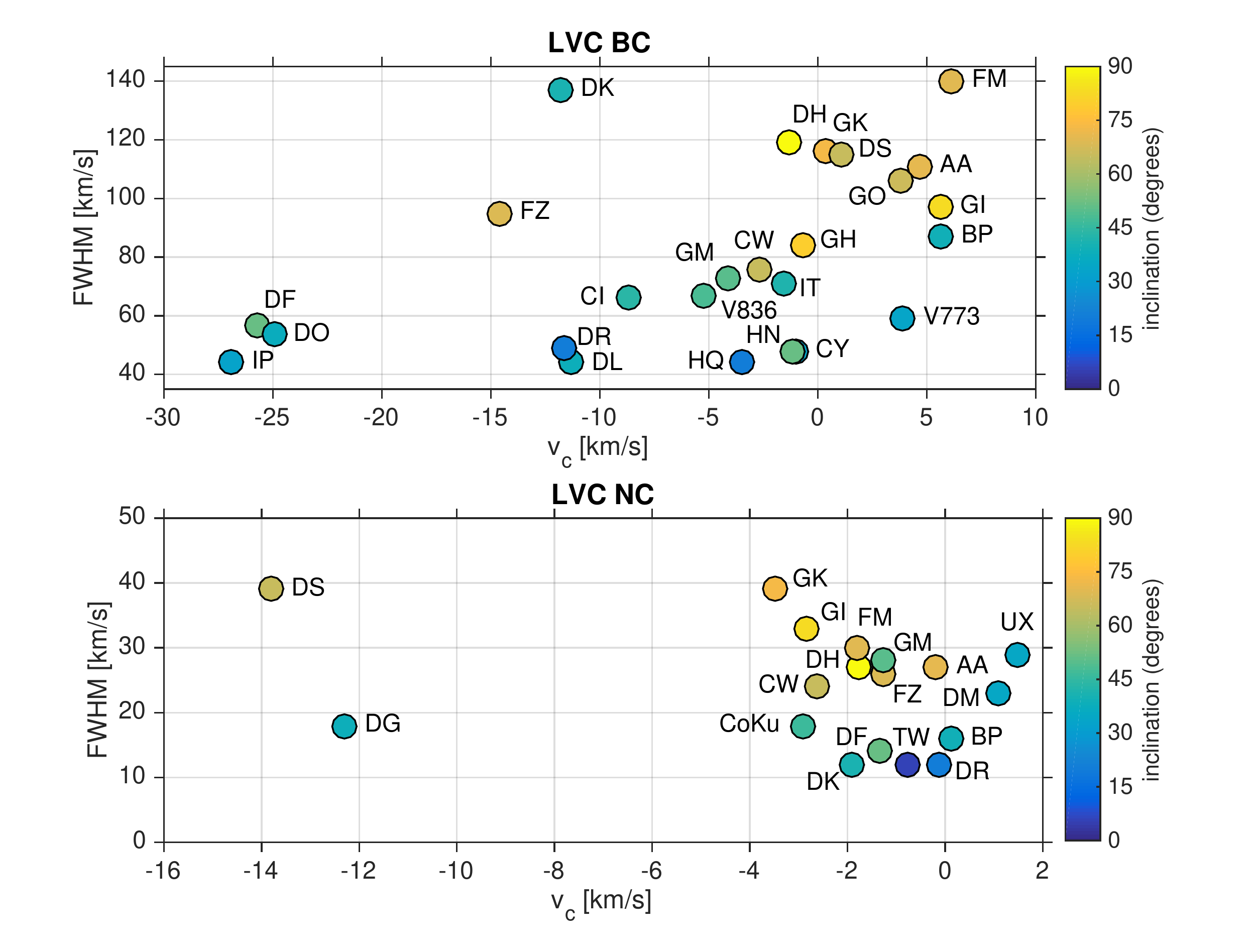}
   \caption{Observed FWHM versus $v_c$ for the LVC BC (upper panel) and NC (lower panel). Colors indicate the observed inclination or, if not observed, the one derived in Section~\ref{sect:inclination}.  Both BC and NC show a tendency for increasing FWHM with increasing inclination but only the BC has decreasing $v_c$ with increasing inclination as expected in wind models.}
   \label{fig:EO_models}
   \end{center}
\end{figure*}

A basic expectation of any wind model is that not only the FWHM but also the centroid $v_c$ of a line formed in the wind will correlate (oppositely) with disk inclination. In a close to face-on configuration Keplerian broadening is minimal and the vertical component of the wind dominates, giving rise to an asymmetric blueshifted profile. On the opposite extreme, for a close to edge-on configuration, Keplerian broadening dominates and because the wind is emerging in the plane of the sky, the resulting profile is symmetric and centered at the stellar velocity (see e.g. \citealt{alexander2008}). We see the expected relations with FWHM and inclination for both the BC and NC LVC but the situation is not so clear when their centroid velocities are also considered. We illustrate the relation between all 3 quantities in Figure~\ref{fig:EO_models}, plotting the observed \oi{} 6300\AA{} FWHM vs $v_c$ for the LVC BC (upper panel) and NC (lower panel) with datapoints color-coded by disk inclination (see Section \ref{sect:inclination}).  

In the case of the BC, we see larger FWHM associated with larger disk inclination (vertical color gradient) coupled with, for most of the sources, larger blueshifts for narrower lines and lower disk inclination (horizontal color gradient). These are the expected behaviors for a wind. A Kendall $\tau$ test gives a probability of only $\sim$ 1\% that the BC FWHM and $v_c$ are uncorrelated. Moreover, the sources with small BC redshifts are mostly in disks seen at high inclination, where, as described in Section 5.4, an extended disk height at large radii may obscure part of the wind from the inner disk while the receding gas (from the observer's perspective) remains unobstructed (Gorti et al. in preparation).As the BC has the general characteristics expected for a simple wind and seems to arise in the innermost disk with maximum velocities of 10 to 27 km/s, we find it likely that the BC is the base of an MHD disk wind, since thermal speeds cannot reach the necessary escape speeds this deep in the stellar gravitational potential field and photoevaporation cannot occur.

In the case of the NC, plotted over a smaller range of FWHM and $v_c$ than the BC, again we see FWHM associated with larger disk inclination (vertical color gradient), however there is no relation between the $v_c$ and either the FWHM or the inclination. A Kendall $\tau$ test lends to a probability of $\sim$ 21\% that FWHM and $v_c$ are uncorrelated, consistent with no correlation. Two sources (DS Tau and DG Tau) stand out with much higher blueshifts than the rest of the NC, but one is close to face on and the other close to edge on. Thus, although half of the NC have blueshifted $v_c$, they do not behave in the manner expected for a simple wind.

Although Figures~\ref{fig:inclination_plot} and Figure~\ref{fig:EO_models} show a relation between FWHM and inclination in the BC and NC there is considerable scatter and a few outliers.  However, we have not taken into account the fact that there will be an additional source of broadening in both the BC and NC if they are formed in outflowing gas.  In the Class II sources, comprising most of our sample, a dusty disk is assumed to occult the receding flow, and the observed FWHM would come both from Keplerian rotation and expanding wind streamlines, producing anomalously high line widths for certain wind geometries (e.g. BC of DK Tau and FZ Tau). In the case of the transition disks, dominated by NC LVC emission, the large dust-free cavities appear to have bound gas, although the gas is not as close in to the star to produce BC emission (Section 5). However, as there would presumably be no opacity source to occult a receding flow, the NC could include a contribution from flowing gas but still be centered on the stellar velocity and the line width would be enhanced by both approaching and receding gas  (see the case of TW~Hya discussed in \citealt{pascucci2011} and modeling by \citealt{ercolano2010}).

\subsection{Comparison with Wind Models}
The past decade has seen significant advancements in the theory of thermal photoevaporative flows and sophisticated models, accounting for heating by stellar X-rays, EUV, and FUV photons, are now available (e.g. \citealt{alexander2014} for a review). We anticipate that these models will be soon providing more rigorous tests of whether the LVC might arise in thermal winds.  Theoretical models for magnetically driven centrifugal disk winds have also grown increasingly sophisticated and recently the relevance of MRI-driven accretion over much of the disk has been challenged (e.g. \citealt{turner2014}) putting MHD disk winds back in the spotlight for extracting disk angular momentum to enable accretion onto the star  (e.g. \citealt{bai16}). Global simulations of these winds have been recently presented by \cite{gressel2015} and thermal-chemical models have been investigated by \cite{Panoglou12}, but predictables that can be directly compared with observations of TTS forbidden lines are still lacking. 

We began this study to investigate the possibility that the LVC in TTS might arise in photoevaporative flows. It seems unlikely this could be the case for the BC of the LVC, which we attribute to the base of an MHD disk wind. However, despite the lack of a convincing trend between centroid velocity and FWHM/disk inclination for the NC of the LVC we note that the {\it range} of FWHM and $v_c$ for the NC are consistent with those predicted for photoevaporative flows by \citealt{ercolano2010, ercolano2016}. Using the radiation-hydrodynamic code of \cite{owen2011}, which includes X-ray and EUV irradiation, \cite{ercolano2016} find that the [O~I] 6300\,\AA{} line is produced by collisional excitation and extends to $\sim$35\,AU above the disk, where EUV photons and soft X-rays are absorbed and the gas is hot. Predicted FWHM range from 8 to 32\,km/s while peak centroids go from 0 to -7\,km/s, with the largest blueshifts occurring for intermediate disk inclinations ($ i \sim 40-60^{\rm o}$), contrary to a simple wind geometry. We do not see this trend either.

Although hydrodynamical models that can predict line profiles of flows driven by FUV photons have not been developed, the expectation is that these flows, being cooler  ($\sim$1,000\,K) and mostly neutral, will have smaller velocities than those driven by X-rays. The sound speed for such cool flows is only $\sim$2\,km/s and the critical radius beyond which the gas would be unbound is $\sim$10\,AU around a solar-mass star. In this scenario, the \oi{} emission will not be thermal and will arise from the dissociation of OH molecules in mostly bound gas inside of 10 AU \citep{gorti2011}, which could explain the very small or absent blueshifts we see in many NC lines. 
%High spectral resolution observations of the \sii{} 4068.6 {\AA} line, which has a critical density of $2.6\times10^6$ cm$^{-3}$, similar to that of the \oi{} line at 6300 {\AA}, would be extremely helpful to clarify if the \oi{} emission is thermal. 
In summary, at the present time the connection of the forbidden line NC of the LVC in TTS to photoevaporative flows remains ambiguous, and we cannot exclude that the NC is also part of a  MHD disk wind.

\section{Conclusions}
\setlength{\parindent}{0cm} We have analyzed optical high-resolution ($\sim$7\,km/s) spectra from a sample of 33 TTS whose disks span a range of evolutionary stages to clarify the origin of the LVC from oxygen and sulfur forbidden emission lines. We detect the [O~I] 6300\,\AA{} line in 30/33 TTS, the [O~I] 5577\,\AA{} line in 16/33,  and the [S~II] 6730\,\AA{} in only 8/33 TTS. After isolating the forbidden LVC emission by removing any HVC contributions, if present, we draw the following  conclusions about the residual LVC component:
\begin{itemize}
\item All TTS with \oi{} detections show LVC emission. Thirteen out of 30 sources with [O~I] 6300\ \AA{} emission have LVC emission that can be described as the combination of a broad (BC) and a narrow (NC) line emitting region. The remaining sources show LVC emission that is either only BC (12/30) or only NC (5/30).
\item The BC of the LVC is very common, seen in 25/30 TTS. The NC of the LVC is less common, seen in 18/25 TTS. Both components are found over the full range of accretion luminosities/disk accretion rates and their luminosities, combined or individually, correlate with the accretion luminosity. LVC that are solely or predominantly NC are usually transition disk sources. 
\item Comparison with spectra from HEG shows that in most stars the LVC is stable over timescales of decades. However, we do find evidence for variations, with the LVC disappearing entirely in one star and only the NC of the LVC disappearing in another star. 
\item The BC shows a relation between the FWHM and either observed or derived disk inclination suggesting it is broadened by Keplerian rotation at disk radii between 0.05 to 0.5\,AU. Also, a significant number of BC have blueshifts in excess of 5 km/s. These larger blueshifts are associated with narrower lines and lower disk inclinations, as expected if the BC includes emission from a wind, in addition to Keplerian broadening. The BC with larger blueshifts also tend to be found in sources with higher accretion luminosity and HVC emission from microjets.  Since the emission likely arises from 0.05 to 0.5 AU, where the gravity of the star and the disk is strong, it is unlikely to trace a photoevaporative flow but rather the slower moving portion of an MHD disk wind.
\item The NC also shows a relation between the FWHM and either observed or derived disk inclination suggesting it is broadened by Keplerian rotation at disk radii between 0.5 to 5\,AU. Half of the NC features are blueshifted between -2 to -5 km/s and the other half have centroids consistent with the stellar velocity. Although the expected relation for a simple wind between disk inclination and centroid velocity is not found, we cannot exclude the possibility that the NC of the LVC arises in photoevaporative flows.
\item Regardless of the disk evolutionary stage, NC profiles consistent with bound gas can be reproduced by gas in Keplerian motion with a surface brightness decreasing as a power law between 0.1\,AU and $\sim$10\,AU, but with 80\% of the emission arising within 5\,AU. The implication for transition disks is that the NC arises from gas inside the dust cavity.
\item If forbidden emission lines are produced by collisional excitation with electrons, the \oi{} 5577/6300 ratios suggest high temperatures ($>$5,000\,K) and large electron densities ($> 10^6$\,cm$^{-3}$). Without the additional constraints on density and temperature that would be provided by high resolution spectra of the \sii{} 4069\,\AA{} line, the possibility remains that the excitation of [O I] is  not thermal. Dissociation of OH molecules in a cool ($\sim$1,000\,K), bound, mostly neutral disk layer could be the source of the [O I] emission in objects with very small and absent shifts in the NC LVC.
%\item {\bf The very small and absent shifts in the \oi{} NC may also result from non thermal  emission from the dissociation of OH molecules in a cool ($\sim 1,000$\,K), mostly neutral disk layer. High-resolution spectra of the \sii{} 4069\,\AA{} lines are key to distinguish between thermal vs non thermal emission.} 
\end{itemize}

\setlength{\parindent}{0cm} Disk winds, both MHD and photoevaporative, deplete material from several scale heights above the midplane. As dust grains grow they settle toward the midplane. The implication is that disk winds mostly deplete the protoplanetary disk of gas, which consequently increases the dust-to-gas ratio with time \citep{gorti15,bai16}. This increase can directly impact the formation of planetesimals, terrestrial planets, and the cores of giant planets. Since disk winds play a significant role in disk dispersal and planet formation, both models and expanded observational data sets need to pursue the origin of the LVC NC, and constrain the rate at which material is lost via disk winds.

\acknowledgements
I. Pascucci, U. Gorti, and D. Hollenbach acknowledge support from a Collaborative NSF Astronomy \& Astrophysics Research Grant (IDs: 1312962 and 1313003). S. Edwards acknowledges support from NASA grant NNG506GE47G issued through the Office of Space Science. The data presented herein were obtained at the W.M. Keck Observatory, which is operated as a scientific partnership among the California Institute of Technology, the University of California and the National Aeronautics and Space Administration. The Observatory was made possible by the generous financial support of the W.M. Keck Foundation.
The authors wish to recognize and acknowledge the very significant cultural role and reverence that the summit of Mauna Kea has always had within the indigenous Hawaiian community.  We are most fortunate to have the opportunity to conduct observations from this mountain.
	
{\it Facilities:} \facility{Keck/HIRES}

\appendix
%\section{Gaussian fitting of the \oi{} 6300\,\AA{} profiles}\label{appendix}
%Our approach to deconstruct forbidden line profiles is discussed in detail in Section~\ref{sect:fitting} and Figure~\ref{fig:Pan} gives examples of Gaussian fits for a few profiles with different degrees of complexity. In Figures~\ref{fig:full_fit1} and \ref{fig:full_fit2} we show all the individual Gaussian and total profile fits for the 30 stars where the \oi{} 6300\,\AA{} line is detected. In these figures the HVC is shaded in green, the LVC BC is marked with a red line, the LVC NC with a blue line, and the total fits are shown with purple dashed lines.

\section{Sky Subtraction and Slit Position Angle}\label{PA_appendix}
\begin{figure}[h] %  figure placement: here, top, bottom, or page (H,T,B,P RESPECTIVELY)
   \begin{center}
   \includegraphics[width=6in]{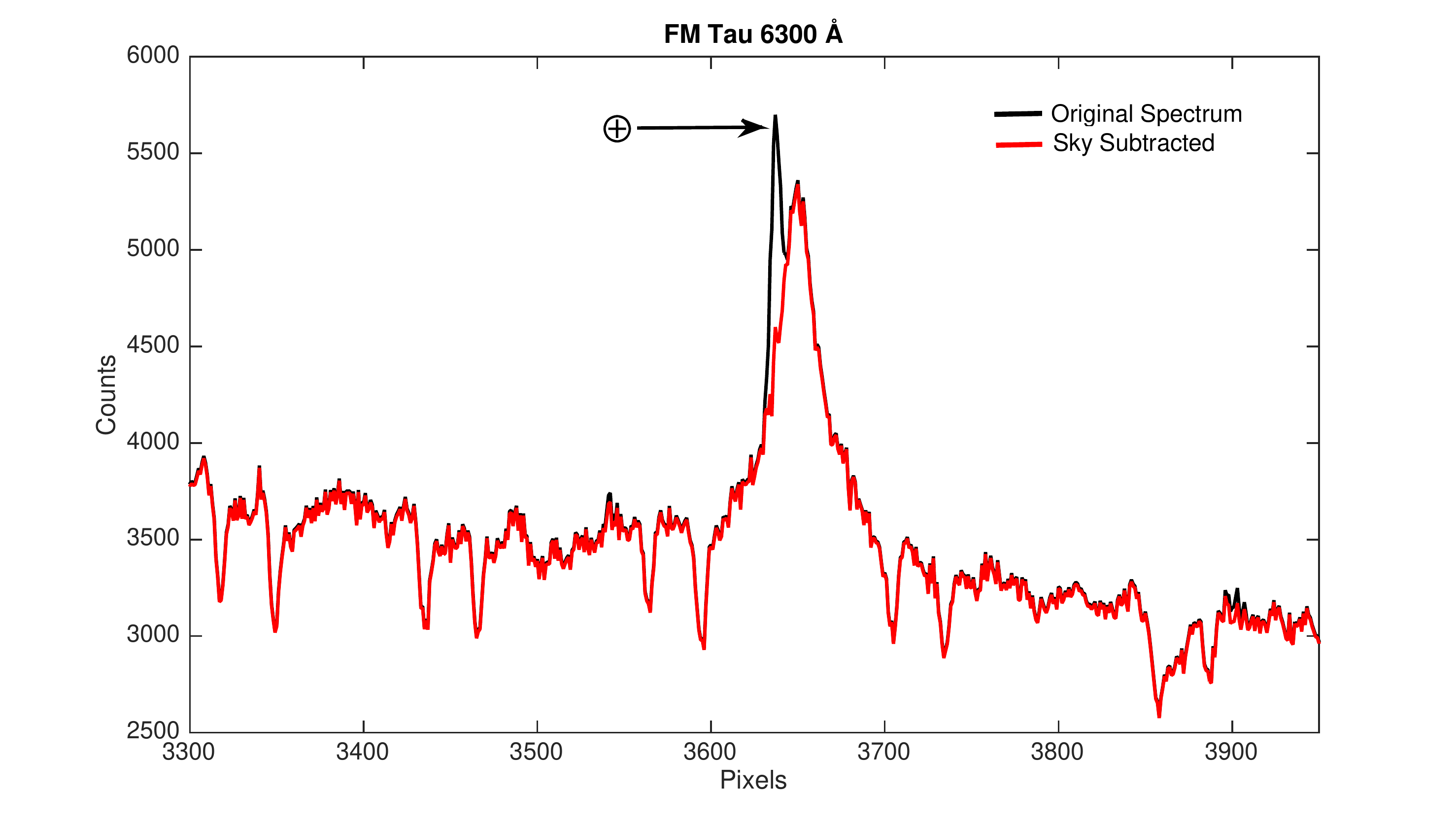}
   \caption{FM~Tau spectrum around the \oi{} 6300 {\AA} before (black) and after (red) sky subtraction.
   Note that the strong \oi{} terrestrial line is well removed by the MAKEE pipeline.}
   \label{fig:sky_sub}
   \end{center}
\end{figure}

As mentioned in Section~\ref{sect:obs} the MAKEE pipeline performs an automatic sky subtraction. We show in Figure~\ref{fig:sky_sub} the spectrum of FM~Tau before (black) and after sky subtraction (red) to highlight that even the strong terrestrial \oi{} emission line at 6300\,\AA{} is well removed by the pipeline.

The Keck spectra presented here were acquired in the standard mode which places the slit along the parallactic angle in order to minimize potential slit losses. This approach was taken because our main interest was to study the LVC, which was known to be compact, rather than the jet emission, which, most likely, extends beyond the slit width. For completeness, we provide  in Table~\ref{tab:PA_table} slit position angles and disk position angles (which should be close to 90$^{\circ}$ of the jet position angles). 
GO Tau, UX Tau A, and DS Tau have the slit most closely aligned of a possible jet, within $\sim$ 10$^{\circ}$, yet none of them show a jet signature in the \oi{} spectra.

\section{Collisional Excitation Model}\label{appendix}
The collisional excitation model described here is used to gain physical insight in the temperature and electron density of the region traced by oxygen forbidden lines. We assume a homogeneous and isothermal slab of gas, where the excitation is due solely to electron collisions. We considered a ground state and four additional excited  states (5-level atom) both for the neutral and ionized oxygen. We computed the relative populations of the levels as a function of gas temperature and density by including the processes of collisional excitation, collisional de-excitation, and spontaneous radiative decay. Einstein coefficients for radiative decay were taken from the NIST database\footnote{$http://physics.nist.gov/PhysRefData/ASD/lines\_form.html$} and electron collision strengths from \cite{draine2011}.  We have not included collisions with neutral hydrogen because the de-excitation cross section of the level $^1$S$_0$ is not known (see e.g. discussion in \citealt{ercolano2010}). However, neutral collisions should be negligible when the electron abundance is larger than $\sim10^{-3}$, as suggested by the same disk models, because the electron rate coefficients ($\sim 10^{-9}$\,cm$^3$/s) are much larger than those for H ($\sim 10^{-12}$\,cm$^3$/s).  Because of the very similar ionization potential of H and O we have taken the ratio of H$^+$/H to be equal to O$^+$/O, equal to 0.5 for an ionization fraction of 0.33, close to the value expected in the \oi{} emitting region in some photoevaporative wind models (see Figure~2 in \citealt{owen2011}).

\clearpage

\begin{table}
%\footnotesize
\begin{center}

\caption{Source Properties}
\label{tab:source_properties}
\vspace{0.3cm}
\begin{tabular}{lcccccccc}
\tableline
\\
Source & SpTy & SED  & $A_{\rm v}$ & log $L_{\ast}$ & $M_{\ast}$ &$i$ & REF  & log $L_{X}$ \\
            &       &   &  &[L$_\odot$] &[M$_\odot$]     &   [deg]       & ($i$) &  [L$_\odot$]      \\          
\\
\tableline
\\
AA~Tau & M0.6 & II & 0.40 & -0.35 & 0.57 & 71 & C13 & -3.49\\ 

BP~Tau & M0.5 & II & 0.45 &-0.38 & 0.62 & 39 & G11 & -3.45 \\ 

CI~Tau & K5.5 & II & 1.90 & -0.20 & 0.90 & 44 & G11 & -4.30  \\

CoKu~Tau 4 & M1.1 & T & 1.75 & -0.50 & 0.54 & 46$^{\dagger}$ & --- & ---  \\

CW~Tau & K3 & II & 1.80 & -0.35 & 1.01 & 65 & P14 & -3.13  \\

CY~Tau & M2.3 & II & 0.35 & -0.58 & 0.41 & 34 & G11 & -4.37  \\

DF~Tau & M2.7 & II & 0.10 & -0.04 & 0.32 & 52$^{\dagger}$ & --- & --- \\

DG~Tau & K7 & II & 1.60 & -0.29 & 0.76 & 38 & G11 & -4.18$^{\ast}$   \\

DH~Tau & M2.3 & II & 0.65 & -0.66 & 0.41 & $\sim$90$^{\dagger}$ & --- & -2.66 \\

DK~Tau & K8.5 & II & 0.70 & -0.27 & 0.68 & 41 & AJ14 & -3.62 \\

DL~Tau & K5.5 & II & 1.80 & -0.30 & 0.92 & 38 & G11 & ---  \\

DM~Tau & M3 & T & 0.10 & -0.89 & 0.35 & 35 & AN11 & -4.33$^{\ast}$  \\

DN~Tau & M0.3 & II & 0.55 & -0.08 & 0.55 & 39 & I09 & -3.52 \\

DO~Tau & M0.3 & II & 0.75 & -0.64 & 0.70 & 37$^{\dagger}$ & ---  & ---\\

DR~Tau & K6 & II & 0.45 & -0.49 & 0.90 & 20 & S09 & ---\\

DS~Tau & M0.4 & II & 0.25 & -0.72 & 0.69 & 65 & AJ14 & ---  \\
          
FM~Tau & M4.5 & II & 0.35 & -1.15 & 0.15 & $\sim$70 & P14 & -3.86 \\

FZ~Tau & M0.5 & II & 3.5 & -0.48 & 0.63  & 69$^{\dagger}$ & --- & -3.78 \\

GH~Tau & M2.3 & II & 0.40 & -0.19 & 0.36 & 80$^{\dagger}$ & --- & -4.55 \\

GI~Tau & M0.4 & II & 2.55 & -0.25 & 0.58 & 83$^{\dagger}$ & --- & -3.66 \\

GK~Tau & K6.5 & II & 1.50 & -0.03 & 0.69 & 73 & AJ14 & -3.42 \\

GM~Aur & K6 & T & 0.30 & -0.31 & 0.88 & 50 & G11 & --- \\

GO~Tau & M2.3 & II & 1.5 & -0.70 & 0.42 & 66 & AW07 & -4.19\\

HN~Tau & K3 & II & 1.15 & -0.77 & 0.70 & 52 & AJ14 & -4.40$^{\ast}$ \\

HQ~Tau & K2 & II & 2.6 & 0.65 & 1.53 & 20$^{\dagger}$ & ---  & -2.86 \\

IP~Tau & M0.6 & II & 0.75 & -0.41 & 0.59 & 33$^{\dagger}$ & --- & ---  \\

IT~Tau & K6 & II & 3.1 & -0.01 & 0.76 & 42 & AJ14 & -2.77  \\

TW~Hya & M0.5 & T & 0.00 & -0.72 & 0.69 & 6 & R12 & -3.85   \\

UX~Tau A  & K0 & T & 0.65 & 0.22 & 1.51 & 35 & AN11 & ---  \\

VY~Tau  & M1.5 & II/III & 0.6 & -0.41 & 0.47 & --- & --- & ---  \\

V710~Tau & M3.3 & II & 0.8 & -0.43 & 0.30 & 44 & AJ14 & -3.45 \\

V773~Tau & K4 & II/III & 0.95 & 0.48 & 0.98 & 34$^{\dagger}$ & ---  & -2.61 \\

V836~Tau & M0.8 & II & 0.6 & -0.52 & 0.58 & 48 & P14 & --- \\
\\
\tableline
\end{tabular}
%\begin{tablenotes}
%% Any table notes must follow the \end{tabular} command.
\tablerefs{\citealt{akjen2014} (AJ14); \citealt{andrews2011} (AN11); \citealt{AW07} (AW07); \citealt{cox2013} (C13); \citealt{gudel2007}; \citealt{guilloteau2011} (G11);
\citealt{HH14}; \citealt{isella09} (I09); \citealt{keane2014}; \citealt{pietu2014} (P14); \citealt{rosenfeld2012} (R12); \citealt{scheg09} (S09)}    
\tablecomments{SED entries are taken from \cite{pascucci2015}. SpT, Av, $L_{\ast}$, and $M_{\ast}$ values were taken from \cite{HH14}. 
A $^{\dagger}$ indicates sources where inclinations were derived in this work as described in Section \ref{sect:inclination}.
L$_X$ values were taken from \cite{gudel2007} who used DEM fits. If the L$_X$ value is marked with an $^{\ast}$ then it was derived using 1-T, 2-T fits instead. The one exception is TW Hya, in this case the L$_X$ value was taken from \cite{SS2004}.}
%\tablenotetext{a}{$^{\dagger}$ Sources where inclinations were derived in this work as described in Section 4.3}
%\tablenotetext{b}{SpT, Av, $L_{\ast}$, and $M_{\ast}$ values were taken from Herczeg \& Hillenbrand (2014). L$_X$ values were taken from G{\"u}del et al. (2007) who used DEM fits. If the L$_X$ value is marked with an $^{\ast}$ then it was derived using 1-T, 2-T fits instead.}
%\tablenotetext{c}{SED entries are taken from Pascucci et al. (2015)}
\end{center}
\end{table}

\begin{table}

\begin{center}
\caption{Forbidden Line Equivalent Widths  \label{tab:OI_5577_6300}}
\vspace{0.3cm}
\begin{tabular}{lccccc}
\tableline
\\
Source & Photospheric Standard & {$r_{6300}$} & EW (\oi{} 6300) & EW (\oi{} 5577) & EW (\sii{} 6731) \\
&&&[{\AA}]&[{\AA}]&[{\AA}]\\
\\
\tableline
\\
AA~Tau & V819 Tau & 0 & 0.74 $\pm$ 0.03 & 0.21 $\pm$ 0.03 & $<$0.005 \\

BP~Tau & V819 Tau & 0.6 & 0.35  $\pm$ 0.05 & 0.20 $\pm$ 0.04 & $<$0.005  \\

CI~Tau & HBC 427 & 0.6 & 0.16 $\pm$ 0.06 & $<$0.016 & $<$0.130 \\

CoKu~Tau 4 & V1321~Tau & 0 & 0.13 $\pm$ 0.02 & $<$0.006 & $<$0.004 \\

CW~Tau & HBC 427 & 1.5 & 1.70  $\pm$ 0.02 & 0.35 $\pm$ 0.05 & 0.18 $\pm$ 0.07 \\

CY~Tau & GL 15a & 0.6 & 0.42  $\pm$ 0.06 & 0.18 $\pm$ 0.07 & $<$0.006 \\

DF~Tau & V819 Tau & 1.6 & 1.51  $\pm$ 0.12 & 0.15 $\pm$ 0.03 & $<$0.004 \\

DG~Tau & V819 Tau & 1.0 & 9.83 $\pm$ 0.19 & 0.62 $\pm$ 0.04 & 1.62 $\pm$ 0.04 \\

DH~Tau & V1321~Tau & 0.5 & 0.75 $\pm$ 0.04 & 0.22 $\pm$ 0.02 & $<$0.007 \\

DK~Tau & V819 Tau & 0.4 & 1.33  $\pm$ 0.09 & 0.31 $\pm$ 0.05 & 0.10 $\pm$ 0.06\\

DL~Tau & HBC 427 & 1.0 & 1.11  $\pm$ 0.06 & $<$0.018 & 0.16 $\pm$ 0.06 \\

DM~Tau & V1321~Tau & 0.1 & 0.56 $\pm$ 0.09 & $<$0.017 & $<$0.026 \\

DN~Tau & V819 Tau & 0 & $<$0.007 & $<$0.007 & $<$0.007 \\

DO~Tau & V819 Tau & 1.5 & 5.24  $\pm$0.06 & 0.49 $\pm$ 0.02 & 0.49 $\pm$ 0.04  \\

DR~Tau & V819 Tau, No Correction & 5.6 & 0.22 $\pm$ 0.05 & $<$0.009 & $<$0.008 \\

DS~Tau & V819 Tau & 0.6 & 0.20 $\pm$ 0.05 & 0.11 $\pm$ 0.02 & $<$0.005 \\

FM~Tau &  V1321 Tau, No Correction & 3.5 & 0.70 $\pm$ 0.04 & 0.28 $\pm$ 0.04 & $<$0.005 \\

FZ~Tau & V1321~Tau & 1.6 & 0.78 $\pm$ 0.04 & 0.12 $\pm$ 0.03 & 0.15 $\pm$ 0.02 \\

GH~Tau & V1321~Tau & 0.1 & 0.33 $\pm$ 0.03 & $<$0.004 & $<$0.004 \\

GI~Tau & V819 Tau & 0.5 & 1.30 $\pm$ 0.07 & 0.36 $\pm$ 0.07 & 0.19 $\pm$ 0.03 \\

GK~Tau & V819 Tau & 0.1 & 0.28  $\pm$ 0.02 & $<$0.004 & $<$0.005 \\

GM~Aur & HBC 427 & 0.6 & 0.35 $\pm$ 0.03 & $<$0.009 & $<$0.011 \\

GO~Tau & V1321 Tau & 0.3 & 0.16 $\pm$ 0.02 & 0.11 $\pm$ 0.04 & $<$0.005 \\

HN~Tau & V819 Tau, No Correction & 1.3 & 5.77 $\pm$ 0.08 & $<$0.025 & 1.28 $\pm$ 0.30  \\

HQ~Tau & HBC 427 & 0.7 & 0.37 $\pm$ 0.05 & $<$0.007 & $<$0.006  \\

IP~Tau & V819 Tau & 0.4, 0.3 & 0.69 $\pm$ 0.07, 0.52 $\pm$ 0.03 & $<$0.008, $<$0.005 & $<$0.010, $<$0.009 \\

IT~Tau & HBC 427 & 0 & 0.30 $\pm$ 0.06 & $<$0.004 & $<$0.004 \\

TW~Hya & V819 Tau & 0.5 &  0.50 $\pm$ 0.04 & 0.09 $\pm$ 0.02 & $<$0.012 \\

UX~Tau A & HR 8832, HBC 427 & 0.2, 0.2 & 0.22 $\pm$ 0.02, 0.09 $\pm$ 0.02 & $<$0.006, $<$0.006 & $<$0.003, $<$0.003  \\

VY~Tau & V1321 Tau & 0 & $<$0.010 & $<$0.008 & $<$0.008 \\

V710~Tau & V1321~Tau & 0.1  & $<$0.008 & $<$0.007 & $<$0.011 \\

V773~Tau & HBC 427 & 0.1 & 0.54 $\pm$ 0.03 & $<$0.003 & $<$0.003 \\

V836~Tau & V819 Tau & 0.2 & 0.58 $\pm$ 0.06 & 0.22 $\pm$ 0.09 & $<$0.007  \\
\\

\tableline
\end{tabular}
%% Any table notes must follow the \end{tabular} command.
\tablecomments{$^{\ast}$ The 3-${\sigma}$ upper limits were computed assuming an unresolved line 
of FWHM 6.6 km/s.}
\end{center}
\end{table}

\pagebreak

\begin{table}

\begin{center}
\begin{threeparttable}
\caption{Accretion Properties \color{black}\label{tab:acc_properties}}
\label{tab:tablenotes}
\vspace{0.3cm}
\begin{tabular}{lccccccc}
\tableline
Source  & EW (H$\alpha$) & log $L_{H\alpha}$ & log $L_{acc}$ (H$\alpha$) & log $\dot M_{acc}$ (H$\alpha$) & log $L_{OI}$ (total)  & log $L_{OI}$ (total LVC) & log $L_{OI}$ (LVC NC)\\
             & [{\AA}]  & [L$_\odot$] & [L$_\odot$]  & [M$_\odot$/year] & [L$_\odot$]  &[L$_\odot$] & [L$_\odot$] \\
\tableline
\\
AA~Tau &10.5 & -3.48 & -2.40 & -9.34 & -4.74 & -4.82 & -5.33 \\

BP~Tau  & 97.0 & -2.34 & -1.12 & -8.17 & -4.89 & -5.00 & -5.83 \\

CI~Tau  & 84.5 & -2.11 & -0.87 & -8.03 & -4.87 & -4.90 & ---\\

CoKu~Tau 4 & 1.16 & -4.62 & -3.67 & -10.7 & -5.68 & -5.72 & -5.72 \\

CW~Tau  & 110 & -1.91 & -0.64 & -8.01 & -3.74 & -3.91 & -4.46\\

CY~Tau  &114 & -2.52 & -1.33 & -8.20 &  -5.09 &  -5.09 & ---\\

DF~Tau  & 46.7 & -2.22 & -0.98 & -7.46 & -3.94 & -4.30 & -5.27 \\

DG~Tau & 63.5 & -2.26 & -1.03 & -8.12 & -3.16 & -4.05 & -4.05  \\

DH~Tau  & 34.5 & -3.14 & -2.02 & -8.93 & -4.99 & -5.01 & -5.50 \\

DK~Tau & 33.5 & -2.67 & -1.50 & -8.53 & -4.17 & -4.28 & -5.69 \\

DL~Tau  & 92.6 & -2.10 & -0.85 & -8.06 & -4.06 & -4.64 & --- \\

DM~Tau  & 104 & -3.14 & -2.02 & -8.94 & -5.65 & -5.74 & -5.74 \\

DN~Tau  & 13.5 & -3.06 & -1.93 & -8.76 & $<$ -5.87 & $<$ -5.87 & ---\\

DO~Tau &136 & -2.23 & -1.00 & -8.21 & -3.75 & -4.37 & --- \\

DR~Tau  & 43.9 & -2.10 & -0.85 & -8.10 & -4.45 & -4.56 & -5.01 \\

DS~Tau  & 49.4 & -2.97 & -1.82 & -9.03 & -5.47 & -5.47 & -6.17 \\
          
FM~Tau  & 78.3 & -3.19 & -2.07 & -8.66 & -5.54 & -5.56 & -5.92   \\

FZ~Tau  & 176 & -1.95 & -0.68 & -7.77 & -4.41 & -4.70 & -5.22 \\

GH~Tau  & 11.8 & -3.28 & -2.18 & -8.80 & -5.02 & -5.04 & ---\\

GI~Tau  & 22.2 & -2.85 & -1.69 & -8.62 & -4.19 & -4.38 & -4.87 \\

GK~Tau  & 14.9 & -2.87 & -1.71 & -8.58 & -4.66 & -4.77 & -5.15\\

GM~Aur  & 92.9 & -2.19 & -0.95 & -8.15 & -4.66 & -4.65 & -4.88\\

GO~Tau  & 45.9 & -3.12 & -2.00 & -8.92 & -5.76 & -5.73 & --- \\

HN~Tau  & 89.2 & -2.47 & -1.27 & -8.69 & -3.67 & -4.33 & --- \\

HQ~Tau  & 2.22 & -2.76 & -1.60 & -8.68 & -3.54 & -3.71 & --- \\

IP~Tau & 10.4  & -3.39 & -2.29 & -9.30 & -4.70 & -5.50 & --- \\

IT~Tau  &16.9 & -2.83 & -1.67 & -8.63 & -4.63 & -4.62& --- \\

TW~Hya  & 230 & -2.32 & -1.10 & -8.34 & -5.09 & -5.13 & -5.13 \\

UX~Tau A  & 10.2 & -2.70 & -1.52 & -8.84 & -4.75 & -4.70 & -4.70 \\

VY~Tau & 4.24 & -3.96 & -2.93 & -9.81 & $<$ -6.13 & $<$ -6.13 & --- \\

V710~Tau  & 1.87 & -4.49 & -3.52 & -10.1 &  $<$ -6.52 & $<$ -6.52 & --- \\

V773~Tau  & 3.02 & -3.02 & -1.88 & -8.79 & -3.79 & -3.86 & --- \\

V836~Tau  & 10.4 & -3.58 & -2.51 & -9.56 & -4.94 & -4.91 & --- \\
\\

\tableline
\end{tabular}
\tablecomments{Values for IP Tau are from 2006, for UX Tau A from 2012 as described in Section \ref{sect:fitting}}
%% Any table notes must follow the \end{tabular} command.
\end{threeparttable}
\end{center}
\end{table}

\pagebreak

\begin{deluxetable}{l ccccccccc}
%\tabletypesize{\tiny}
\tablewidth{0pt}
\tablecaption{High Velocity Component Fit Parameters \label{tab:HVC_params}}	
\tablehead{ \colhead{ } &\colhead{ } & \multicolumn{2}{c}{{\bf HVC blue 1} } & \multicolumn{2}{c}{{\bf HVC blue 2} } & \multicolumn{2}{c}{{\bf HVC red 1} } & \multicolumn{2}{c}{{\bf HVC red 2} } \\
\\
 	\colhead{Source} & \colhead{Line}  & \colhead{$v_c$}  & \colhead{FWHM} & \colhead{$v_c$}  & \colhead{FWHM} &  \colhead{$v_c$} & \colhead{FWHM} & \colhead{$v_c$}  & \colhead{FWHM} \\
         \colhead{ } & \colhead{[\AA]} & \colhead{[km/s]} & \colhead{[km/s]} & \colhead{[km/s]} & \colhead{[km/s]} & \colhead{[km/s]} & \colhead{[km/s]} & \colhead{[km/s]} & \colhead{[km/s]} \\
         }

\startdata
\\
AA Tau     	& 	\oi{} $\lambda$ 6300 &	-33 &	25 &	--- &	--- &	--- &	--- & --- &	---\\
\\
BP Tau &  	\oi{} $\lambda$ 6300 &	--- &	--- &	--- &	--- &	124 & 46	& ---& ---\\
\\
CW Tau &  	\oi{} $\lambda$ 6300 &	-112 &	89 &	---&	--- &	--- &	--- & --- & ---\\
&	\oi{} $\lambda$ 5577 &	-112 &	89 &	--- &	--- &	--- &	---	& --- & ---\\
&	\sii{} $\lambda$ 6731 &	-116 &	38 &	--- &	--- &	--- &	--- & --- & ---	\\
\\
DF Tau &  	\oi{} $\lambda$ 6300 &	-114 &	56 &	--- &	--- &	82 &	102	& --- & ---\\
\\
DG Tau &  	\oi{} $\lambda$ 6300 &	-144 &	118 &	-38 &	68 &	--- &	--- & --- & ---	\\
&	\oi{} $\lambda$ 5577 &	-144 &	118 &	-38 &	68 &	--- &	--- & --- & ---	\\
&	\sii{} $\lambda$ 6731 &	-103 &	171 &	-34 &	46 &	--- &	--- & --- & ---	\\
\\
DK Tau & 	\oi{} $\lambda$ 6300 &	-126 &	37 &	-42 &	51 &	--- &	--- & --- & ---	\\
&	\sii{} $\lambda$ 6731 &	-56 &	38 &	---- &	--- &	--- &	--- & --- & ---	\\
\\
DL Tau &  	\oi{} $\lambda$ 6300 &	-138 &	115 &	--- &	--- &	--- &	---& --- & ---	\\
&	\sii{} $\lambda$ 6731 &	-133 &	70 &	--- &	--- &	--- &	--- & --- & ---	\\
\\
DO Tau &  	\oi{} $\lambda$ 6300 &	-97 &	51 &	-95 &	19 &	--- &	--- & --- & ---	\\
&	\oi{} $\lambda$ 5577 &	-85 &	74 &	--- &	--- &	--- &	--- & --- & ---	\\
&	\sii{} $\lambda$ 6731 &	-93 &	27 &	--- &	--- &	--- &	--- & --- & ---	\\
\\
FZ Tau &  	\oi{} $\lambda$ 6300 &	-124 &	32 &	--- &	--- &	76 &	70 & 125 & 29	\\
&	\sii{} $\lambda$ 6731 &	--- &	--- &	--- &	--- &	130 & 23 & --- & ---	\\
\\
GI Tau &  	\oi{} $\lambda$ 6300 &	-71 &	49 &	--- &	--- &	40 &	23 & --- & ---	\\
&	\sii{} $\lambda$ 6731 &	-61 &	31 &	--- &	--- &	39 &	22 & --- & ---	\\
\\
HN Tau & 	\oi{} $\lambda$ 6300 &	-66 &	130 &	--- &	--- &	--- &	--- & --- & ---	\\
&	\sii{} $\lambda$ 6731 &	-84 &	113 &	--- &	--- &	--- &	--- & --- & ---\\
\\
HQ Tau & 	\oi{} $\lambda$ 6300 &	-41 &	27 &	--- &	--- &	--- &	--- & --- & ---	\\
\\
IP Tau &	\oi{} $\lambda$ 6300 &	-38 &	124 &	--- &	--- &	--- &	--- & --- & ---	\\
\enddata
\tablecomments{Centroid velocities and FWHMs of Gaussian fits for high velocity components. These have been separated into blue-shifted HVC and red-shifted HVC, both of which there are occasionally two.}
\end{deluxetable}

\pagebreak

\begin{deluxetable}{l ccc ccc ccc ccc c}
%\tabletypesize{\tiny}
\tablewidth{0pt}
\tablecaption{\oi{} LVC parameters \label{tab:2_gaus_params}}																

%\begin{tabular}

\tablehead{ \colhead{ } & \multicolumn{6}{c}{6300 {\AA}} & \multicolumn{6}{c}{5577 {\AA}}\\
\\
		  \colhead{Source} & \multicolumn{3}{c}{Narrow Component} & \multicolumn{3}{c}{Broad Component} & \multicolumn{3}{c}{Narrow Component} & \multicolumn{3}{c}{Broad Component}  \\
		  \\
		  \colhead{ } & \colhead{FWHM} & \colhead{$v_c$} & \colhead{EW} & \colhead{FWHM} & \colhead{$v_c$} & \colhead{EW} & \colhead{FWHM} & \colhead{$v_c$} & \colhead{EW} & \colhead{FWHM} & \colhead{$v_c$} & \colhead{EW} \\
		   \colhead{ } & \colhead{[km/s]} & \colhead{[km/s]} & \colhead{[\AA]} & \colhead{[km/s]} & \colhead{[km/s]} & \colhead{[\AA]} & \colhead{[km/s]} & \colhead{[km/s]} & \colhead{[\AA]} & \colhead{[km/s]} & \colhead{[km/s]} & \colhead{[\AA]} & \colhead{ } \\
		   }

\startdata
\\
AA Tau	&	27	&	-0.19	&	0.19	&	111	&	4.67	&	0.42	&	27	&	-0.19	&	0.02	&	111	&	4.67	&	 0.18	\\
BP Tau	&	16	&	0.11	&	0.04	&	87	&	5.65	&	0.23	&	16	&	0.11	&	0.01	&	87	&	5.65	&	0.15	\\
CI Tau	&	---	&	---	&	---	&	66	&	-8.70	&	0.15	&	---	&	---	&	---	&	---	&	---	&	---	\\
CoKu Tau 4	&	18	&	-2.90	&	0.12	&	---	&	---	&	---	&	---	&	---	&	---	&	---	&	---	&	---	\\
CW Tau	&	24	&	-2.63 	&	0.32	&	76	&	-2.70	&	0.83	&	23	&	0.92 	&	0.05	&	73	& 1.61	&	0.25	\\
CY Tau	&	---	&	---	&	---	&	48	&	-1.02	&	 0.47	&	---	&	---	&	---	&	61	&	-0.98	&	0.22	\\
DF Tau	&	14	&	-1.34	&	0.07	&	57	&	-25.7	&	0.58	&	14	&	0.93	&	0.03	&	49	&	-14.0 	&	0.09	\\
DG Tau	&	18	&	-12.3	&	1.25	&	---	&	---	&	---	&	19	&	-10.6	&	0.26	&	---	&	---	&	---	\\
DH Tau	&	27	&	-1.75	&	0.23	&	119	&	-1.31	&	0.48	&	27	&	-1.75	&	0.08	&	119	&	-1.31	&	0.15	\\
DK Tau	&	12	&	-1.90	&	0.04	&	137	&	-11.8	&	0.98	&	12	&	-1.90	&	0	&	137	&	-11.8	&	0.30	\\
DL Tau	&	---	&	---	&	---	&	44	&	-11.3	&	0.29	&	---	&	---	&	---	&	---	&	---	&	---	\\
DM Tau	&	23	&	1.10	&	0.45	&	---	&	---	&	---	&	---	&	---	&	---	&	---	&	---	&	---	\\
DO Tau	&	---	&	---	&	---	&	54	&	-24.9	&	1.24	&	---	&	---	&	---	&	58	&	-17.2	&	0.31	\\
DR Tau	&	12	&	-0.11	&	0.06	&	49	&	-11.6	&	0.11	&	---	&	---	&	---	&	---	&	---	&	---	\\
DS Tau	&	39	&	-13.8	&	0.04	&	115	&	1.10	&	0.16	&	39	&	-13.8	&	0.02	&	115	&	1.10	&	0.09	\\
FM Tau	&	30	&	-1.82	&	0.29	&	140	&	6.15	&	0.38	&	30	&	-1.82	&	0.12	&	140	&	6.15	&	0.13	\\
FZ Tau	&	26	&	-1.28	&	0.12	&	95	&	-14.6	&	0.28	&	26	&	-1.28	&	0.06 &	95	&	-14.6	&	0.09	\\
GH Tau	&	---	&	---	&	---	&	84	&	-0.66	&	0.31	&	---	&	---	&	---	&	---	&	---	&	---	\\
GI Tau	&	33	&	-2.85	&	0.27	&	97	&	5.61	&	0.57	&	33	&	-2.85	&	0.09	&	97	&	5.61	&	0.23	\\
GK Tau	&	39	&	-3.50	&	0.09	&	116	&	0.40	&	0.13	&	---	&	---	&	---	&	---	&	---	&	---	\\
GM Aur	&	28	&	-1.26	&	0.21	&	73	&	-4.11	&	0.15	&	---	&	---	&	---	&	---	&	---	&	---	\\
GO Tau	&	---	&	---	&	---	&	106	&	3.84	&	0.17	&	---	&	---	&	---	&	106	&	3.84	&	0.11	\\
HN Tau	&	---	&	---	&	---	&	48	&	-1.18	&	1.27	&	---	&	---	&	---	&	---	&	---	&	---	\\
HQ Tau	&	---	&	---	&	---	&	44	&	-3.47	&	0.25	&	---	&	---	&	---	&	---	&	---	&	---	\\
IP Tau	&	---	&	---	&	---	&	44	&	-26.9	&	0.11	&	---	&	---	&	---	&	---	&	---	&	---	\\
IT Tau	&	---	&	---	&	---	&	71	&	-1.54	&	0.31	&	---	&	---	&	---	&	---	&	---	&	---	\\
TW Hya	&	12	&	-0.78	&	0.46	&	---	&	---	&	---	&	11	&	0.37	&	0.07	&	---	&	---	&	---	\\
UX Tau A	&	29	&	1.50	&	0.10	&	---	&	---	&	---	&	---	&	---	&	---	&	---	&	---	&	---	\\
V773 Tau	&	---	&	---	&	---	&	59	&	3.91	&	0.46	&	---	&	---	&	---	&	---	&	---	&	---	\\
V836 Tau	&	---	&	---	&	---	&	67	&	-5.23	&	0.63	&	---	&	---	&	---	&	67	&	-5.23	&	0.12	\\
\enddata

%\end{tabular}
\end{deluxetable}

\begin{table}

\begin{center}
\caption{\sii{} 6731 {\AA} LVC Parameters \label{tab:SII_properties}}

\begin{tabular}{lccc}
\tableline \\
& FWHM  & $v_c$  & EW \\
& [km/s] & [km/s] & [\AA]\\
\\
\tableline \\
CW~Tau &  24 & -5.51 & 0.10 \\
\\
HN~Tau &  55 & -9.43 & 0.45 \\
\\
\tableline
\end{tabular}
\end{center}
\end{table}

\begin{deluxetable}{ccccccccccc}
\vspace{0.3cm}

\tablecaption{Results from Modeling the 9 Sources with Bound LVC NC \label{tab:model_results}}

\tablehead{\colhead{Source} & \colhead{$M_{\ast}$} & \colhead{$R_{in}$} & \colhead{$R_{out}$} & \colhead{$R_{80\%}$} & \colhead {$\alpha$}\\
			\colhead{} & \colhead{[M$_\odot$]} & \colhead{[AU]} & \colhead{[AU]} & \colhead{[AU]} 
			}
			
\startdata
 \multicolumn{6}{l}{\bf{Transition Disks}} \\
 \\
 
UX Tau A & 1.51 & 0.50 & 18.0 & 1.83 & 2.22 \\
\\
GM Aur & 0.88 & 0.07 & 10.9 & 0.40 & 1.89 \\
\\
TW Hya & 0.69 & 0.05 & 11.1 & 0.18 & 2.23 \\
\\
DM Tau & 0.35 & 0.02 & 3.00 & 0.35 & 1.47 \\
\\
\hline
\\
 \multicolumn{6}{l}{\bf{Full Disks}} \\
 \\

 DR Tau & 0.90 & 0.46 & 18.2 & 3.39 & 1.65 \\
 \\
 BP Tau & 0.62 & 0.08 & 9.20 & 3.90 & 0.94 \\
 \\
 FZ Tau & 0.63 & 0.17 & 19.7 & 1.50  & 1.66 \\
 \\
 AA Tau & 0.57 & 0.22 & 11.4 & 2.04 & 1.56 \\
 \\
 DF Tau & 0.32 & 0.20 & 8.61 & 6.92 & 0.00 \\
\enddata

\end{deluxetable}

\begin{table}

\begin{center}
\caption{Slit and disk position angles.  \label{tab:PA_table}}
\vspace{0.3cm}
\begin{tabular}{lccccc}
\tableline
\\
Source & Slit PA & Disk PA & REF \\
&[$^{\circ}$]&[$^{\circ}$]& (Disk PA)\\
\\
\hline
\\
AA Tau & 105 & 97 & C13 \\
BP Tau & 246 & 107 & G11 \\
CI Tau & 84.8 & 285 & G11 \\
CoKu Tau 4 & 223 & --- & --- \\
CW Tau & 260 & 332 & P14 \\
CY Tau & 251 & 63 & G11 \\
DF Tau & 102 & --- & --- \\
DG Tau & 102 & 43 & G11 \\
DK Tau & 190 & 15 & AJ14 \\
DL Tau & 84.1 & 141 & G11 \\
DM Tau & 95.6 & 155 & AN11 \\
DN Tau & 236 & 86 & I09 \\
DO Tau & 270 & --- & --- \\
DR Tau & 78.8 & 108 & I09 \\
DS Tau & 252 & 165 & AJ14 \\
FM Tau & 260 & 83 & P14 \\
FZ Tau & 256 & --- & --- \\
GH Tau & 91.7 & --- & --- \\
GI Tau & 260 & --- & --- \\
GK Tau & 263 & 93 & AJ14 \\
GM Aur & 92.1 & 144 & G11 \\
GO Tau & 99.2 & 0 & AW07 \\
HN Tau & 79.5 & 65 & AJ14 \\
HQ Tau & 273 & --- & --- \\
IP Tau & 112 & --- & --- \\
IT Tau & 254 & 106 & AJ14 \\
TW Hya & 337 & 332 & PD08 \\
UX Tau A & 278 & 176 & AN11 \\
VY Tau & 274 & --- & --- \\
V710 Tau & 96.4 & 82 & AJ14 \\
V773 Tau & 270 & --- & --- \\
V836 Tau & 92.5 & -122 & P14 \\
\\

%\tableline
\end{tabular}
%\begin{tablenotes}
%% Any table notes must follow the \end{tabular} command.
\tablerefs{\citealt{akjen2014} (AJ14); \citealt{andrews2011} (AN11); \citealt{AW07} (AW07); \citealt{cox2013} (C13); \citealt{guilloteau2011} (G11);
 \citealt{isella09} (I09); \citealt{pietu2014} (P14); \citealt{PO08} (PD08) }    
 %\tablecomments{Values for Disk Position (Disk PA) and slit orientation (SKY PA) are presented here for the 22 sources with spatially resolved disk inclinations from the literature.}

%\tablenotetext{a}{$^{\dagger}$ Sources where inclinations were derived in this work as described in Section 4.3}
%\tablenotetext{b}{SpT, Av, $L_{\ast}$, and $M_{\ast}$ values were taken from Herczeg \& Hillenbrand (2014). L$_X$ values were taken from G{\"u}del et al. (2007) who used DEM fits. If the L$_X$ value is marked with an $^{\ast}$ then it was derived using 1-T, 2-T fits instead.}
%\tablenotetext{c}{SED entries are taken from Pascucci et al. (2015)}
\end{center}
\end{table}

\end{document}